\documentclass[paper]{JHEP3}
\usepackage{epsfig}
\usepackage{cancel}

\def\beq{\begin{equation}}
\def\beqn{\begin{eqnarray}}
\def\eeq{\end{equation}}
\def\eeqn{\end{eqnarray}}
\def\abs#1{\left|#1\right|}
\def\ket#1{|#1\rangle}
\def\bra#1{\langle #1|}
\def\brat#1{\langle #1}
\def\eik#1#2{\Big[#1,#2\Big]}

\newcommand\sss{\scriptscriptstyle}
\newcommand\mydot{\!\cdot\!}

\newcommand\half{\frac{1}{2}}

\newcommand\gs{g_{\sss S}}

\newcommand\blambda{\bar{\lambda}}
\newcommand\amp{{\cal A}}

\newcommand\ampn{\amp^{(n)}}
\newcommand\ampnpo{\amp^{(n+1)}}
\newcommand\ampQGR{\amp^{(2q;n+1)}}
\newcommand\ampQGBgg{\amp^{(2q;n)}}
\newcommand\ampQGBqq{\amp^{(2(q-1);n+2)}}
\newcommand\RED{EXT}
\newcommand\ampR{{\cal A}_{\rm\sss \RED}}
\newcommand\ampRnpo{\ampR^{(n)}}
\newcommand\ampRQGR{\ampR^{(2q;n)}}
\newcommand\ampRQGRqq{\ampR^{(2(q-1);n+2)}}
\newcommand\ampCS{\widehat{\cal A}}

\newcommand\ampCSn{\ampCS^{(n)}}
\newcommand\ampCSnpo{\ampCS^{(n+1)}}
\newcommand\ampCSQGR{\ampCS^{(2q;n+1)}}
\newcommand\ampCSQGBgg{\ampCS^{(2q;n)}}
\newcommand\ampCSQGBqq{\ampCS^{(2(q-1);n+2)}}
\newcommand\ampsq{{\cal M}}
\newcommand\ampsqL{{\cal M}_{\rm\sss L}}
\newcommand\ampsqS{{\cal M}_{\rm\sss SOFT}}
\newcommand\ampsqC{{\cal M}_{\rm\sss COLL}}
\newcommand\ampsqSC{{\cal M}_{\rm\sss SC}}

\newcommand\ampsqn{\ampsq^{(n)}}
\newcommand\ampsqnpo{\ampsq^{(n+1)}}
\newcommand\ampsqnpoS{\ampsq^{(n+1)}_{\rm\sss SOFT}}
\newcommand\ampsqnpoC{\ampsq^{(n+1)}_{\rm\sss COLL}}
\newcommand\ampsqnpoSC{\ampsq^{(n+1)}_{\rm\sss SC}}
\newcommand\ampsqQGR{\ampsq^{(2q;n+1)}}
\newcommand\ampsqQGRS{\ampsq^{(2q;n+1)}_{\rm\sss SOFT}}
\newcommand\ampsqQGRC{\ampsq^{(2q;n+1)}_{\rm\sss COLL}}

\newcommand\ampsqQGBgg{\ampsq^{(2q;n)}}
\newcommand\ampsqQGBqq{\ampsq^{(2(q-1);n+2)}}
\newcommand\wampsq{\ampsq}
\newcommand\wampsqn{\wampsq^{(n)}}
\newcommand\wampsqQGBgg{\wampsq^{(2q;n)}}
\newcommand\wampsqQGBqq{\wampsq^{(2(q-1);n+2)}}
\newcommand\ampsqR{{\cal M}_{\rm\sss \RED}}

\newcommand\ampsqRn{\ampsqR^{(n)}}

\newcommand\ampsqRQGBgg{\ampsqR^{(2q;n)}}
\newcommand\wampsqRpm{\ampsq_{\rm\sss \RED+-}}
\newcommand\wampsqRpmn{\wampsqRpm^{(n)}}
\newcommand\wampsqRpmQGBgg{\wampsqRpm^{(2q;n)}}
\newcommand\wampsqRmp{\ampsq_{\rm\sss \RED-+}}
\newcommand\wampsqRmpn{\wampsqRmp^{(n)}}
\newcommand\wampsqRmpQGBgg{\wampsqRmp^{(2q;n)}}
\newcommand\wampsqRll{\ampsq_{\rm\sss \RED\lambda\blambda}}
\newcommand\wampsqRlln{\wampsqRll^{(n)}}
\newcommand\wampsqRllQGBgg{\wampsqRll^{(2q;n)}}
\newcommand\bampsq{\overline{\cal M}}
\newcommand\bampsqn{\bampsq^{(n)}}
\newcommand\zampsqn{\bampsqn}
\newcommand\zampsqQGBggs{\bampsq^{(2q;n)}_s}
\newcommand\zampsqQGBggk{\bampsq^{(2q;n)}_k}
\newcommand\zampsqQGBqqqq{\bampsq^{(2(q-1);n+2)}_{qq}}
\newcommand\wzampsqn{\zampsqn}
\newcommand\wzampsqQGBggsmp{\bampsq^{(2q;n)}_{s,-+}}
\newcommand\wzampsqQGBqqqqmp{\bampsq^{(2(q-1);n+2)}_{qq,-+}}
\newcommand\wzampsqQGBqqqqpm{\bampsq^{(2(q-1);n+2)}_{qq,+-}}
\newcommand\wzampsqQGBqqqqll{\bampsq^{(2(q-1);n+2)}_{qq,\lambda\blambda}}
\newcommand\ampX{{\cal X}}
\newcommand\ampXCS{\widehat{\cal X}}
\newcommand\ampY{{\cal Y}}
\newcommand\ampYCS{\widehat{\cal Y}}
\newcommand\ampZ{{\cal Z}}
\newcommand\ampZCS{\widehat{\cal Z}}
\newcommand\seta{\{a_i\}}

\newcommand\setan{\{a_i\}_{i=1}^n}
\newcommand\setanpo{\{a_i\}_{i=1}^{n+1}}
\newcommand\setaQGR{\{a_i\}_{i=-2q}^{n+1}}
\newcommand\setaQGBgg{\{a_i\}_{i=-2q}^n}
\newcommand\setc{\{c_i\}}
\newcommand\setcQGBgg{\{c_i\}_{i=-2q}^n}
\newcommand\knpo{k_{n+1}}
\newcommand\kmq{k_{-q}}
\newcommand\kmtq{k_{-2q}}
\newcommand\anpo{a_{n+1}}
\newcommand\anpt{a_{n+2}}
\newcommand\amtq{a_{-2q}}
\newcommand\amoq{a_{-q}}
\newcommand\amor{a_{-r}}
\newcommand\amtqpo{a_{-2q+1}}
\newcommand\sigmap{\sigma^\prime}
\newcommand\isigma{\sigma^{-1}}
\newcommand\isigmap{\sigma^{\prime^{-1}}}
\newcommand\Sigmap{\Sigma^\prime}
\newcommand\iSigma{\Sigma^{-1}}
\newcommand\iSigmap{\Sigma^{\prime^{-1}}}
\newcommand\gammap{\gamma^\prime}
\newcommand\igamma{\gamma^{-1}}
\newcommand\igammap{\gamma^{\prime^{-1}}}
\newcommand\Gammap{\Gamma^\prime}
\newcommand\iGamma{\Gamma^{-1}}
\newcommand\iGammap{\Gamma^{\prime^{-1}}}
\newcommand\gammaY{\gamma^{\sss (Y)}}
\newcommand\gammaZ{\gamma^{\sss (Z)}}
\newcommand\pp{p^\prime}
\newcommand\NC{N}
\newcommand\CA{C_{\sss A}}
\newcommand\CF{C_{\sss F}}
\newcommand\TF{T_{\sss F}}

\newcommand\xicut{\xi_{cut}}

\newcommand\deltaO{\delta_{\sss O}}

\newcommand\qb{\bar{q}}
\newcommand\ub{\bar{u}}
\newcommand\db{\bar{d}}
\newcommand\flowR{{\cal F}_{2q;n+1}}
\newcommand\flowBgg{{\cal F}_{2q;n}}
\newcommand\flowBqq{{\cal F}_{2(q-1);n+2}}
\newcommand\flowRJ{{\cal F}_{2q;n+1}^{(J)}}
\newcommand\flowRK{{\cal F}_{2q;n+1}^{(K)}}

\newcommand\flowRA{{\cal F}_{2q;n+1}^{(A)}}
\newcommand\flowRB{{\cal F}_{2q;n+1}^{(B)}}
\newcommand\ident{{\cal I}}
\newcommand\Mp{-\!p}
\newcommand\Mr{-\!r}
\newcommand\Mq{-\!q}
\newcommand\iJ{J^{-1}}
\newcommand\iK{K^{-1}}
\newcommand\iKi{\iK_i}
\newcommand\iKj{\iK_j}
\newcommand\eikint{{\cal E}}
\newcommand\Sfun{{\cal S}}
\newcommand\Sfunij{\Sfun_{ij}}

\preprint{
 CERN-PH-TH/2011-124
 }
\title{Colourful FKS subtraction}

\author{Stefano Frixione%
  \thanks{On leave of absence from INFN, Sezione di Genova, Italy.}\\
  PH Department, TH Unit, CERN, CH-1211 Geneva 23, Switzerland\\
  ITPP, EPFL, CH-1015 Lausanne, Switzerland\\
  E-mail: \email{Stefano.Frixione@cern.ch}
}
\abstract{
I formulate in a colour-friendly way the FKS method for the computation 
of QCD cross sections at the next-to-leading order accuracy. This is
achieved through the definition of subtraction terms for squared matrix
elements, constructed with single colour-dressed or pairs of colour-ordered 
amplitudes. The latter approach relies on the use of colour flows, 
is exact to all orders in $N$, and is thus particularly suited to 
being organized as a systematic expansion in $1/N$.
}

\begin{document}

\section{Introduction\label{sec:intro}}
The computation of amplitudes in QCD is a problem whose complexity
grows factorially with the number of particles even at tree level. 
Such complexity stems from the extremely large number of Feynman
diagrams that contribute to many-particle processes and which, 
apart from the inherent complication of the Lorentz structure,
induce a proliferation of mutually-independent colour factors,
that give rise to an involved colour algebra. Lorentz and colour
structures can be separated by expressing scattering amplitudes as 
sums of products of dual (or colour-ordered) amplitudes times colour 
factors (see e.g.~ref.~\cite{Mangano:1990by} and references therein). 
The problem of the efficient computation of dual amplitudes has 
attracted considerable attention, and nowadays several solutions
exist (Berends-Giele recursion relations~\cite{Berends:1987me}, CSW 
relations~\cite{Cachazo:2004kj}, BCF recursion relations~\cite{Britto:2004ap}).
In order to predict observable cross sections, however, the colour
algebra must be performed. Its factorial growth has been bypassed
by working in the colour-configuration space~\cite{Caravaglios:1998yr},
and by re-expressing it as an ordinary integral~\cite{Draggiotis:2002hm};
these approaches avoid the use of Feynman diagrams~\cite{Caravaglios:1995cd},
and are essentially equivalent to colour-dressed recursion 
relations~\cite{Duhr:2006iq}; by these means, the Feynman-diagram 
factorial complexity is reduced to an exponential one.

The problem posed by the factorially-growing complexity of the
colour algebra was simply irrelevant for the calculation of observables 
at the next-to-leading order (NLO) and beyond, because several other 
issues limited anyhow the applicability of NLO techniques to small-multiplicity
processes. This is not the case any longer, thanks to the progress recently
achieved in the automation of the two essential ingredients in an
NLO computation: the subtraction of real-emission
singularities~\cite{Gleisberg:2007md,Bevilacqua:2009zn,Frederix:2009yq,
Frederix:2010cj} (where use has been made of the universal subtraction 
formalisms known as FKS~\cite{Frixione:1995ms,Frixione:1997np} and
dipole~\cite{Catani:1996vz}), and the computation of one-loop matrix 
elements~\cite{Ellis:2008qc,vanHameren:2009dr,Berger:2010zx,
Mastrolia:2010nb,Hirschi:2011pa} (based on generalized 
unitarity~\cite{Bern:1994zx,Ellis:2007br,Ellis:2008ir} and integrand 
reduction~\cite{delAguila:2004nf,Ossola:2006us,Mastrolia:2010nb}
techniques). Therefore, the question becomes relevant of how to
best organize an NLO computation, in order to be able to exploit
the solutions that work well for tree-level matrix elements,
and to use new tree-level approaches which may become available in the
future. As far as the one-loop contribution is concerned, the problem
is conceptually analogous to that relevant to tree-level amplitudes (but
technically more complicated, see e.g.~ref.~\cite{Bern:1996je}), 
since it consists in finding the optimal representation of an amplitude
in terms of dual amplitudes and colour structures. On the other hand,
the subtraction of real-emission singularities is complicated by
the fact that it must be performed at the level of amplitudes squared,
while it is the amplitudes (not squared) that display the simplest
factorization properties in the soft and collinear regions (see 
e.g.~ref~\cite{Kosower:1997zr}). As a result, in the subtraction
terms defined in the FKS and dipole formalisms, the colour structure
is not factorized, since it appears in both the universal kernels
and the short-distance, process-dependent reduced matrix elements.
Such a convolution is avoided in the antenna subtraction
method~\cite{Giele:1991vf,Kosower:1997zr,Campbell:1998nn},
which is based on the use of the squares of, or the interferences
between, dual amplitudes as elementary quantities whose singularities
have to be subtracted.

The aim of this paper is that of formulating the FKS subtraction
in terms of colour-dressed and colour-ordered amplitudes.
This will allow one to use, with only a few trivial modifications,
the techniques developed at tree level to deal with the problem
of the factorially-growing complexity. In order to be more precise,
let me briefly digress and introduce the tenets of the FKS method.

The basic idea of the FKS subtraction formalism is that of treating
in an independent manner the singularities present in a multi-parton
matrix element squared. This is achieved by damping all singularities
except one soft and one collinear, which are pre-determined, and by
repeating this procedure for all singularities in turn. 
One introduces a set of functions $\Sfunij$ such that:
\beq
\sum_{ij}\Sfunij=1\,.
\label{sumSij}
\eeq
The sum in eq.~(\ref{sumSij}) can be thought of as extending to all
strongly-interacting pairs of particles, although in practice significant
simplifications are possible (see ref.~\cite{Frederix:2009yq} for an
exhaustive discussion). For a given $(i,j)$ pair, the function $\Sfunij$
is equal to zero in all soft and collinear limits, except those
associated with $k_i\to 0$, and with $k_i\!\parallel\! k_j$. Given a
real-emission matrix element squared $\ampsq$, one exploits
eq.~(\ref{sumSij}) by writing
\beq
\ampsq=\sum_{ij}\ampsq_{ij}\,,\;\;\;\;\;\;\;\;
\ampsq_{ij}=\Sfunij\ampsq\,.
\eeq
The matrix elements $\ampsq_{ij}$ are independent from each other and,
thanks to the properties of the $\Sfunij$ functions,
have one soft and one collinear singularity at most (depending
on the identities of particles $i$ and $j$). These singularities
are subtracted as follows: be $E_i$ and $\theta_{ij}$ the energy
of parton $i$ and the angle between partons $i$ and $j$, in the rest
frame of the incoming particles. The (divergent) integration of 
$\ampsq_{ij}$ over the phase-space $d\phi$ is replaced by its
(convergent) subtracted form:
\beq
\ampsq_{ij}\,d\phi\;\longrightarrow\;
\left(\frac{1}{E_i}\right)_+
\left(\frac{1}{1-\cos\theta_{ij}}\right)_+
\Big[E_i^2(1-\cos\theta_{ij})\ampsq_{ij}\Big]\,
\frac{d\phi}{E_i}\,.
\label{FKSsubtr}
\eeq
If one neglects, for the sake of this discussion, the factors
that multiply $\ampsq$ in the square brackets on the
r.h.s.~of eq.~(\ref{FKSsubtr}), one sees that the result of the
two plus prescriptions is that of constructing the linear combination:
\beq
\ampsq-\ampsqS-\ampsqC+\ampsqSC\,.
\label{subt}
\eeq
Here, the second, third, and fourth terms are the limits (in the sense
of asymptotic behaviour, as customary) of $\ampsq$ when $E_i\to 0$,
$\theta_{ij}\to 0$, and $(E_i,\theta_{ij})\to (0,0)$, respectively.

As was already mentioned, so far the FKS procedure has been formulated 
by assuming that all contributions to eq.~(\ref{subt}) be colour-summed. 
In the present paper, $\ampsq$ will be decomposed
into a sum of terms, each of which has an immediate interpretation
in terms of colours. Furthermore, this sum will have to commute 
with the subtraction procedure. In other words, each of
the terms in the sum must have well-defined soft, collinear, and
soft-collinear limits, so as for each such individual term a linear
combination identical to that of eq.~(\ref{subt}) can
be defined, which is finite locally in the phase space.
I shall consider two options: the sum over colour configurations,
and the sum over colour flows. In the former case, the relevant
matrix elements will be the colour-dressed amplitudes squared;
in the latter case, they will be the squares of, or the interferences
between, colour-ordered amplitudes.

This paper is organized as follows. In sections~\ref{sec:gluons}
and~\ref{sec:quarks} I shall carry out the programme described 
above for pure-gluon amplitudes and for generic quark-gluon
amplitudes respectively. The reader who is not interested in 
the technical details of the derivation may skip these two
sections, and go directly to section~\ref{sec:summ}, where I summarize 
the results for the limits of the matrix elements. In section~\ref{sec:Borns}
I consider the case of the contributions to the NLO cross section
which do not originate from the subtraction procedure of
eq.~(\ref{FKSsubtr}). Section~\ref{sec:disc} presents a discussion
on the findings of this paper and a very brief comparison with antenna 
and dipole methods, and section~\ref{sec:concl} reports my conclusions. 
The conventions adopted for the colour matrices are given in 
appendix~\ref{sec:conv}, while appendix~\ref{sec:appa} contains a 
few technicalities, whose role is important for the understanding 
of the derivations presented in the main text. Finally, in
appendix~\ref{sec:ex} some of the findings of this paper are
applied to a quark-gluon process.

\section{Gluon amplitudes\label{sec:gluons}}
Given that the aim of this paper is the re-organization of the 
subtraction procedure in a colour-friendly way, it is convenient
to consider all particles entering a hard scattering as outgoing,
in order to simplify the notation. The Born and
real-emission processes will therefore be
\beqn
0&\longrightarrow&n\,,
\label{ggBproc}
\\
0&\longrightarrow&n+1\,,
\label{ggRproc}
\eeqn
respectively. Furthermore, it is not 
restrictive to assign labels $n+1$ and $n$ to the FKS parton and
to its sister, respectively\footnote{I remind the reader that, within
one given region of the FKS dynamic phase-space partition achieved
by means of eq.~(\ref{sumSij}), the FKS parton is defined to be 
the only one that can give rise to soft singularities (hence, $i$), 
while the pair composed of the FKS parton and its sister is the only 
pair that can give rise to collinear singularities (hence, $(i,j)$).}. 
The final results which will be obtained
here will be easily cast into the usual form adopted in the context
of FKS subtraction 
elsewhere~\cite{Frixione:1995ms,Frixione:1997np,Frederix:2009yq} --
processes with a physical four-momentum configuration $2\to n-2$ or
$2\to n+1-2$ will be obtained by crossing, while partons (and in 
particular the FKS parton and its sister) may be relabeled.
In the introductory part of this section, where I shall discuss the
general features of the scattering amplitudes, I shall consider
the Born process of eq.~(\ref{ggBproc}) to be definite. The case
of the real-emission process of eq.~(\ref{ggRproc}) can simply
be obtained with the formal replacement $n\to n+1$.

Given an $n$-gluon colour configuration in SU$(N)$
\beq
\left\{a_i\right\}_{i=1}^n\,,\;\;\;\;\;\;\;
a_i\,\in\,\left\{1,\ldots N^2-1\right\}
\label{coladj}
\eeq
the corresponding scattering amplitude can be written as follows,
where I adopt the representation used in ref.~\cite{Mangano:1990by}:
\beqn
\ampn(a_1,\ldots a_n)&=&\sum_{\sigma\in P_n^\prime}
{\rm Tr}\Big(\lambda^{a_{\sigma(1)}}\ldots\lambda^{a_{\sigma(n)}}\Big)
\ampCSn\left(\sigma(1),\ldots\sigma(n)\right)
\nonumber\\*&\equiv&\sum_{\sigma\in P_n^\prime}
\Lambda\left(\seta,\sigma\right)\ampCSn(\sigma)\,,
\label{mgCDamp}
\eeqn
where I have introduced the shorthand notation:
\beq
\Lambda\left(\seta,\sigma\right)=
{\rm Tr}\Big(\lambda^{a_{\sigma(1)}}\ldots\lambda^{a_{\sigma(n)}}\Big).
\label{Lambdashort}
\eeq
When no ambiguity is possible, the first argument of $\Lambda$
(i.e., the set of colour indices) will be dropped. Here,
$\lambda^a$ are the Gell-Mann matrices, whose normalization conventions are
given in appendix~\ref{sec:conv}, and $P_n^\prime$ is the set of non-cyclic 
permutations of the first $n$ integers. The non-cyclicity condition 
can be imposed by simply requiring
\beq
\sigma(1)=1\,.
\label{noncycl}
\eeq
The quantity on the l.h.s.~of eq.~(\ref{mgCDamp}) is the colour-dressed 
amplitude, and $\ampCSn$ is the dual (or colour-ordered) amplitude. 
Each term in the sum on the r.h.s.~of eq.~(\ref{mgCDamp}) corresponds
to a colour flow (in which the colour of gluon $\sigma(i)$ is connected
with the anticolour of gluon $\sigma(i+1)$). In a technical sense,
I shall identify the flow with the ordered set
\beq
\left(\sigma(1),\ldots\sigma(n)\right)\,.
\eeq
It will be convenient to regard each
colour configuration as a vector $\ket{a_1,\ldots a_n}$; the set
of all colour configurations can therefore be made equivalent
to an ortho-normal base in a vector space (called colour space
henceforth):
\beqn
&&\brat{b_1,\ldots b_n}\ket{a_1,\ldots a_n}=
\prod_{i=1}^n\delta_{a_ib_i}\,,
\label{basisnorm}
\\
&&\sum_{\setan}\ket{a_1,\ldots a_n}\bra{a_1,\ldots a_n}=I\,.
\label{basisproj}
\eeqn
Equation~(\ref{mgCDamp}) suggests to define the following vectors
in the colour space:
\beqn
&&\ket{\ampn(a_1,\ldots a_n)}=\ampn(a_1,\ldots a_n)\ket{a_1,\ldots a_n}\,,
\label{Acoldef}
\\
&&\ket{\ampn(\sigma)}=\sum_{\setan}
\Lambda\left(\seta,\sigma\right)
\ampCSn\left(\sigma\right)
\ket{a_1,\ldots a_n}\,,
\label{Aflowdef}
\eeqn
which can be used to construct a vector that corresponds to the 
physical amplitude (i.e. the amplitude obtained by summing over 
all colour configurations or flows):
\beqn
\ket{\ampn}&=&\sum_{\setan}\ket{\ampn(a_1,\ldots a_n)}\,,
\label{ampCD}
\\
\ket{\ampn}&=&\sum_{\sigma\in P_n^\prime}\ket{\ampn(\sigma)}\,.
\label{ampflow}
\eeqn
What is relevant to cross-section computations is the amplitude
squared\footnote{In refs.~\cite{Frixione:1995ms,Frixione:1997np,
Frederix:2009yq}, the notation $\ampsqn$ is used for an amplitude squared, 
times the flux factor, times spin and colour average factors. All these 
factors are omitted here, being irrelevant for the present discussion.}:
\beq
\ampsqn=\brat{\ampn}\ket{\ampn}\,.
\eeq
Equations~(\ref{ampCD}) and~(\ref{ampflow}) can be used to rewrite
the amplitude squared in two different ways:
\beqn
\ampsqn&=&\sum_{\setan}\ampsqn(a_1,\ldots a_n)
\label{sumCD}
\\*
&=&\sum_{\sigma,\sigmap\in P_n^\prime}\ampsqn(\sigmap,\sigma)\,,
\label{sumflow}
\eeqn
where I defined
\beqn
&&\ampsqn(a_1,\ldots a_n)=
\brat{\ampn(a_1,\ldots a_n)}\ket{\ampn(a_1,\ldots a_n)}
\label{MmCD}
\\*
&&\phantom{\ampsqn(a_1,\ldots a_n)}
\equiv\abs{\ampn(a_1,\ldots a_n)}^2,
\label{amportho}
\\
&&\ampsqn(\sigmap,\sigma)=
\brat{\ampn(\sigmap)}\ket{\ampn(\sigma)}\,.
\label{Mmflow}
\eeqn
In terms of scalar quantities, eq.~(\ref{Mmflow}) is:
\beq
\ampsqn(\sigmap,\sigma)=
\ampCSn(\sigmap)^\star \,C(\sigmap,\sigma)\,\ampCSn(\sigma)\,,
\label{Mmflowsc}
\eeq
where I defined the colour-flow matrix element:
\beq
C(\sigmap,\sigma)=\sum_{\setan}\Lambda\left(\seta,\sigmap\right)^\star
\Lambda\left(\seta,\sigma\right)\,.
\label{CFmatdef}
\eeq
The hermiticity of the Gell-Mann matrices and eq.~(\ref{sumijkl})
imply that $C$ is a real, symmetric matrix:
\beq
C(\sigma,\sigmap)\in {\mathbb R}\,,\;\;\;\;\;\;\;\;
C(\sigmap,\sigma)=C(\sigma,\sigmap)\,.
\eeq
Note that while the amplitudes at fixed colour configurations are
orthogonal (a condition which is formally enforced here by
eq.~(\ref{basisnorm})), this is only
true for $\ket{\ampn(\sigma)}$ in the large-$\NC$ limit. In
fact, it is well known that
\beq
\ampsqn(\sigmap,\sigma)=c_\sigma\delta_{\sigmap\sigma}\,
\NC^{n-2}\left(\NC^2-1\right)+{\cal O}\left(\frac{1}{\NC^2}\right)\,,
\label{orthoflow}
\eeq
with $c_\sigma$ a suitable real number. From the definition, one
also sees that
\beq
\ampsqn(\sigmap,\sigma)=\ampsqn(\sigma,\sigmap)^\star\,.
\label{Mflowcplx}
\eeq

The quantities defined in eqs.~(\ref{MmCD}) and~(\ref{Mmflow}), or
their analogues for quark-gluon amplitudes
to be introduced later, will be the basic building blocks 
for the definition of FKS subtraction at fixed colour configurations 
and flows respectively. As far as the latter is concerned, a final
comment is in order. Colour flows are an important ingredient in the
context of event generators, where they are used to determine, on 
statistical basis, the colour connections amongst the hard partons 
which initiate the showers. This determination is driven by
$\ampsqn(\sigmap,\sigma)$ with $\sigmap=\sigma$ (since such quantities
are positive-definite), while the matrix elements with $\sigmap\ne\sigma$ 
are taken into account only in an averaged sense~\cite{Odagiri:1998ep}.
Therefore, as far as event generators go, a flow may actually be better 
defined as the pair of identical permutations $(\sigma,\sigma)$, the pair 
understanding a quantity relevant to the amplitude-squared level. In order 
to generalize this idea in a way consistent with the terminology typically 
used when dealing with amplitudes, I shall call the pair
\beq
(\sigmap,\sigma)
\label{closeddef}
\eeq
a {\em closed flow}. This reminds one of the fact that the pair
in eq.~(\ref{closeddef}) corresponds to a set of colour loops;
the counting of loops is a very simple way to estimate the largest 
possible power of $N$ which appears in $\ampsqn(\sigmap,\sigma)$.
To further this, as one sees from eq.~(\ref{CFmatdef}) the two
flows that define a closed flow play a similar, but not identical,
role. When I shall need to distinguish them, I shall call
the ordered sets
\beqn
\left(\sigmap(1),\ldots\sigmap(n)\right)\,,\;\;\;\;\;\;\;\;
\left(\sigma(1),\ldots\sigma(n)\right)
\eeqn
as L-flow and R-flow respectively, with the understanding that
they enter the amplitudes $\bra{\ampn(\sigmap)}$ and $\ket{\ampn(\sigma)}$.
This naming convention stems from the simple interpretation of 
eq.~(\ref{Mmflow}) in terms of cut diagrams, where one (arbitrarily) 
associates bra vectors $\langle .|$ with the left side of the cut.

\subsection{Subtraction of colour-summed matrix elements\label{sec:glusum}}
According to the conventions introduced at the beginning of
sect.~\ref{sec:gluons}, the quantities that enter eq.~(\ref{subt})
are defined as follows:
\beqn
\lim_{k_{n+1}\to 0}\ampsqnpo&=&\ampsqnpoS\,,
\label{limS}
\\
\lim_{k_{n+1}\parallel k_n}\ampsqnpo&=&\ampsqnpoC\,.
\label{limC}
\eeqn
The counterterm $\ampsqnpoSC$, which is responsible
for removing the double counting due to $\ampsqnpoS$
and $\ampsqnpoC$ in the soft-collinear region, is by
construction:
\beq
\ampsqnpoSC=
\lim_{k_{n+1}\parallel k_n}\ampsqnpoS\equiv
\lim_{k_{n+1}\to 0}\ampsqnpoC\,.
\label{MSCdef}
\eeq
The last identity in eq.~(\ref{MSCdef}) can, and will,
be used as a check of self-consistency when constructing $\ampsqnpoS$
and $\ampsqnpoC$.

The result for the soft limit of real-emission matrix elements
can be taken e.g.~from ref.~\cite{Frederix:2009yq}:
\beq
\ampsqnpoS=\half\gs^2
\sum_{k,l=1}^n\eik{k}{l}\ampsqn_{kl}\,,
\label{Msoftdef}
\eeq
with
\beqn
\eik{k}{l}&=&\frac{k_k\mydot k_l}{k_k\mydot\knpo\, k_l\mydot\knpo}
(1-\delta_{kl})\,,
\label{eikdef}
\\
\ampsqn_{kl}&=&-2\bra{\ampn}\,\sum_b Q^b(k)Q^b(l)\;\ket{\ampn}\,.
\label{Mkldef}
\eeqn
The term in round brackets on the r.h.s.~of eq.~(\ref{eikdef}) is
redundant, since $k_k^2=0$, but it is useful when one formally
manipulates eikonal factors.
The operators $Q^b(k)$ describe the way in which 
the colour of gluon $k$ is affected when a soft gluon (which here is 
always labelled by $n+1$) of colour $b$ is emitted by it. In the present
context, it is useful to regard the result of $Q^b(k)$ as twofold:
it changes the colour state of gluon $k$, and it creates gluon $n+1$
with colour $b$. This can be written as follows:
\beqn
\bra{a_k a_{n+1}} Q^b(k)\ket{c_k}&=&
\delta_{ba_{n+1}}\left(Q^b(k)\right)_{a_kc_k}=
\delta_{ba_{n+1}}\left(T^b\right)_{a_kc_k}\,,
\label{Qdef}
\\
\left(T^b\right)_{ac}&=&-if^{bac}\,.
\label{Qreprglu}
\eeqn
The definition given here implies that 
\mbox{$Q^b(l)\ket{\ampn}$} is a vector in the colour space
of $n+1$ gluons. In the colour subspace of gluon $k$, the operator
$Q^b(k)$ is hermitian.

The collinear limit of real-emission matrix elements can also
be taken from ref.~\cite{Frederix:2009yq}. It reads:
\beq
\ampsqnpoC=\frac{\gs^2}{k_n\mydot\knpo}\Bigg\{
P_{gg}(z)\ampsqn
+Q_{gg^\star}(z)\,
\Re\Bigg(\frac{\langle k_n k_{n+1}\rangle}{[k_n k_{n+1}]}
\wampsqn_{-+}\Bigg)\Bigg\},
\label{Mcolldef0}
\eeq
where
\beq
\wampsqn_{-+}=\brat{\ampn_-}\ket{\ampn_+}\,,
\eeq
with $\ampn_\pm$ the $n$-gluon amplitude at fixed $\pm$ helicity of
the $n^{th}$ gluon (i.e.~of the gluon that branches). The momentum 
fraction $z$ is defined according to:
\beq
k_n=z(k_n+\knpo)\;\;\;\;{\rm when}\;\;\;\;k_{n+1}\parallel k_n\,.
\label{zdef}
\eeq
The symbols $P_{gg}(z)$ and $Q_{gg^\star}(z)$ denote the Altarelli-Parisi
kernel and its azimuthal counterpart, respectively, relevant to 
$g\to gg$ branchings. The latter kernel has been introduced in 
ref.~\cite{Frixione:1995ms}, and I refer the reader to that paper 
for further details.
By using the identity
\beq
\sum_b Q^b(n)Q^b(n)=\CA\,I\,,
\label{Casimir}
\eeq
with $I$ being the identity operator in the colour space, I now 
rewrite eq.~(\ref{Mcolldef0}) in a slightly different (but completely 
equivalent) form, which is better suited to the manipulations 
I shall carry out in the rest of this paper:
\beq
\ampsqnpoC=\frac{\gs^2}{k_n\mydot\knpo}\Bigg\{
\hat{P}_{gg}(z)\zampsqn
+\hat{Q}_{gg^\star}(z)\,
\Re\Bigg(\frac{\langle k_n k_{n+1}\rangle}{[k_n k_{n+1}]}
\wzampsqn_{-+}\Bigg)\Bigg\},
\label{Mcolldef}
\eeq
where
\beqn
\zampsqn&=&\bra{\ampn}\,\sum_b Q^b(n)Q^b(n)\;\ket{\ampn}\,,
\label{Mcolldefz}
\\
\wzampsqn_{-+}&=&\bra{\ampn_-}\,\sum_b Q^b(n)Q^b(n)\;\ket{\ampn_+}\,,
\label{Mtcolldefz}
\eeqn
and I have introduced the quantities
\beq
\hat{P}_{gg}(z)=\frac{1}{\CA}P_{gg}(z)\,,\;\;\;\;\;\;\;\;\;
\hat{Q}_{gg^\star}(z)=\frac{1}{\CA}Q_{gg^\star}(z)\,.
\label{APred}
\eeq

The consistency between eqs.~(\ref{Msoftdef}) and~(\ref{Mcolldef}),
in the sense of eq.~(\ref{MSCdef}), is easy to prove. Let me start
from computing the collinear limit of $\ampsqnpoS$. Since the 
only dependence on $\knpo$ is in the eikonal factors, the relevant
quantity is:
\beq
\lim_{k_{n+1}\parallel k_n}\eik{k}{l}=
\frac{1}{1-z}\frac{1}{k_n\mydot\knpo}
\left(\delta_{kn}+\delta_{ln}\right)
(1-\delta_{kl})\,,
\label{eiklim}
\eeq
having used eq.~(\ref{zdef}). Therefore
\beq
\lim_{k_{n+1}\parallel k_n}\ampsqnpoS=
\frac{\gs^2}{2}\frac{1}{1-z}\frac{1}{k_n\mydot\knpo}
\left\{\sum_{l=1}^{n-1}\ampsqn_{nl}+
\sum_{k=1}^{n-1}\ampsqn_{kn}\right\}\,.
\eeq
As is proved in appendix~\ref{sec:appa}, the colour-conservation
condition is
\beq
\sum_{k=1}^n Q^b(k)\ket{\ampn}=0\;\;\;\;\;\;
\Longrightarrow\;\;\;\;\;\;
\sum_{k=1}^{n-1} Q^b(k)\ket{\ampn}=-Q^b(n)\ket{\ampn}
\label{colcons}
\eeq
for any colour index $b$. Therefore, by using the definitions of the
colour-linked Born's given in eq.~(\ref{Mkldef}), and the ortho-normality
of the colour-vector basis of eq.~(\ref{basisnorm}), one obtains:
\beq
\lim_{k_{n+1}\parallel k_n}\ampsqnpoS=
\gs^2\frac{2}{1-z}\frac{1}{k_n\mydot\knpo}
\bra{\ampn}\,\sum_b Q^b(n)Q^b(n)\;\ket{\ampn}\,.
\eeq
This expression is manifestly identical to the soft limit
of eq.~(\ref{Mcolldef}), since
\beq
\hat{P}_{gg}(z)\;\stackrel{z\to 1}{\longrightarrow}\;\frac{2}{1-z}\,,
\;\;\;\;\;\;\;\;
\hat{Q}_{gg^\star}(z)\;\stackrel{z\to 1}{\longrightarrow}\; 0\,.
\label{APsoft}
\eeq

A comment, which will apply throughout the paper, is
necessary here. In general, the reduced kinematic configurations
(that enter the Born amplitudes) that one obtains by taking the soft
or the collinear limit of a fully-resolved configuration (that enter
the real-emission amplitudes) need not coincide. However, as explained
in ref.~\cite{Frederix:2009yq}, in the context of FKS subtraction
is possible and convenient to adopt phase-space parametrizations
for which the two do coincide. This justifies the fact that I have
used the same symbol $\ampn$ e.g.~in eqs.~(\ref{Mkldef}) 
and~(\ref{Mcolldefz}), which would in general understand 
different kinematics. I shall use a unique notation for the 
soft- and collinear-induced Born amplitudes in the rest of the paper.
Having said that, I should also like to stress that, even if the two
kinematics were different, the results would be unchanged, since in the
FKS method soft and collinear singularities are treated separately. The
only exception would be in sect.~\ref{sec:disc}, where the choice
of kinematic configurations in the two limits becomes relevant.

\subsection{Subtraction at fixed colour configurations\label{sec:gluCD}}
In this section, I shall consider the problem of defining
the subtractions of eq.~(\ref{subt}) for a given colour
configuration. This implies that the quantity whose soft and
collinear limits one needs to construct is
\beq
\ampsqnpo(a_1,\ldots\anpo)=\abs{\ampnpo(a_1,\ldots\anpo)}^2\,.
\label{MnpoCD}
\eeq
The relevant limits can be obtained directly from $\ampsqnpoS$ and 
$\ampsqnpoC$, given in eqs.~(\ref{Msoftdef}) and~(\ref{Mcolldef})
respectively. Formally, 
one proceeds as follows. One starts from eqs.~(\ref{limS}) and~(\ref{limC}),
and then fixes the colour configuration $\setanpo$ on both sides
of those equations by inserting there the projector
\beq
\ket{a_1,\ldots\anpo}\bra{a_1,\ldots\anpo}\,.
\label{projanpo}
\eeq
In the l.h.s.'s, such an insertion indeed results in singling out
the colour-dressed amplitude squared of eq.~(\ref{MnpoCD}):
\beqn
&&\ampsqnpo=\brat{\ampnpo}\ket{\ampnpo}
\nonumber\\*&&\phantom{aaaaaaaa}\longrightarrow\;\;
\brat{\ampnpo}\ket{a_1,\ldots\anpo}
\brat{a_1,\ldots\anpo}\ket{\ampnpo}
\equiv\ampsqnpo(a_1,\ldots\anpo)\,.
\phantom{aaaaaa}
\eeqn
When performing the insertion of the projector of eq.~(\ref{projanpo})
in the r.h.s.~of eq.~(\ref{limS}), by using eq.~(\ref{Msoftdef}) one obtains:
\beq
\ampsqnpoS(a_1,\ldots\anpo)=\half\gs^2
\sum_{k,l=1}^n\eik{k}{l}
\ampsqn_{kl}(a_1,\ldots\anpo)\,,
\label{MsoftCD}
\eeq
where
\beq
\ampsqn_{kl}(a_1,\ldots\anpo)=-2\bra{\ampn}\,
Q^{\anpo}(k)\ket{a_1,\ldots\anpo}\bra{a_1,\ldots\anpo}
Q^{\anpo}(l)\;\ket{\ampn}\,,
\label{MklCD}
\eeq
having taken the Kronecker delta of eq.~(\ref{Qdef}) into account.
Using eqs.~(\ref{ampCD}) and~(\ref{Qdef}), eq.~(\ref{MklCD}) 
can be easily re-expressed in terms of scalar quantities best
suited to numerical computations:
\beqn
\label{Mklscalar}
&&\ampsqn_{kl}(a_1,\ldots\anpo)=
\\*&&\phantom{aaa}
-2\sum_{b_k^\prime b_l}
\ampn(a_1,\ldots b_k^\prime\,\ldots a_l\,\ldots a_n)^\star
{\cal Q}_{gg}(\anpo;a_k,a_l)_{b_k^\prime b_l}
\ampn(a_1,\ldots a_k\,\ldots b_l\,\ldots a_n),
\nonumber
\eeqn
where I have introduced the matrices
\beq
{\cal Q}_{gg}(\anpo;a_k,a_l)_{bc}=
\left(T^{\anpo}\right)_{ba_k}\left(T^{\anpo}\right)_{a_lc}=
f^{\anpo a_k b}f^{\anpo a_l c}\,.
\label{Qmatdef}
\eeq
As the notation suggests, I stress that the colour index $\anpo$ is not 
summed over on the r.h.s.~of eq.~(\ref{Qmatdef}). By construction, in
SU(3) there are $8^3$ ${\cal Q}_{gg}$ matrices, of dimension $8\!\times\!8$.

The computation of the collinear limit is performed along the same lines.
One obtains:
\beqn
\ampsqnpoC(a_1,\ldots\anpo)&=&\frac{\gs^2}{k_n\mydot\knpo}\Bigg\{
\hat{P}_{gg}(z)\zampsqn(a_1,\ldots\anpo)
\nonumber\\*&&\phantom{\frac{\gs^2}{k_n\mydot\knpo}}
\!+\hat{Q}_{gg^\star}(z)\,
\Re\Bigg(\frac{\langle k_n k_{n+1}\rangle}{[k_n k_{n+1}]}
\wzampsqn_{-+}(a_1,\ldots\anpo)\Bigg)\Bigg\},\phantom{aaaa}
\label{McollCD}
\eeqn
with
\beqn
\zampsqn(a_1,\ldots\anpo)&=&\bra{\ampn}\,
Q^{\anpo}(n)\ket{a_1,\ldots\anpo}\bra{a_1,\ldots\anpo}
Q^{\anpo}(n)\;\ket{\ampn}\,,\phantom{aaa}
\\
\wzampsqn_{-+}(a_1,\ldots\anpo)&=&\bra{\ampn_-}\,
Q^{\anpo}(n)\ket{a_1,\ldots\anpo}\bra{a_1,\ldots\anpo}
Q^{\anpo}(n)\;\ket{\ampn_+}\,.\phantom{aaa}
\eeqn
These can be re-expressed in terms of scalar quantities, as done
for their soft counterparts in eq.~(\ref{Mklscalar}):
\beqn
&&\zampsqn(a_1,\ldots\anpo)=
\nonumber\\*&&\phantom{aaaaaa}
\sum_{b_n^\prime b_n}
\ampn(a_1,\ldots a_{n-1},b_n^\prime)^\star
{\cal Q}_{gg}(\anpo;a_n,a_n)_{b_n^\prime b_n}
\ampn(a_1,\ldots a_{n-1},b_n),
\label{M0scalar}
\\
&&\wzampsqn_{-+}(a_1,\ldots\anpo)=
\nonumber\\*&&\phantom{aaaaaa}
\sum_{b_n^\prime b_n}
\ampn_-(a_1,\ldots a_{n-1},b_n^\prime)^\star
{\cal Q}_{gg}(\anpo;a_n,a_n)_{b_n^\prime b_n}
\ampn_+(a_1,\ldots a_{n-1},b_n).
\label{wM0scalar}
\eeqn
The consistency between eqs.~(\ref{MsoftCD}) and~(\ref{McollCD})
in the sense of eq.~(\ref{MSCdef}) can be proved in the same way
as was done at the end of sect.~\ref{sec:glusum} for eqs.~(\ref{Msoftdef}) 
and~(\ref{Mcolldef}). In fact, apart from eq.~(\ref{eiklim}) which 
holds independently of the treatment of the colours, the key point 
there is colour conservation as given in eq.~(\ref{colcons}).
As is discussed in appendix~\ref{sec:appa}, colour conservation is a property
that holds at fixed $(n+1)$-gluon colour configurations, which is
sufficient to prove the point.
Equations~(\ref{Qmatdef}) and~(\ref{ffsum}) imply that
\beq
\sum_{\anpo}\sum_{a_n}{\cal Q}_{gg}(\anpo;a_n,a_n)_{bc}
=\CA\delta_{bc}\,,
\label{Casimir3}
\eeq
which is nothing but the matrix form of eq.~(\ref{Casimir}).

\subsection{Subtraction at fixed flows\label{sec:gluflow}}
I now turn to the definition of subtraction at given flows.
The problem is formally more involved than that of
sect.~\ref{sec:gluflow}, because of the lack of the analogue
of the projector in eq.~(\ref{projanpo}). The basic idea is
however still the same. To be definite, I consider the soft
limit, the collinear limit being fully analogous. One starts 
from eq.~(\ref{limS}), and expresses both sides as sums over
$(n+1)$-gluon L- and R-flows:
\beqn
\lim_{k_{n+1}\to 0}\left(\sum_{\Sigma,\Sigmap\in P_{n+1}^\prime}
\ampsqnpo(\Sigmap,\Sigma)\right)&\equiv&
\sum_{\Sigma,\Sigmap\in P_{n+1}^\prime}\lim_{k_{n+1}\to 0}
\ampsqnpo(\Sigmap,\Sigma)
\nonumber\\*&=&
\sum_{\Sigma,\Sigmap\in P_{n+1}^\prime}
\ampsqnpoS(\Sigmap,\Sigma)\,.
\label{limS2}
\eeqn
At this point, since the representation in terms of flows is unique, 
the terms with the same $\Sigma$ and $\Sigmap$ on the two sides 
of eq.~(\ref{limS2}) will be the analogues, at fixed flows,
of the first two terms of eq.~(\ref{subt}) --
they will play the same roles as eqs.~(\ref{MnpoCD}) 
and~(\ref{MsoftCD}) played in the case of subtraction
at fixed colour configurations.

This procedure is completely trivial as far as the l.h.s.~of
eq.~(\ref{limS2}) is concerned, since it amounts to using the definition 
of colour vectors at fixed flows, given in eq.~(\ref{Mmflow}):
\beqn
\ampsqnpo(\Sigmap,\Sigma)&=&
\brat{\ampnpo(\Sigmap)}\ket{\ampnpo(\Sigma)}
\nonumber\\*&=&
\ampCSnpo(\Sigmap)^\star \,C(\Sigmap,\Sigma)\,\ampCSnpo(\Sigma)\,,
\label{MnpoSS}
\eeqn
with the colour-flow matrix $C$ defined as in eq.~(\ref{CFmatdef}),
with $n\to n+1$ there.
What is less trivial is the r.h.s.~of eq.~(\ref{limS2}). The existence
of such a decomposition is guaranteed by the fact that the amplitudes
relevant to the reduced matrix element $\ampsqnpoS$ live in the 
$(n+1)$-gluon colour space. However, their colour structures are obtained 
by means of the operators $Q^b$ that act on $n$-gluon colour 
vectors (see eq.~(\ref{Mkldef})), and one therefore obtains the contributions 
at given $\Sigma$ and $\Sigmap$ in a rather indirect way. The
easiest way to perform the computation of $\ampsqnpoS(\Sigmap,\Sigma)$
is obviously that of using the decomposition of the $n$-gluon colour vectors 
in terms of $n$-gluon flows, eq.~(\ref{ampflow}), as the starting point.
This implies that when computing $\ampsqnpoS$ one will end up
dealing with quantities such as:
\beq
\sum_{k,l}\eik{k}{l}\,
\bra{\ampn(\sigma^\prime)}\,\sum_b Q^b(k)Q^b(l)\;\ket{\ampn(\sigma)}\,.
\label{Mklss}
\eeq
As is shown in appendix~\ref{sec:appa} (see in particular eq.~(\ref{QvsII})), 
for a given index $l$ and an $n$-gluon R-flow $\sigma$, the vector
\mbox{$Q^b(l)\ket{\ampn(\sigma)}$} corresponds to the two 
$(n+1)$-gluon R-flows
\beq
I_+(\isigma(l))\,\sigma\,,\;\;\;\;\;\;\;\;I_-(\isigma(l))\,\sigma\,,
\label{npoRF}
\eeq
with $I_\pm$ defined in eqs.~(\ref{Ipdef}) and~(\ref{Imdef}).
In the case of L-flows, one finds instead:
\beq
I_+(\isigmap(k))\,\sigmap\,,\;\;\;\;\;\;\;\;I_-(\isigmap(k))\,\sigmap\,.
\label{npoLF}
\eeq
This is equivalent to saying that eq.~(\ref{npoRF}) defines
two maps:
\beq
(l,\sigma)\;\stackrel{I_+}{\longrightarrow}\;\Sigma_+\,,
\;\;\;\;\;\;\;\;
(l,\sigma)\;\stackrel{I_-}{\longrightarrow}\;\Sigma_-\,,
\label{mapR}
\eeq
relevant to R-flows. The situation is obviously identical
for L-flows:
\beq
(k,\sigmap)\;\stackrel{I_+}{\longrightarrow}\;\Sigmap_+\,,
\;\;\;\;\;\;\;\;
(k,\sigmap)\;\stackrel{I_-}{\longrightarrow}\;\Sigmap_-\,.
\label{mapL}
\eeq
In order to find the representation that appears on the r.h.s.~of 
eq.~(\ref{limS2}), one has therefore to fix $\Sigma_\pm$ and $\Sigmap_\pm$,
and to invert the maps of eqs.~(\ref{mapR}) and~(\ref{mapL}).

I shall explicitly carry out this procedure in sect.~\ref{sec:gluflowR}.
Before turning to that, it is worth remarking that by construction
the gluon-insertion operators $Q^b$ do not affect the
dual amplitudes (i.e., they leave the Lorentz structure invariant).
Hence, both $Q^b\ket{\ampn(\sigma)}$ and $\ket{\ampn(\sigma)}$ 
will contain the {\em same} $n$-gluon dual amplitude $\ampCSn(\sigma)$. 
This suggests a procedure alternative to that implied by eq.~(\ref{limS2}). 
Namely, one may simply fix the $n$-gluon flows that appear in 
eq.~(\ref{Mklss}), and therefore choose not to invert the maps 
of eqs.~(\ref{mapR}) and~(\ref{mapL}). Rather, these maps will be
used to identify all $\ampsqnpo(\Sigmap,\Sigma)$ whose sum
has a soft limit proportional to the quantity in eq.~(\ref{Mklss}). 
I shall discuss this approach in sect.~\ref{sec:gluflowB}.

The physical meaning of the two procedures sketched above is obvious. 
While in eq.~(\ref{limS2}) one fixes the flows at the level of 
real-emission matrix elements, in the other case the fixed
flows are those of the Born matrix elements.
Therefore, I shall refer to these two viewpoints as fixed real 
flows and fixed Born flows respectively.

\subsubsection{Fixed real flows\label{sec:gluflowR}}
As was anticipated above, I start by expressing the soft matrix element 
as a sum over $n$-gluon flows:
\beqn
\ampsqnpoS&=&\sum_{\sigma,\sigmap\in P_n^\prime}
\ampsqnpoS(\sigmap,\sigma)\,,
\label{QAeqAs}
\\
\ampsqnpoS(\sigmap,\sigma)&=&\half\gs^2
\sum_{k,l=1}^n\eik{k}{l}
\ampsqn_{kl}(\sigmap,\sigma)\,,
\label{Msoftdef2}
\\
\ampsqn_{kl}(\sigmap,\sigma)&=&-2
\bra{\ampn(\sigma^\prime)}\,\sum_b Q^b(k)
Q^b(l)\;\ket{\ampn(\sigma)}\,.
\label{Mkldef2}
\eeqn
In order to compute explicitly the colour-linked Born's that
appear in eq.~(\ref{Mkldef2}), I introduce the following $(n+1)$-gluon 
colour vectors associated with an underlying $n$-body dynamics:
\beq
\ket{\ampRnpo(\Sigma)}=
\sum_{\setanpo}
\Lambda\left(\seta,\Sigma\right)
\ampCSn\left(\Sigma_{\cancel{n+1}}\right)
\ket{a_1,\ldots \anpo}\,,
\eeq
where $\Sigma\in P_{n+1}^\prime$, and I defined
\beq
\Sigma_{\cancel{n+1}}=
\left(\Sigma(1),\ldots\cancel{n+1},\ldots\Sigma(n+1)\right)\,.
\eeq
By using the results of appendix~\ref{sec:appa}, and as already 
anticipated in eq.~(\ref{npoRF}), one obtains:
\beq
\sum_b Q^b(l)\,\ket{\ampn(\sigma)}=
\ket{\ampRnpo(I_+(\isigma(l))\,\sigma)}-
\ket{\ampRnpo(I_-(\isigma(l))\,\sigma)}.
\label{QonA}
\eeq
I can now use eq.~(\ref{QonA}) in eq.~(\ref{Mkldef2}) since,
for any two $n$-gluon colour vectors $\bra{v^\prime}$ and $\ket{v}$,
the follow identity holds:
\beq
\sum_b \bra{v^\prime}Q^b(k)Q^b(l)\ket{v}=
\sum_{b,c}\delta_{bc} \bra{v^\prime}Q^b(k)Q^c(l)\ket{v}=
\sum_{b,c}\delta_{bc} \brat{w^\prime b}\ket{w c}=
\sum_{b,c}\brat{w^\prime b}\ket{w c}\,,
\label{doublesum}
\eeq
since $\delta_{bc}$ is automatically enforced by the dot product of the
colour vectors, which receives the contribution $\brat{b}\ket{c}=\delta_{bc}$ 
from the $(n+1)^{th}$-gluon subspace. Then:
\beqn
\ampsqn_{kl}(\sigmap,\sigma)&=&-2\,\Big\{
\bra{\ampRnpo(I_+(\isigmap(k))\,\sigmap)}-
\bra{\ampRnpo(I_-(\isigmap(k))\,\sigmap)}\Big\}
\nonumber\\*&&\phantom{-2}\,
\Big\{
\ket{\ampRnpo(I_+(\isigma(l))\,\sigma)}-
\ket{\ampRnpo(I_-(\isigma(l))\,\sigma)}\Big\}\,.
\label{Mklss2}
\eeqn
One can now change the labellings in the sums that appear on the r.h.s.~of
eq.~(\ref{Msoftdef2}), $\sum_{kl}f(k,l)=\sum_{ij}f(\sigmap(i),\sigma(j))$.
Furthermore, one exploits eq.~(\ref{ImeqIp}), which can also be 
conveniently extended:
\beq
I_-(1)\sigma=I_+(0)\sigma\,,\;\;\;\;\;\;\;\;
\ket{\ampRnpo(I_+(0)\sigma)}\equiv \ket{\ampRnpo(I_+(n)\sigma)}\,,
\label{ImeqIp0}
\eeq
where the last condition can be imposed because of the cyclicity of 
the trace. One then obtains:
\beqn
&&\ampsqnpoS(\sigmap,\sigma)=-\gs^2
\sum_{i,j=1}^n\eik{\sigmap(i)}{\sigma(j)}\Bigg\{
\brat{\ampRnpo(I_+(i)\,\sigmap)}
\ket{\ampRnpo(I_+(j)\,\sigma)}
\nonumber\\*&&\phantom{aaa}
-\brat{\ampRnpo(I_+(i)\,\sigmap)}
\ket{\ampRnpo(I_+(j-1)\,\sigma)}
-\brat{\ampRnpo(I_+(i-1)\,\sigmap)}
\ket{\ampRnpo(I_+(j)\,\sigma)}
\nonumber
\\*&&\phantom{aaa}
+\brat{\ampRnpo(I_+(i-1)\,\sigmap)}
\ket{\ampRnpo(I_+(j-1)\,\sigma)}\Bigg\}\,.
\label{Mkltmp}
\eeqn
By relabelling $j-1\to j$ in the second and fourth terms on the r.h.s.~of 
eq.~(\ref{Mkltmp}), and $i-1\to i$ in the third and fourth terms, and
by using the second equality in eq.~(\ref{ImeqIp0}), one finally gets:
\beqn
\ampsqnpoS(\sigmap,\sigma)&=&-\gs^2
\sum_{i,j=1}^n\Big\{
\eik{\sigmap(i)}{\sigma(j)}
-\eik{\sigmap(i)}{\sigma(j+1)}
\nonumber\\*&&\phantom{-\gs^2\sum_{i,j=1}^n}
-\eik{\sigmap(i+1)}{\sigma(j)}
+\eik{\sigmap(i+1)}{\sigma(j+1)}\Big\}
\nonumber\\*&&\phantom{-\gs^2\sum_{i,j=1}^n}
\times\brat{\ampRnpo(I_+(i)\,\sigmap)}
\ket{\ampRnpo(I_+(j)\,\sigma)},
\label{Mppdef}
\eeqn
where, as a consequence of the relabellings mentioned above, 
one must understand
\beq
\sigma(n+1)=\sigma(1)\,.
\label{srange}
\eeq
At variance with eq.~(\ref{Mkldef2}), eq.~(\ref{Mppdef}) contains a 
single $(n+1)$-gluon closed flow for a given $(i,j)$ pair. 
This is what renders it easy to cast
the sum over $\sigma$ and $\sigmap$ (see eq.~(\ref{QAeqAs}))
of the quantities in eq.~(\ref{Mppdef}) in the same form
as the r.h.s.~of eq.~(\ref{limS2}). One can start by using the following
identity, which I write for R-flows to be definite (the case of L-flows
being identical):
\beq
\sum_{\sigma\in P_n^\prime}\sum_{j=1}^n f(\sigma,j)=
\sum_{\sigma\in P_n^\prime}\sum_{j=1}^n\;
\sum_{\Sigma\in P_{n+1}^\prime}
\!\!\delta(\Sigma,I_+(j)\,\sigma)\,
f(\sigma,j)\,,
\label{siSident}
\eeq
with $\delta$ on the r.h.s.~of eq.~(\ref{siSident}) the Kronecker
delta. The identity above holds true since
\beq
1=\sum_{\Sigma\in P_{n+1}^\prime} 
\!\!\delta(\Sigma,I_+(j)\,\sigma)
\phantom{aaaaaa}{\rm for~given~}\phantom{aa}\sigma,\;j\,,
\label{Ndef}
\eeq
which is a consequence of the fact that, at fixed $\sigma$ and $j$, 
the $(n+1)$-gluon flow $I_+(j)\sigma$ always exists and is unique.
Now, the sum over $\Sigma$ in eq.~(\ref{siSident})
will play the same role as that on the r.h.s.~of eq.~(\ref{limS2}). 
Hence, one can exploit the $\delta$ in eq.~(\ref{siSident}) 
to get rid of the sums over $\sigma$ and $j$. This implies
solving the equation:
\beqn
&&I_+(j)\,\sigma=\Sigma\;\;\;\;\Longleftrightarrow\;\;\;\;
\label{svsS}
\\*&&\phantom{aaaaaa}
(\sigma(1),\ldots \sigma(j),n+1,\sigma(j+1),\ldots\sigma(n))
\nonumber
\\*&&\phantom{aaa}\;=
(\Sigma(1),\ldots \Sigma(j),\Sigma(j+1),\Sigma(j+2),\ldots\Sigma(n+1))
\label{sjvsS}
\eeqn
for a given $\Sigma$. It is immediate to see that this solution
always exists:
\beqn
\sigma&=&\Sigma_{\cancel{n+1}}\,,
\\
j&=&\iSigma(n+1)-1\,,
\label{jsol}
\eeqn
and is unique. In other words, there is
one-to-one correspondence between pairs composed of one $P_n^\prime$
permutation and one integer, and $P_{n+1}^\prime$ permutations.
Equation~(\ref{siSident}) then becomes:
\beq
\sum_{\sigma\in P_n^\prime}\sum_{j=1}^n f(\sigma,j)=
\sum_{\Sigma\in P_{n+1}^\prime} 
f\left(\Sigma_{\cancel{n+1}},\iSigma(n+1)-1\right)\,.
\label{sseqSS}
\eeq
Equation~(\ref{sseqSS}) is consistent with simple counting: the
$n!$ $P_{n+1}^\prime$ permutations can be obtained by inserting
the integer $n+1$ in all the $n$ possible ways (not $n+1$, because
of non-cyclicity) in each of the $(n-1)!$ $P_n^\prime$ permutations:
$n!=n\times (n-1)!$. 

These arguments, when applied to L-flows, give
\beqn
\sigmap&=&\Sigmap_{\cancel{n+1}}\,,
\\
i&=&\iSigmap(n+1)-1\,.
\label{isol}
\eeqn
The results above can now be used in eqs.~(\ref{QAeqAs}) and~(\ref{Mppdef}).
One obtains:
\beqn
\ampsqnpoS&=&\sum_{\Sigma,\Sigmap\in P_{n+1}^\prime}
\ampsqnpoS(\Sigmap,\Sigma)\,,
\label{QAeqAsRF}
\\
\ampsqnpoS(\Sigmap,\Sigma)&=&-\gs^2
\Big\{
\eik{\Sigmap(\iSigmap(n+1)-1)}{\Sigma(\iSigma(n+1)-1)}
\label{MssMpp2}
\\*&&\phantom{-\gs^2\!\!\!}
-\eik{\Sigmap(\iSigmap(n+1)-1)}{\Sigma(\iSigma(n+1)+1)}
\nonumber\\*&&\phantom{-\gs^2\!\!\!}
-\eik{\Sigmap(\iSigmap(n+1)+1)}{\Sigma(\iSigma(n+1)-1)}
\nonumber\\*&&\phantom{-\gs^2\!\!\!}
+\eik{\Sigmap(\iSigmap(n+1)+1)}{\Sigma(\iSigma(n+1)+1)}\Big\}
\ampsqRn(\Sigmap,\Sigma),
\nonumber
\\
\ampsqRn(\Sigmap,\Sigma)&=&
\brat{\ampRnpo(\Sigmap)}
\ket{\ampRnpo(\Sigma)}.
\label{Mppdef2}
\eeqn
In the evaluation of the eikonal factors that appear in eq.~(\ref{MssMpp2}),
one must take the analogues of eqs.~(\ref{noncycl}) and~(\ref{srange}) into 
account, namely
\beq
\Sigma(1)=1\,,\;\;\;\;\;\;\;\;
\Sigma(n+2)=\Sigma(1)\,,
\label{Sconds}
\eeq
and similarly for $\Sigmap$. Note that the second identity 
in eq.~(\ref{Sconds}) is enforced by the solution of eq.~(\ref{sjvsS})
in the case when $n+1$ occupies the rightmost position in $I_+(j)\sigma$.
Equations~(\ref{MssMpp2}) and~(\ref{Mppdef2}) only depend 
on $\Sigma$ and $\Sigmap$, and are therefore
in the form suited to be used in eq.~(\ref{limS2}). In terms of scalar
quantities, eq.~(\ref{Mppdef2}) reads as follows:
\beq
\ampsqRn(\Sigmap,\Sigma)=
\ampCSn(\Sigmap_{\cancel{n+1}})^\star\,
C(\Sigmap,\Sigma)\,
\ampCSn(\Sigma_{\cancel{n+1}})\,.
\label{Mred}
\eeq

I now address the case of the collinear limit, which I deal with as was done 
for the soft limit. Namely, in eqs.~(\ref{Mcolldef})--(\ref{Mtcolldefz})
I use the representation of the amplitude in terms of flows,
eq.~(\ref{ampflow}), and its analogue for L-flows:
\beq
\bra{\ampn}=\sum_{\sigmap\in P_n^\prime}\bra{\ampn(\sigmap)}\,.
\eeq
In doing that, I explicitly express the real part that appears in the term
proportional to $Q_{gg^\star}$ in eq.~(\ref{Mcolldef}), by writing it as
the sum of its argument plus the complex conjugate of the latter; this 
is necessary in order to identify $\sigma$ with the R-flow, and $\sigmap$ 
with the L-flow. I thus obtain:
\beq
\ampsqnpoC=\sum_{\sigma,\sigmap\in P_n^\prime}
\ampsqnpoC(\sigmap,\sigma)\,,
\label{QAeqAc}
\eeq
where
\beqn
\ampsqnpoC(\sigmap,\sigma)&=&\frac{\gs^2}{k_n\mydot\knpo}\Bigg\{
\hat{P}_{gg}(z)\zampsqn(\sigmap,\sigma)
\label{McollFB}
\\*&&\phantom{aaaa}
\!+\half\hat{Q}_{gg^\star}(z)\,
\Bigg(\frac{\langle k_n k_{n+1}\rangle}{[k_n k_{n+1}]}
\wzampsqn_{-+}(\sigmap,\sigma)
+\frac{[k_n k_{n+1}]}{\langle k_n k_{n+1}\rangle}
\wzampsqn_{+-}(\sigmap,\sigma)\Bigg)\Bigg\},
\nonumber
\eeqn
and
\beqn
\zampsqn(\sigmap,\sigma)&=&\bra{\ampn(\sigmap)}\,
\sum_b Q^b(n)Q^b(n)\;\ket{\ampn(\sigma)}\,,
\label{amp0}
\\
\wzampsqn_{\lambda\blambda}(\sigmap,\sigma)&=&
\bra{\ampn_{\lambda}(\sigmap)}\,
\sum_b Q^b(n)Q^b(n)\;\ket{\ampn_{\blambda}(\sigma)}\,.
\label{wamp0}
\eeqn
Given that eq.~(\ref{wamp0}) is essentially identical to eq.~(\ref{amp0}),
I shall deal only with $\zampsqn$ in the following; the term 
proportional to the azimuthal kernel $\hat{Q}_{gg^\star}(z)$ will be
reinstated at the end. By comparing eq.~(\ref{amp0}) with 
eq.~(\ref{Mkldef2}) one gets:
\beq
\zampsqn(\sigmap,\sigma)=-\half\ampsqn_{nn}(\sigmap,\sigma)\,.
\label{zerovsnn}
\eeq
Hence, from eq.~(\ref{Mklss2}) one obtains:
\beqn
\zampsqn(\sigmap,\sigma)&=&
\brat{\ampRnpo(I_+(\isigmap(n))\,\sigmap)}
\ket{\ampRnpo(I_+(\isigma(n))\,\sigma)}
\nonumber\\*&-&
\brat{\ampRnpo(I_+(\isigmap(n))\,\sigmap)}
\ket{\ampRnpo(I_-(\isigma(n))\,\sigma)}
\nonumber\\*&-&
\brat{\ampRnpo(I_-(\isigmap(n))\,\sigmap)}
\ket{\ampRnpo(I_+(\isigma(n))\,\sigma)}
\nonumber\\*&+&
\brat{\ampRnpo(I_-(\isigmap(n))\,\sigmap)}
\ket{\ampRnpo(I_-(\isigma(n))\,\sigma)}\,.
\label{M0nn}
\eeqn
The $(n+1)$-gluon L- and R-flows that appear on the r.h.s.~of
eq.~(\ref{M0nn}) are such that:
\beqn
&&I_+(\isigmap(n))\sigmap\,,\;\;\;\;\;\;
I_+(\isigma(n))\sigma
\;\;\;\longrightarrow\;\;\;
(\ldots n,n+1,\ldots)\,,
\label{c1}
\\
&&I_-(\isigmap(n))\sigmap\,,\;\;\;\;\;\;
I_-(\isigma(n))\sigma
\;\;\;\longrightarrow\;\;\;
(\ldots n+1,n,\ldots)\,,
\label{c2}
\eeqn
that is, $n$ and $n+1$ must be contiguous. This implies that all
flows for which this condition is not satisfied correspond to
matrix elements that do not diverge in the collinear limit.
In order to proceed as was done in the case of the soft limit, 
I can now impose the analogues of eq.~(\ref{svsS}). By considering 
only R-flows to be definite, there are two possible cases as shown
in eqs.~(\ref{c1}) and~(\ref{c2}), and therefore one has either
\beq
I_+(\isigma(n))\sigma\equiv\Sigma=(\ldots n,n+1,\ldots)\,,
\label{c3}
\eeq
or
\beq
I_-(\isigma(n))\sigma\equiv\Sigma=(\ldots n+1,n,\ldots)\,.
\label{c4}
\eeq
These equations impose the following constraints on $\Sigma$:
\beq
\iSigma(n)=\iSigma(n+1)-1
\label{c5}
\eeq
in the case of eq.~(\ref{c3}), and
\beq
\iSigma(n)=\iSigma(n+1)+1
\label{c6}
\eeq
in the case of eq.~(\ref{c4}). The conditions of eqs.~(\ref{c5}) 
and~(\ref{c6}) are obviously mutually exclusive. By repeating the same
exercise for L-flows, and by using eq.~(\ref{M0nn}), one finally arrives at:
\beq
\sum_{\sigma,\sigmap\in P_n^\prime}\zampsqn(\sigmap,\sigma)=
\sum_{\Sigma,\Sigmap\in P_{n+1}^\prime}
\delta(\Sigmap,\Sigma)\ampsqRn(\Sigmap,\Sigma)\,,
\label{zreqpp}
\eeq
where $\ampsqRn$ is given in eq.~(\ref{Mred}), and
\beq
\delta(\Sigmap,\Sigma)=\sum_{\alpha=-1,1}\sum_{\beta=-1,1} \alpha\beta\,
\delta\Big(\iSigmap(n),\iSigmap(n+1)+\alpha\Big)\,
\delta\Big(\iSigma(n),\iSigma(n+1)+\beta\Big).
\label{deltadef}
\eeq
The $\delta$ symbols that appear on the r.h.s.~of eq.~(\ref{deltadef}) 
are the ordinary Kronecker delta's. Note that only one of the four
terms in the sum in eq.~(\ref{deltadef}) can be different from
zero at given $\Sigma$ and $\Sigmap$. These four terms corresponds
to those on the r.h.s.~of eq.~(\ref{M0nn}); their signs there are
equivalent to the factor $\alpha\beta$ in eq.~(\ref{deltadef}). Furthermore,
$\delta(\Sigmap,\Sigma)$ is equal to zero when $n$ is not contiguous
to $n+1$ in either the L- or R-flow. These properties of 
eq.~(\ref{deltadef}) ensure that eq.~(\ref{zreqpp}) holds.
Putting all together, one obtains:
\beqn
\ampsqnpoC&=&\sum_{\Sigma,\Sigmap\in P_{n+1}^\prime}
\ampsqnpoC(\Sigmap,\Sigma)\,,
\label{McollFRsum}
\\
\ampsqnpoC(\Sigmap,\Sigma)&=&\frac{\gs^2}{k_n\mydot\knpo}
\,\delta(\Sigmap,\Sigma)\Bigg\{
\hat{P}_{gg}(z)\ampsqRn(\Sigmap,\Sigma)
\label{McollFR}
\\*&&\phantom{a}
\!+\half\hat{Q}_{gg^\star}(z)\,
\Bigg(\frac{\langle k_n k_{n+1}\rangle}{[k_n k_{n+1}]}
\wampsqRmpn(\Sigmap,\Sigma)
+\frac{[k_n k_{n+1}]}{\langle k_n k_{n+1}\rangle}
\wampsqRpmn(\Sigmap,\Sigma)
\Bigg)\Bigg\},
\nonumber
\eeqn
where
\beq
\wampsqRlln(\Sigmap,\Sigma)=
\brat{\ampn_{\rm\sss \RED\lambda}(\Sigmap)}
\ket{\ampn_{\rm\sss \RED\blambda}(\Sigma)},
\label{wMcollFR}
\eeq
or, in terms of scalar quantities:
\beq
\wampsqRlln(\Sigmap,\Sigma)=
\ampCSn_\lambda(\Sigmap_{\cancel{n+1}})^\star\,
C(\Sigmap,\Sigma)\,
\ampCSn_{\blambda}(\Sigma_{\cancel{n+1}})\,.
\label{wMcollFRsc}
\eeq
The consistency between eqs.~(\ref{MssMpp2}) and~(\ref{McollFR})
in the sense of eq.~(\ref{MSCdef}) can be proved by direct computation,
using eq.~(\ref{eiklim}). One can verify that the Kronecker delta's that
appear on the r.h.s.~of the latter equation combine effectively to give
$\delta(\Sigmap,\Sigma)$ defined in eq.~(\ref{deltadef}); the 
quickest way to see this is that of observing that the linear 
combination of the four eikonals in eq.~(\ref{MssMpp2}) can
be rewritten as follows:
\beq
\sum_{\alpha=-1,1}\sum_{\beta=-1,1} \alpha\beta\,
\eik{\Sigmap(\iSigmap(n+1)+\alpha)}{\Sigma(\iSigma(n+1)+\beta)},
\label{eiksum}
\eeq
i.e.~a similar form as eq.~(\ref{deltadef}). It is interesting to
notice that the collinear limit of an eikonal, eq.~(\ref{eiklim}),
constrains $n$ to be contiguous to $n+1$ in either the R- or the
L-flow, but not necessarily in both. However, when $n$ and $n+1$
are contiguous  only in one of the two flows, the linear combination in
eq.~(\ref{eiksum}) contains two divergent eikonals with opposite
signs, and therefore is ultimately non-singular. This is the reason
why eq.~(\ref{deltadef}) forces $n$ and $n+1$ to be contiguous
on both sides of the cut.

I point out that the treatment of the collinear limit as done above
is quite similar to that of the soft limit owing to the use of 
eq.~(\ref{Mcolldef}) rather than of eq.~(\ref{Mcolldef0}) as
a starting point, and in particular to the presence of the 
$Q^b$ operators in the reduced $n$-gluon matrix elements that
appear in the former expression. The same technical trick will
be adopted when treating the case of quark-gluon amplitudes.

\subsubsection{Fixed Born flows\label{sec:gluflowB}}
I now discuss the second of the strategies outlined at the
beginning of sect.~\ref{sec:gluflow}. I shall make extensive
use of the results obtained in sect.~\ref{sec:gluflowR},
and I shall start from considering the soft limit. 

Fixing Born flows is equivalent to the following interpretation
of eq.~(\ref{siSident}): at a given $\sigma$, one considers
all $(n+1)$-gluon flows $\Sigma=I_+(j)\sigma$ that arise when 
performing the sum over $j$. Equation~(\ref{Mppdef}) then implies that such 
$\Sigma$'s will constitute the set:
\beq
\zeta(\sigma)=\bigcup_{l=1}^n\,\Big\{I_+(\isigma(l))\,\sigma\Big\}\equiv
\bigcup_{j=1}^n\,\Big\{I_+(j)\,\sigma\Big\}\,.
\label{zetaset}
\eeq
The discussion that follows eq.~(\ref{siSident}) means that:
\beqn
&&\zeta(\sigma_1)\,\bigcap\,\zeta(\sigma_2)=\emptyset
\;\;\;\;\;\;\;\;
{\rm if}\;\;\;\;\sigma_1\ne\sigma_2\,,
\label{zetaI}
\\
&&\bigcup_{\sigma\in P_n^\prime}\zeta(\sigma)=P_{n+1}^\prime\,.
\label{zetaU}
\eeqn
Therefore, the sets $\zeta(\sigma)$ achieve a non-overlapping
partition of the space of the $(n+1)$-gluon flows
into subsets associated with $n$-gluon flows.
One is thus led to define $(n+1)$-gluon matrix elements
at fixed $n$-gluon flows:
\beqn
\ampsqnpo(\sigmap,\sigma)&=&
\sum_{\Sigma\in\zeta(\sigma)}\sum_{\Sigmap\in\zeta(\sigmap)}
\ampsqnpo(\Sigmap,\Sigma)\,,
\label{MnpoBorn}
\\
\ampsqnpo&=&\sum_{\sigma,\sigmap\in P_n^\prime}\ampsqnpo(\sigmap,\sigma)\,,
\label{Mnposum}
\eeqn
where eq.~(\ref{Mnposum}) follows from eqs.~(\ref{zetaI}) and~(\ref{zetaU}).
The derivation carried out in sect.~\ref{sec:gluflowR} then implies
that the soft limit of $\ampsqnpo(\sigmap,\sigma)$ is equal to 
$\ampsqnpoS(\sigmap,\sigma)$ defined in eq.~(\ref{Msoftdef2}).
The colour-linked Born's of eq.~(\ref{Mkldef2}) read, in terms 
of scalar quantities:
\beqn
\ampsqn_{kl}(\sigmap,\sigma)&=&-2\,\ampCSn(\sigmap)^\star\Big[
C\left(I_+(\isigmap(k))\,\sigmap,\,I_+(\isigma(l))\,\sigma\right)
\nonumber\\*&&\phantom{-2\,\ampCSn(\sigmap)}\!\!
-C\left(I_+(\isigmap(k))\,\sigmap,\,I_-(\isigma(l))\,\sigma\right)
\nonumber\\*&&\phantom{-2\,\ampCSn(\sigmap)}\!\!
-C\left(I_-(\isigmap(k))\,\sigmap,\,I_+(\isigma(l))\,\sigma\right)
\nonumber\\*&&\phantom{-2\,\ampCSn(\sigmap)}\!\!
+C\left(I_-(\isigmap(k))\,\sigmap,\,I_-(\isigma(l))\,\sigma\right)
\Big]\ampCSn(\sigma)\,.
\label{Mkldef2sc}
\eeqn

The case of the collinear limit is analogous to that of the soft one.
The relevant matrix element, $\ampsqnpoC(\sigmap,\sigma)$, has already
been defined, see eq.~(\ref{McollFB}). Since the $n$-gluon flows
are kept fixed here, the explicit expansion performed in eq.~(\ref{M0nn})
is not needed, and one can exploit eq.~(\ref{Casimir}) to rewrite:
\beqn
\ampsqnpoC(\sigmap,\sigma)&=&\frac{\gs^2}{k_n\mydot\knpo}\Bigg\{
P_{gg}(z)\ampsqn(\sigmap,\sigma)
\label{McollFB2}
\\*&&\phantom{aaaa}
\!+\half Q_{gg^\star}(z)\,\Bigg(
\frac{\langle k_n k_{n+1}\rangle}{[k_n k_{n+1}]}
\wampsqn_{-+}(\sigmap,\sigma)+
\frac{[k_n k_{n+1}]}{\langle k_n k_{n+1}\rangle}
\wampsqn_{+-}(\sigmap,\sigma)
\Bigg)\Bigg\}\,.
\nonumber
\eeqn
In terms of scalar quantities, $\ampsqn(\sigmap,\sigma)$ is given
in eq.~(\ref{Mmflowsc}), and
\beq
\wampsqn_{\lambda\blambda}(\sigmap,\sigma)=
\ampCSn_{\lambda}(\sigmap)^\star \,C(\sigmap,\sigma)\,
\ampCSn_{\blambda}(\sigma)\,.
\label{wMmflowsc}
\eeq
On the other hand, the expansion of eq.~(\ref{M0nn}) serves to define
the analogue of the set $\zeta(\sigma)$, namely:
\beq
\zeta_{\rm\sss C}(\sigma)=\Big\{I_+(\isigma(n))\,\sigma\,,\;
I_-(\isigma(n))\,\sigma\Big\}\,.
\label{zetasetC}
\eeq
At this point, the quantity whose collinear limit is given by
eq.~(\ref{McollFB2}) can be defined in the same way as that
in eq.~(\ref{MnpoBorn}), namely:
\beq
\sum_{\Sigma\in\zeta_{\rm\sss C}(\sigma)}
\sum_{\Sigmap\in\zeta_{\rm\sss C}(\sigmap)}
\ampsqnpo(\Sigmap,\Sigma)\,.
\label{MnpoBornC}
\eeq
However, this is not particularly convenient, since the soft
limit of eq.~(\ref{MnpoBornC}) is not eq.~(\ref{Msoftdef2}),
which renders it impossible to properly define the subtractions
of eq.~(\ref{subt}). On the other hand, one observes
that
\beq
\zeta_{\rm\sss C}(\sigma)\,\subseteq\,\zeta(\sigma)\,,
\eeq
and that all flows belonging to the (generally non empty) set
\beq
\zeta(\sigma)\,\setminus\,\zeta_{\rm\sss C}(\sigma)
\eeq
correspond to matrix elements which are non-singular in the
collinear limit. These two facts imply that not only the quantity
in eq.~(\ref{MnpoBornC}), but also $\ampsqnpo(\sigmap,\sigma)$ 
defined in eq.~(\ref{MnpoBorn}) has the collinear limit given
by eq.~(\ref{McollFB2}), or by its fully equivalent form
eq.~(\ref{McollFB}). One can then conclude that, 
at fixed $\sigma$ and $\sigmap$, the quantities
$\ampsqnpo(\sigmap,\sigma)$, $\ampsqnpoS(\sigmap,\sigma)$, and 
$\ampsqnpoC(\sigmap,\sigma)$ have all the properties required
to play the same roles as the first three matrix elements 
in eq.~(\ref{subt}). As far
as the soft-collinear counterterm is concerned, it is well defined
thanks to the consistency of eqs.~(\ref{Msoftdef2}) and~(\ref{McollFB})
in the sense of eq.~(\ref{MSCdef}). This can be proved exactly as
done in sect.~\ref{sec:glusum}, thanks to eq.~(\ref{colcons3}), i.e.
to colour conservation at fixed Born flows.

\section{Quark-gluon amplitudes\label{sec:quarks}}
I now turn to discussing the case of amplitudes where both quarks
and gluons are present. The real-emission process will contain
$q$ quark-antiquark pairs, and $n+1$ gluons:
\beq
0\;\longrightarrow\; 2q+(n+1)\,.
\label{QGRproc}
\eeq
The $i^{th}$ parton entering this process will be labelled
according to the following conventions\footnote{The label $i=0$
is not associated with any parton. In the following, I shall not
bother to exclude explicitly $i=0$ when summing over parton labels;
this condition will be understood.}:
\beqn
-2q\le i\le -q-1
\;\;\;\;&\longrightarrow&\;\;\;\;{\rm antiquarks}\,,
\label{qbars}
\\
-q\le i\le -1\phantom{-q}\;\;
\;\;\;\;&\longrightarrow&\;\;\;\;{\rm quarks}\,,
\label{qs}
\\
1\le i\le n+1\phantom{-}
\;\;\;\;&\longrightarrow&\;\;\;\;{\rm gluons}\,.
\label{glus}
\eeqn
An equal-flavour $q\qb$ pair will correspond to indices $i=-p$ and $i=-p-q$; 
for the moment, I shall limit myself to discussing the case of one pair
per quark flavour. The quarks may be massless of massive; there is no
need to distinguish these two cases, since the FKS subtraction formulae
do not explicitly depend on quark masses (with the obvious exception that
a massive $q\qb$ pair cannot induce a collinear singularity -- in the 
following, I shall always understand that a quark or an antiquark is 
massless when is involved in a collinear branching).
The structure of the underlying Born amplitudes will 
depend on the type of limit considered. When one of the gluons becomes soft,
or two gluons become collinear, or one quark/antiquark becomes collinear
to a gluon, the Born process is as follows:
\beq
0\;\longrightarrow\; 2q+n\,.
\label{QGBprocgg}
\eeq
In other words, this is the situation in which the FKS parton is
a gluon; consistently with what was done in sect.~\ref{sec:gluons}, 
such a gluon will have label $n+1$. Its FKS sisters will be labelled by $n$,
$-q$, and $-2q$ for $g\to gg$, $q\to qg$, and $\qb\to\qb g$ branchings
respectively. At the Born level, gluon number $n+1$ in eq.~(\ref{glus})
will not appear. The analogues of the labelling of 
eqs.~(\ref{qbars})--(\ref{glus}) will thus be:
\beqn
-2q\le i\le -q-1
\;\;\;\;&\longrightarrow&\;\;\;\;{\rm antiquarks}\,,
\label{B0qbars}
\\
-q\le i\le -1\phantom{-q}\;\;
\;\;\;\;&\longrightarrow&\;\;\;\;{\rm quarks}\,,
\label{B0qs}
\\
1\le i\le n\phantom{-q-\;\,\,}
\;\;\;\;&\longrightarrow&\;\;\;\;{\rm gluons}\,.
\label{B0glus}
\eeqn
On the other hand, in the case when the FKS parton
is a quark, i.e. for a $g\to q\qb$ branching, the underlying Born
will be a process:
\beq
0\;\longrightarrow\; 2(q-1)+(n+2)\,.
\label{QGBprocqq}
\eeq
The analogues of the labelling of eqs.~(\ref{qbars})--(\ref{glus})
will thus be:
\beqn
-2q+1\le i\le -q-1
\;\;\;\;&\longrightarrow&\;\;\;\;{\rm antiquarks}\,,
\label{Bqbars}
\\
-q+1\le i\le -1\phantom{-q}\;\;
\;\;\;\;&\longrightarrow&\;\;\;\;{\rm quarks}\,,
\label{Bqs}
\\
1\le i\le n+2\phantom{-}
\;\;\;\;&\longrightarrow&\;\;\;\;{\rm gluons}\,.
\label{Bglus}
\eeqn
The FKS parton will have label $-q$, and its sister label $-2q$
(which implies that, for this case to be non trivial, these
quarks must be massless).
At the Born level, the gluon with label $n+2$ will be identified
with the one branching into the $q\qb$ pair.
When describing the properties of a generic amplitude in the 
remainder of this section, I shall use the labelling of
eqs.~(\ref{B0qbars})--(\ref{B0glus}) to be definite.

The colours of gluons will be denoted as in eq.~(\ref{coladj}),
whereas in the case of quarks and antiquarks:
\beq
\left\{a_i\right\}_{i=-2q}^{-1}\,,\;\;\;\;\;\;\;
a_i\,\in\,\left\{1,\ldots N\right\}\,.
\label{colfund}
\eeq
An amplitude at a given colour configuration, i.e. the analogue
of eq.~(\ref{mgCDamp}), will be written in the following way:
\beq
\ampQGBgg(\amtq,\ldots a_n)=
\sum_{\gamma\in\flowBgg}\,\Lambda\left(\seta,\gamma\right)\,
\ampCSQGBgg(\gamma)\,,
\label{mgCDampQG}
\eeq
where $\gamma$ and $\flowBgg$ play the same role as $\sigma$ and $P_n^\prime$
in eq.~(\ref{mgCDamp}); namely, they denote a flow and the set of all flows 
relevant to the process of eq.~(\ref{QGBprocgg}) respectively. 
In order to determine the
precise forms of these quantities, I use again the representation of 
ref.~\cite{Mangano:1990by}. In particular, one sees that the
generic form of the colour structure that multiplies the dual
amplitude is: 
\beqn
\Lambda\left(\seta,\gamma\right)&=&
N^{-\rho(\gamma)}\,
\Big(\lambda^{a_{\sigma(t_0+1)}}\ldots
\lambda^{a_{\sigma(t_1)}}\Big)_{a_{-1}a_{\mu(-1-q)}}
\nonumber\\*&&\phantom{N^{aa}\;}\times
\Big(\lambda^{a_{\sigma(t_1+1)}}\ldots
\lambda^{a_{\sigma(t_2)}}\Big)_{a_{-2}a_{\mu(-2-q)}}
\nonumber\\*&&\phantom{N^{aa}\;}\times\ldots
\nonumber\\*&&\phantom{N^{aa}\;}\times
\Big(\lambda^{a_{\sigma(t_{q-1}+1)}}\ldots
\lambda^{a_{\sigma(t_q)}}\Big)_{\amoq a_{\mu(-2q)}}\,,
\label{Lambdagq}
\eeqn
with
\beq
\rho(\gamma)=\min\Big\{q-1,\,
\sum_{p=1}^q\delta\Big(\!\Mp-q,\mu(-p-q)\Big)\Big\}\,.
\label{rhodef}
\eeq
In eq.~(\ref{Lambdagq}), $\sigma$ and $\mu$ denote permutations 
(including cyclic ones) of the first $n$ and $q$ integers
respectively, $\sigma\in P_n$ and $\mu\in P_q$. The range of
the arguments of, and the values assumed by, the latter have been 
extended, so as:
\beqn
&&\mu(p)=-\mu(-p-q)-q\,,\;\;\;\;\;\;\;\;
-2q\le p\le -q-1\,,
\\
&&\Longrightarrow\phantom{aaaaaa}\mu(p)\in\{-q-1,\ldots -2q\}\,.
\eeqn
Finally, the set of $q+1$ integers $t_p$
\beqn
&&t=\left\{t_0,\ldots t_q\right\}\,,
\label{part1}
\\*&&
0=t_0\le t_1\le\ldots t_{q-1}\le t_q=n\,,
\label{part2}
\eeqn
achieves a partition of the ordered set of the first $n$ integers into
$q$ subsets of ordered integers, with the $p^{th}$ cell of the partition
defined to be:
\beq
(t_{p-1}+1,t_{p-1}+2,\ldots t_p-1,t_p)\,,\;\;\;\;\;\;\;\;1\le p\le q\,.
\label{pcell}
\eeq
The ensemble of such partitions I shall denote by
\beq
T_{n|q}\,.
\eeq
It should be stressed that the $p^{th}$ cell of the partition,
eq.~(\ref{pcell}), may coincide with the empty set, which one can 
formally write by setting
\beq
t_{p-1}=t_p\,.
\label{emptyline}
\eeq
This case of a zero-length cell corresponds to defining:
\beq
\Big(\lambda^{a_{\sigma(t_{p-1}+1)}}\ldots
\lambda^{a_{\sigma(t_p)}}\Big)_{a_{-p}a_{\mu(-p-q)}}
\equiv\delta_{a_{-p}a_{\mu(-p-q)}}
\eeq
as prescribed in ref.~\cite{Mangano:1990by}. Note that all cells of $t$ (in 
the case of amplitudes that feature only quarks), or all cells except one
(when there is at least one gluon entering the process) can have zero length.
Equation~(\ref{Lambdagq}) implies the following definition of flow:
\beqn
\gamma&=&\bigcup_{p=1}^q \gamma_p\,,
\label{qgflow}
\\
\gamma_p&=&\Big(\Mp\,;\sigma(t_{p-1}+1),\ldots\sigma(t_p);\mu(-p-q)\Big)\,.
\label{qgflowp}
\eeqn
The physical interpretation of these equations is straightforward.
Equations~(\ref{qgflowp}) represents a colour antenna that connects
the colour of quark $-p$ with the anticolour of antiquark $\mu(-p-q)$;
attached to this line there are $t_p-t_{p-1}$ gluons (therefore, in
the case of eq.~(\ref{emptyline}), there will be no gluons).
Then, the integer $\rho(\gamma)$, defined in eq.~(\ref{rhodef}), 
is the number of times when a quark-antiquark colour line coincides 
with a flavour line (minus one, when there is a maximal coincidence).
The flow, eq.~(\ref{qgflow}), is the set of all colour lines. 
By convention, the $p^{th}$ colour line begins with
quark $-p$. The construction above implies that
\beq
\flowBgg=\left(P_n,P_q,T_{n|q}\right)\,.
\label{Bflowdef}
\eeq
At this point, one can trivially extend the definitions of colour
vectors and the corresponding amplitudes given in sect.~\ref{sec:gluons}.
One has just to formally replace \mbox{$1\ldots n$} with 
\mbox{$-2q\ldots n$}, and use the definition of flows given
here rather than that of sect.~\ref{sec:gluons}. So for example
\beqn
&&\ket{\ampQGBgg(\amtq,\ldots a_n)}=
\ampQGBgg(\amtq,\ldots a_n)\,\ket{\amtq,\ldots a_n}\,,
\label{QGAcoldef}
\\
&&\ket{\ampQGBgg(\gamma)}=\sum_{\setaQGBgg}
\Lambda\left(\seta,\gamma\right)\,
\ampCSQGBgg(\gamma)\,\ket{\amtq,\ldots a_n}\,,
\label{QGAflowdef}
\eeqn
and so forth.

The case of amplitudes that feature more than one $q\qb$ pair 
per (at least) one flavour is essentially identical to what has been
discussed so far, and only requires additional information on flavour 
lines, which now are not unique. This information can be given e.g.~in
the following form:
\beqn
&&f=\bigcup_{p=1}^q \Big(\!-p,f_{-p}\Big)\,,
\label{flavconn}
\\
&&\{f_{-1},\ldots f_{-q}\}=
\{-q-1,\ldots -2q\}\,,
\eeqn
where $(-p,f_{-p})$ represents the flavour line that connects
quark $-p$ with antiquark $f_{-p}$. Clearly, this notation 
encompasses the equal-flavour case as well,
for which $f_{-p}=-p-q$ for all $p$'s. At this point, one can
define amplitudes at a fixed flavour configuration $f$, which
are gauge invariant, and then sum over all these configurations
to obtain the physical amplitude. The definition of the colour
structure given in eq.~(\ref{Lambdagq}) is unchanged, but
eq.~(\ref{rhodef}) needs be generalized to read:
\beq
\rho(\gamma)=\min\Big\{q-1,\,
\sum_{p=1}^q\delta\Big(\!f_{-p},\mu(-p-q)\Big)\Big\}\,.
\label{rhodefunf}
\eeq
The treatment of equal-flavour amplitudes does not pose any problems,
but slightly complicates the notation. For this reason, in the following
I shall deal explicitly only with the unequal-flavour case; the 
equal-flavour case can be easily recovered by adding a sum over the
flavour configurations given in eq.~(\ref{flavconn}).

\subsection{Subtraction of colour-summed matrix elements\label{sec:QGsum}}
I shall extend here what was done in sect.~\ref{sec:glusum} to the case
of quark-gluon amplitudes squared. The formulae for the relevant soft and
collinear limits are of course well known,
and thus I shall limit myself here to casting them in a form suited
to the calculations that I shall perform in the rest of this section.
As already pointed out before, for quark-gluon amplitudes one needs
distinguish the cases of the FKS parton being a gluon or a quark.

I start from the former case. The underlying Born dynamics is therefore
that of eq.~(\ref{QGBprocgg}). The formulae for the soft limit of the
amplitude squared are given in eqs.~(\ref{Msoftdef})--(\ref{Mkldef}),
with only formal changes due to relabelling, and to the fact that 
self-eikonals need not vanish any longer, owing to the possible presence
of massive quarks:
\beqn
\ampsqQGRS&=&\half\gs^2
\sum_{k,l=-2q}^n\eik{k}{l}\ampsqQGBgg_{kl}\,,
\label{MQGsoftdef}
\\
\eik{k}{l}&=&\frac{k_k\mydot k_l}{k_k\mydot\knpo\, k_l\mydot\knpo}
(1-\delta_{kl}\delta_{0m_k})\,,
\label{eikMdef}
\\
\ampsqQGBgg_{kl}&=&-2\bra{\ampQGBgg}\,\sum_b Q^b(k)Q^b(l)\;\ket{\ampQGBgg}\,.
\label{MQGkldef}
\eeqn
The colour operators $Q^b(k)$ are defined in eqs.~(\ref{Qdef})
and~(\ref{Qreprglu}) in the case when \mbox{$k>0$}, i.e.~when $k$ is
a gluon. This definition needs be extended to the case of
quarks and antiquarks. We have (see e.g. ref.~\cite{Frederix:2009yq}):
\beqn
&&\bra{a_k a_{n+1}} Q^b(k)\ket{c_k}=
\delta_{ba_{n+1}}\left(Q^b(k)\right)_{a_kc_k}\,,
\label{Qmatdef2}
\\
&&\left(Q^b(k)\right)_{a_kc_k}=\lambda^b_{a_kc_k}
\;\;\;\;\;\;\;\;\;~~~-q\le k\le -1
\;\;\;\;\;\;\;\;\;\;\,\Leftrightarrow\;\;\;\;k~{\rm is~a~quark}\,,
\label{Qqdef}
\\
&&\left(Q^b(k)\right)_{a_kc_k}=-\lambda^b_{c_ka_k}
\;\;\;\;\;\;\;\;-2q\le k\le -q-1
\;\;\;\;\Leftrightarrow\;\;\;\;k~{\rm is~an~antiquark}\,.\phantom{aaaa}
\label{Qaqdef}
\eeqn
With this, one can generalize eq.~(\ref{Casimir}) to read:
\beqn
\sum_b Q^b(k)Q^b(k)&=&C(k)\,I\,,
\label{Casimir2}
\\
C(k)&=&\CA\;\;\;\;\;\;\;\;\;\;\;\;\;\;\;\;k>0\,,
\\
C(k)&=&\CF\;\;\;\;\;\;\;\;\;\;\;\;\;\;\;\;k<0\,.
\eeqn
The collinear limit can then be cast in the same form
as in eq.~(\ref{Mcolldef}):
\beqn
\ampsqQGRC&=&\frac{\gs^2}{k_s\mydot\knpo}\Bigg\{
\hat{P}_{\ident_s\ident_s}(z)\zampsqQGBggs
+\hat{Q}_{\ident_s\ident_s^\star}(z)\,
\Re\Bigg(\frac{\langle k_s k_{n+1}\rangle}{[k_s k_{n+1}]}
\wzampsqQGBggsmp\Bigg)\Bigg\},\phantom{aaa}
\label{MQGcolldef}
\\
\zampsqQGBggs&=&\bra{\ampQGBgg}\,\sum_b Q^b(s)Q^b(s)\;\ket{\ampQGBgg}\,,
\label{MQGcolldefz}
\\
\wzampsqQGBggsmp&=&\bra{\ampQGBgg_-}\,\sum_b Q^b(s)Q^b(s)\;
\ket{\ampQGBgg_+}\,,
\label{MtQGcolldefz}
\eeqn
where $z$ is defined as in eq.~(\ref{zdef}), with $k_n\to k_s$ there.
As was discussed at the beginning of sect.~\ref{sec:quarks}, I shall
take
\beq
s=n,-q,-2q
\eeq
as representatives of the cases of the $g\to gg$, $q\to qg$, and $\qb\to\qb g$ 
branchings respectively. I denoted by $\ident_s$ the identity of the
relevant branching parton (i.e., a gluon, a quark, and an antiquark)
and in eq.~(\ref{MQGcolldef}), consistently with eq.~(\ref{APred}),
I have defined:
\beq
\hat{P}_{\ident_s\ident_s}(z)=\frac{1}{C(s)}P_{\ident_s\ident_s}(z)
\,,\;\;\;\;\;\;\;\;\;
\hat{Q}_{\ident_s\ident_s^\star}(z)=
\frac{1}{C(s)}Q_{\ident_s\ident_s^\star}(z)\,.
\label{APredall}
\eeq
The consistency between eqs.~(\ref{MQGsoftdef}) and~(\ref{MQGcolldef}),
in the sense of eq.~(\ref{MSCdef}), can be proved in exactly the same
way as was done in sect.~\ref{sec:gluons}, thanks to the fact that the
colour-conservation condition is fulfilled in the case of quark-gluon
amplitudes as well (see appendix~\ref{sec:appa}):
\beq
\sum_{k=-2q}^n Q^b(k)\ket{\ampQGBgg}=0\;\;\;\;\;\;
\Longrightarrow\;\;\;\;\;\;
\mathop{\sum_{k=-2q}}_{k\ne s}^{n}
Q^b(k)\ket{\ampQGBgg}=-Q^b(s)\ket{\ampQGBgg}\,,
\label{colconsQG}
\eeq
and because eq.~(\ref{APsoft}) generalizes to read:
\beq
\hat{P}_{\ident_s\ident_s}(z)\;
\stackrel{z\to 1}{\longrightarrow}\;
\frac{2}{1-z}\,,
\;\;\;\;\;\;\;\;
\hat{Q}_{\ident_s\ident_s^\star}(z)\;
\stackrel{z\to 1}{\longrightarrow}\; 0\,,
\label{APsoftgen}
\eeq
that is, it also holds in the case of $q\to qg$ and $\qb\to\qb g$ 
branchings\footnote{For the branchings of quarks and antiquarks, the 
kernels $Q(z)$ are actually identical to zero~\cite{Frixione:1995ms}.}.

I shall now discuss the case of the $g\to q\qb$ branchings. Here,
the underlying Born is the process of eq.~(\ref{QGBprocqq}), and
only the collinear limit is relevant (there are no soft singularities
when the FKS parton is a quark). The analogues of 
eqs.~(\ref{MQGcolldef})--(\ref{MtQGcolldefz}) read:
\beqn
\ampsqQGRC&=&\frac{\gs^2}{\kmq\mydot\kmtq}\Bigg\{
\hat{P}_{qg}(z)\zampsqQGBqqqq
\nonumber\\*&&\phantom{\frac{\gs^2}{\kmq\mydot\kmtq}}
+\hat{Q}_{qg^\star}(z)\,
\Re\Bigg(\frac{\langle\kmq\kmtq\rangle}{[\kmq\kmtq]}
\wzampsqQGBqqqqmp\Bigg)\Bigg\},\phantom{aaa}
\label{MQGcollqqdef}
\\
\zampsqQGBqqqq&=&\bra{\ampQGBqq}\,\sum_{bc} G_{bc}^\star G_{bc}\;
\ket{\ampQGBqq}\,,
\label{amps0qqdef}
\\
\wzampsqQGBqqqqmp&=&\bra{\ampQGBqq_-}\,\sum_{bc} G_{bc}^\star G_{bc}\;
\ket{\ampQGBqq_+}\,,
\label{tamps0qqdef}
\eeqn
where
\beq
\hat{P}_{qg}(z)=\frac{1}{\TF}P_{qg}(z)\,,\;\;\;\;\;\;\;\;\;
\hat{Q}_{qg^\star}(z)=\frac{1}{\TF}Q_{qg^\star}(z)\,.
\label{APqgred}
\eeq
The (non-hermitian) operators $G_{bc}$ are the analogues of the
operators $Q^b$ introduced before. In colour space, their 
action is defined as follows:
\beq
\bra{\amtq\amoq}\,G_{bc}\,\ket{\anpt}=
\delta_{b\amoq}\delta_{c\amtq}
\lambda^{\anpt}_{\amoq\amtq}\,.
\label{Gdef}
\eeq
In other words, the operators $G_{bc}$ annihilate gluon number $n+2$,
and create the $q\qb$ pair with labels $-q$ and $-2q$ and colours
$b$ and $c$ respectively. Note that
eq.~(\ref{MQGcollqqdef}) does coincide with the usual form of the
collinear limit since, as is easy to see from eq.~(\ref{Gdef}),
one has:
\beq
\sum_{bc}G_{bc}^\star G_{bc}=\TF\,I\,,
\label{Casimir5}
\eeq
which is the analogue of eq.~(\ref{Casimir2}).

\subsection{Subtraction at fixed colour configurations\label{sec:QGCD}}
The procedure here is identical to that followed in 
sect.~\ref{sec:gluCD}, except for trivial changes in notation.
Its essence is that of inserting into the matrix element squared
the projector onto a given colour configuration, that in this
case reads:
\beq
\ket{\amtq,\ldots\anpo}\bra{\amtq,\ldots\anpo}\,.
\label{projanpoQG}
\eeq
After doing that, one will also have to exploit the definitions
of the $Q^b$ and $G_{bc}$ operators given in eqs.~(\ref{Qmatdef2})
and~(\ref{Gdef}). The quantities whose soft and collinear limits
one needs to construct are therefore
\beq
\ampsqQGR(\amtq,\ldots\anpo)=
\abs{\ampQGR(\amtq,\ldots\anpo)}^2\,.
\label{MQGCD}
\eeq
Its soft limit can be obtained from eq.~(\ref{MQGsoftdef}) through
the procedure described above:
\beqn
&&\ampsqQGRS(\amtq,\ldots\anpo)=\half\gs^2
\sum_{k,l=-2q}^n\eik{k}{l}
\ampsqQGBgg_{kl}(\amtq,\ldots\anpo)\,,
\label{MQGsoftCD}
\\
&&\ampsqQGBgg_{kl}(\amtq,\ldots\anpo)=
\label{MQGklCD}
\\*&&\phantom{aaaaaaaa}
-2\bra{\ampQGBgg}\,
Q^{\anpo}(k)\ket{\amtq,\ldots\anpo}\bra{\amtq,\ldots\anpo}
Q^{\anpo}(l)\;\ket{\ampQGBgg}\,.
\nonumber
\eeqn
which are fully analogous to eqs.~(\ref{MsoftCD}) and~(\ref{MklCD}).
In terms of scalar quantities, eq.~(\ref{MQGklCD}) gets rewritten
as follows:
\beqn
\label{MQGklscalar}
&&\ampsqQGBgg_{kl}(\amtq,\ldots\anpo)=
\\*&&\phantom{aaaaa}
-2\sum_{b_k^\prime b_l}
\ampQGBgg(\amtq\,\ldots b_k^\prime\,\ldots a_l\,\ldots a_n)^\star
{\cal Q}_{\ident_k\ident_l}(\anpo;a_k,a_l)_{b_k^\prime b_l}
\nonumber\\*&&\phantom{aaaaa}\phantom{-2\sum_{b_k^\prime b_l}}\times
\ampQGBgg(\amtq\,\ldots a_k\,\ldots b_l\,\ldots a_n),
\nonumber
\eeqn
with
\beq
{\cal Q}_{\ident_k\ident_l}(\anpo;a_k,a_l)_{bc}=
\left(Q^{\anpo}(k)\right)_{ba_k}
\left(Q^{\anpo}(l)\right)_{a_lc}\,.
\label{Qmatdef3}
\eeq
This manifestly coincides with eq.~(\ref{Qmatdef}) when $k$ and $l$
are both gluons. In the case of $k$ being a gluon and $l$ being a
quark, in SU(3) there are $3\mydot 8^2$ ${\cal Q}$ matrices, 
of dimension $8\!\times\!3$ (for the case when the quark is 
on the left of the cut, and the gluon is on the right, the matrices
have dimensions $3\!\times\!8$). When $k$ and $l$ are both quarks,
the are $3^2\mydot 8$ ${\cal Q}$ matrices,  of dimension $3\!\times\!3$.
If quark(s) are replaced by antiquark(s), the number of matrices and
their dimensions will not change, but their forms will
(see eqs.~(\ref{Qqdef}) and~(\ref{Qaqdef})).

The case of the collinear limit is very similar. For $g\to gg$, $q\to qg$,
and $\qb\to\qb g$ branchings, from eq.~(\ref{MQGcolldef}) one obtains:
\beqn
\ampsqQGRC(\amtq,\ldots\anpo)&=&\frac{\gs^2}{k_s\mydot\knpo}\Bigg\{
\hat{P}_{\ident_s\ident_s}(z)\zampsqQGBggs(\amtq,\ldots\anpo)
\label{MQGcollCD}
\\*&&\phantom{aaaaaa}
\!+\hat{Q}_{\ident_s\ident_s^\star}(z)\,
\Re\Bigg(\frac{\langle k_s k_{n+1}\rangle}{[k_s k_{n+1}]}
\wzampsqQGBggsmp(\amtq,\ldots\anpo)\Bigg)\Bigg\},
\nonumber
\eeqn
with
\beqn
\zampsqQGBggs(\amtq,\ldots\anpo)&=&
\bra{\ampQGBgg}\,Q^{\anpo}(s)\ket{\amtq,\ldots\anpo}
\nonumber\\*&\times&
\bra{\amtq,\ldots\anpo}Q^{\anpo}(s)\;\ket{\ampQGBgg}\,,
\\
\wzampsqQGBggsmp(\amtq,\ldots\anpo)&=&
\bra{\ampQGBgg_-}\,Q^{\anpo}(s)\ket{\amtq,\ldots\anpo}
\nonumber\\*&\times&
\bra{\amtq,\ldots\anpo}Q^{\anpo}(s)\;\ket{\ampQGBgg_+}\,.
\eeqn
In terms of scalar quantities:
\beqn
&&\zampsqQGBggs(\amtq,\ldots\anpo)=
\label{MQG0scalar}
\\*&&\phantom{aaaaa}
\sum_{b_s^\prime b_s}
\ampQGBgg(\{a_i\}_{i\ne s},b_s^\prime)^\star
{\cal Q}_{\ident_s\ident_s}(\anpo;a_s,a_s)_{b_s^\prime b_s}
\ampQGBgg(\{a_i\}_{i\ne s},b_s)\,,
\nonumber
\\
&&\wzampsqQGBggsmp(\amtq,\ldots\anpo)=
\label{wMQG0scalar}
\\*&&\phantom{aaaaa}
\sum_{b_s^\prime b_s}
\ampQGBgg_-(\{a_i\}_{i\ne s},b_s^\prime)^\star
{\cal Q}_{\ident_s\ident_s}(\anpo;a_s,a_s)_{b_s^\prime b_s}
\ampQGBgg_+(\{a_i\}_{i\ne s},b_s)\,.
\nonumber
\eeqn
The colour matrices that enter this equation are such that
\beq
\sum_{\anpo}\sum_{a_s}
{\cal Q}_{\ident_s\ident_s}(\anpo;a_s,a_s)_{bc}
=C(s)\,\delta_{bc}\,,
\label{Casimir4}
\eeq
which extends eq.~(\ref{Casimir3}), and is the matrix form
of eq.~(\ref{Casimir2}).

The case of the $g\to q\qb$ collinear splitting can be derived
starting from eq.~(\ref{MQGcollqqdef}). One obtains:
\beqn
\ampsqQGRC(\amtq,\ldots\anpo)&=&\frac{\gs^2}{\kmq\mydot\kmtq}\Bigg\{
\hat{P}_{qg}(z)\zampsqQGBqqqq(\amtq,\ldots\anpo)
\label{MQGcollqqCD}
\\*&&
+\hat{Q}_{qg^\star}(z)\,
\Re\Bigg(\frac{\langle\kmq\kmtq\rangle}{[\kmq\kmtq]}
\wzampsqQGBqqqqmp(\amtq,\ldots\anpo)\Bigg)\Bigg\},
\nonumber
\eeqn
where
\beqn
\zampsqQGBqqqq(\amtq,\ldots\anpo)&=&
\bra{\ampQGBqq}\,G_{\amoq\amtq}^\star \ket{\amtq,\ldots\anpo}
\nonumber\\*&\times&
\bra{\amtq,\ldots\anpo}G_{\amoq\amtq}\;\ket{\ampQGBqq}\,,
\label{MQGqqzqq}
\\
\wzampsqQGBqqqqmp(\amtq,\ldots\anpo)&=&
\bra{\ampQGBqq_-}\,G_{\amoq\amtq}^\star \ket{\amtq,\ldots\anpo}
\nonumber\\*&\times&
\bra{\amtq,\ldots\anpo}G_{\amoq\amtq}\;\ket{\ampQGBqq_+}\,.
\label{tMQGqqzqq}
\eeqn
In terms of scalar quantities, eqs.~(\ref{MQGqqzqq}) and~(\ref{tMQGqqzqq}) 
read:
\beqn
&&\zampsqQGBqqqq(\amtq,\ldots\anpo)=
\label{MQG0qqscalar}
\\*&&\phantom{aaaaa}
\sum_{b_{n+2}^\prime b_{n+2}}
\ampQGBqq(\amtqpo\,\ldots\cancel{\amoq},a_{-q+1}\ldots
\anpo,b_{n+2}^\prime)^\star
{\cal G}(\amoq,\amtq)_{b_{n+2}^\prime b_{n+2}}
\nonumber\\*&&\phantom{aaaaa}\phantom{-2\sum_{b_k^\prime b_l}}\times
\ampQGBqq(\amtqpo\,\ldots\cancel{\amoq},a_{-q+1}\ldots
\anpo,b_{n+2}),
\nonumber
\\
&&\wzampsqQGBqqqqmp(\amtq,\ldots\anpo)=
\label{wMQG0qqscalar}
\\*&&\phantom{aaaaa}
\sum_{b_{n+2}^\prime b_{n+2}}
\ampQGBqq_-(\amtqpo\,\ldots\cancel{\amoq},a_{-q+1}\ldots
\anpo,b_{n+2}^\prime)^\star
{\cal G}(\amoq,\amtq)_{b_{n+2}^\prime b_{n+2}}
\nonumber\\*&&\phantom{aaaaa}\phantom{-2\sum_{b_k^\prime b_l}}\times
\ampQGBqq_+(\amtqpo\,\ldots\cancel{\amoq},a_{-q+1}\ldots
\anpo,b_{n+2}),
\nonumber
\eeqn
with
\beq
{\cal G}(\amoq,\amtq)_{bc}=
\lambda_{\amtq\amoq}^b\lambda_{\amoq\amtq}^c\,.
\label{Gmatdef}
\eeq
As is clear from eq.~(\ref{Gmatdef}), there are $3^2$ ${\cal G}$ matrices,
of dimensions $8\!\times\!8$ in SU(3). They are such that
\beq
\sum_{\amoq}\sum_{\amtq}{\cal G}(\amoq,\amtq)_{bc}=
\TF\,\delta_{bc}\,,
\label{Casimir6}
\eeq
which is the matrix form of eq.~(\ref{Casimir5}).

\subsection{Subtraction at fixed flows\label{sec:QGflow}}
Although slightly more complicated than in the case of gluon
amplitudes owing to the presence of quarks, the derivation
of the subtraction at fixed flows for quark-gluon amplitudes
proceeds exactly as that carried out in sect.~\ref{sec:gluflow}.
More specifically, I shall use the expressions of the soft
and collinear limits of the colour-summed matrix elements squared, 
and expand them in terms of either the real or the Born flows, using 
fixed-flows amplitudes and the properties of the colour operators 
$Q^b$ and $G_{bc}$. All the necessary technical ingredients are reported in
appendix~\ref{sec:appQG}. There, I extend (if necessary) 
the definitions and derivations of appendix~\ref{sec:appglu},
relevant to gluon amplitudes, in a way which renders it
straightforward the use of arguments by analogy.

There is however one type of singular limit of the quark-gluon
amplitudes that does not have an analogue in the case of gluon 
amplitudes, namely that due to $g\to q\qb$ branchings. I shall 
deal with these singularities in sect.~\ref{sec:flowqgg}.

\subsubsection{Fixed real flows\label{sec:QGflowR}}
The quantity whose limits at given real flows I seek to
construct is the analogue of that in eq.~(\ref{MnpoSS}), namely:
\beqn
\ampsqQGR(\Gammap,\Gamma)&=&\brat{\ampQGR(\Gammap)}\ket{\ampQGR(\Gamma)}
\nonumber\\*&=&
\ampCSQGR(\Gammap)^\star \,C(\Gammap,\Gamma)\,\ampCSQGR(\Gamma)\,,
\label{MQGRSS}
\eeqn
where (see eq.~(\ref{Bflowdef})):
\beq
\Gamma,\Gammap\;\in\;\flowR=\left(P_{n+1},P_q,T_{n+1|q}\right)\,,
\label{Cflowdef}
\eeq
and the colour-flow matrix elements are defined analogously to 
eq.~(\ref{CFmatdef}):
\beq
C(\Gammap,\Gamma)=\sum_{\setaQGR}\Lambda\left(\seta,\Gammap\right)^\star
\Lambda\left(\seta,\Gamma\right)\,.
\label{CFQGmatdef}
\eeq
As was done in sect.~\ref{sec:gluflowR}, one begins by introducing 
real-emission-level colour vectors associated with an underlying
Born dynamics:
\beq
\ket{\ampRQGR(\Gamma)}=
\sum_{\setaQGR}
\Lambda\left(\seta,\Gamma\right)
\ampCSQGBgg\left(\Gamma_{\cancel{n+1}}\right)
\ket{\amtq,\ldots \anpo}\,.
\label{ampRQGggdef}
\eeq
The Born-level flow $\Gamma_{\cancel{n+1}}$ is obtained from the
real-level one $\Gamma$ by eliminating from it the $(n+1)^{th}$ gluon.
More precisely, if
\beq
\Gamma=(\Sigma,M,T)\in\flowR\,,\;\;\;\;\;\;\;\;
T=\left\{T_0,\ldots T_q\right\}\,,
\eeq
then
\beq
\Gamma_{\cancel{n+1}}=(\sigma,\mu,t)\,\in\,\flowBgg\,,
\eeq
where
\beqn
\sigma&\equiv&\left(\sigma(1),\ldots\sigma(n)\right)=
\left(\Sigma(1),\ldots\cancel{n+1},\ldots\Sigma(n+1)\right)\,,
\label{removenpo}
\\
\mu&=&M\,,
\label{samemu}
\\
t&=&\left\{t_0,\ldots t_q\right\}\,,
\label{tvsTa}
\eeqn
with
\beqn
t_p&=&T_p\phantom{-1aaaaaaaa}\forall\;p~~~{\rm s.t.}~~~T_p+1\le\iSigma(n+1)\,,
\label{tvsTb}
\\
t_p&=&T_p-1\phantom{aaaaaaaa}{\rm otherwise}\,.
\label{tvsTc}
\eeqn
Equation~(\ref{removenpo}) is the same as in the case of gluon amplitudes --
the relative ordering of the gluons has nothing to do with their belonging
to a given colour line. Eq.~(\ref{removenpo}) states the obvious but
crucial fact that by removing one gluon the structure of the colour
lines is not affected. Eqs.~(\ref{removenpo})--(\ref{tvsTc}) imply that 
all colour lines in $\Gamma_{\cancel{n+1}}$ contain exactly the same
particles and in the same order as those in $\Gamma$. The exception
is the colour line derived from the colour line $\Gamma_r\in\Gamma$ that
contains gluon $n+1$, since the latter gluon must be removed by definition;
the remaining particles on that line, however, have the same relative
ordering as in $\Gamma_r$.

At this point, as was done in sect.~\ref{sec:gluflowR} I start by expressing
the soft limit of the matrix element squared as a sum over Born-level 
flows. The analogues of eqs.~(\ref{QAeqAs})--(\ref{Mkldef2}) 
read as follows:
\beqn
\ampsqQGRS&=&\sum_{\gamma,\gammap\in\flowBgg}
\ampsqQGRS(\gammap,\gamma)\,,
\label{QAeqAsQG}
\\
\ampsqQGRS(\gammap,\gamma)&=&\half\gs^2
\sum_{k,l=-2q}^n\eik{k}{l}
\ampsqQGBgg_{kl}(\gammap,\gamma)\,,
\label{MQGsoftdef2}
\\
\ampsqQGBgg_{kl}(\gammap,\gamma)&=&-2
\bra{\ampQGBgg(\gamma^\prime)}\,\sum_{b} Q^{b}(k)
Q^{b}(l)\;\ket{\ampQGBgg(\gamma)}\,.
\label{MQGkldef2}
\eeqn
The colour-linked Born's of eq.~(\ref{MQGkldef2}) can be computed
using the results of appendix~\ref{sec:appQG}. In particular, from 
eqs.~(\ref{QCglut})--(\ref{QCaqrkt}) one obtains:
\beqn
\sum_{b}Q^{b}(l)\ket{\ampQGBgg(\gamma)}&=&
\ket{\ampRQGR(I_+(\igamma(l))\,\gamma)}-
\ket{\ampRQGR(I_-(\igamma(l))\,\gamma)}\,,
\phantom{aaaa}
\label{QonAg}
\\
\sum_{b}Q^{b}(l)\ket{\ampQGBgg(\gamma)}&=&
\ket{\ampRQGR(I_+(\igamma(l))\,\gamma)}\,,
\label{QonAq}
\\
\sum_{b}Q^{b}(l)\ket{\ampQGBgg(\gamma)}&=&
-\ket{\ampRQGR(I_-(\igamma(l))\,\gamma)}\,,
\label{QonAa}
\eeqn
for the cases when $l$ is a gluon ($1\le l\le n$), 
a quark ($-q\le l\le -1$), or an antiquark ($-2q\le l\le -q-1$)
respectively. I have denoted by $\igamma(l)$ the position of particle
$l$ in the list that defines the colour flow (see appendix~\ref{sec:appQG},
also for the definitions of the operators $I_\pm$).
One can now plug eqs.~(\ref{QonAg})--(\ref{QonAa})
into eq.~(\ref{MQGkldef2}) and proceed to the explicit expansion
of that equation. This calculation can be performed using
the very same method as in appendix~\ref{sec:appQG}, since the
colour structure of the sums over $k$ or $l$ is the same as that
relevant to colour conservation. More explicitly, one starts by
splitting the sum over particle labels into sums relevant to
single colour lines, and then converts sums over particle labels
into sums over particle positions. This implies the following
manipulations (where I consider the sum over $l$ to give a definite
example):
\beqn
\sum_{l=-2q}^n f(l)&=&
\sum_{p=1}^q\sum_{l\in\gamma_p}f(l)=
\sum_{p=1}^q\left(
f(-p)+\sum_{l\in\gamma_p}f(l)\delta_{\ident_lg}
+f(\mu(-p-q))
\right)
\label{sumQGa}
\\
&=&\sum_{p=1}^q\left(
\delta_{j(-p)}f(\gamma(j))
+\sum_{j=t_{p-1}+1}^{t_p}f(\gamma(j))
+\delta_{j(-p-q)}f(\gamma(j))\right)\,,
\label{sumQGb}
\eeqn
where the three terms on the r.h.s.'s of eqs.~(\ref{sumQGa}) 
and~(\ref{sumQGb}) correspond to the quark, gluons, and antiquark
contributions respectively, relevant to the $p^{th}$ colour line.
The arguments of $f()$ in eq.~(\ref{sumQGb}) imply that the operators
$I_\pm(\igamma(\gamma(j)))=I_\pm(j)$ will result from 
eqs.~(\ref{QonAg})--(\ref{QonAa}). For each colour line, there will
be $t_p-t_{p-1}+1$ contributions due to $I_+$ operators (from the gluons
and the quark), and $t_p-t_{p-1}+1$ contributions due to $I_-$ operators 
(from the gluons and the antiquark). Furthermore, the $I_-$ operators
can be rewritten in terms of $I_+$ operators by using the identity
in eq.~(\ref{ImeqIpQG}). After some trivial algebra, fully analogue
to that of eqs.~(\ref{Ypres})--(\ref{mcont}) bar the presence of
the eikonal factors in eq.~(\ref{MQGsoftdef2}), one obtains:
\beqn
\ampsqQGRS(\gammap,\gamma)&=&-\gs^2
\sum_{\pp,p=1}^q
\sum_{i=t_{\pp-1}+1\ominus 1}^{t_{\pp}}
\sum_{j=t_{p-1}+1\ominus 1}^{t_p}
\Big\{
\eik{\gammap(i)}{\gamma(j)}
-\eik{\gammap(i)}{\gamma(j\oplus 1)}
\nonumber\\*&&
\phantom{-\gs^2\sum_{i,j=1}^n aaaaaaaaaaa}
-\eik{\gammap(i\oplus 1)}{\gamma(j)}
+\eik{\gammap(i\oplus 1)}{\gamma(j\oplus 1)}\Big\}
\nonumber\\*&&
\phantom{-\gs^2\sum_{i,j=1}^n aaaaaaa}
\times\brat{\ampRQGR(I_+(i)\,\gammap)}
\ket{\ampRQGR(I_+(j)\,\gamma)}.
\label{MQGppdef}
\eeqn
This result is basically identical to that of eq.~(\ref{Mppdef}),
which can be easily understood as follows (for the sake of this
argument, it is sufficient to consider the case of flows with
one colour line). In a gluon amplitude, the $(n+1)^{th}$ gluon 
appears in a given position in a real-level flow as the result
of the actions of the operators $I_+$ and $I_-$ associated with
the gluon that precedes and follows it in the flow respectively.
When the $(n+1)^{th}$ gluon occupies the first (last) position,
the relevant $I_+$ ($I_-$) operator is that associated with the
last (first) gluon in the flow, as a consequence of the cyclicity
of the trace. In the case of quark-gluon amplitudes, the colour
structure is not cyclic, so the latter argument is not valid. 
However, when the $(n+1)^{th}$ gluon is the left-most (right-most)
gluon of the flow, there exists a relevant $I_+$ ($I_-$) operator, associated
with the quark (antiquark) rather than with the last (first) gluon in
the flow. It should finally be stressed that the fact that the 
quantities $i+1$ and $j+1$ which appear in eq.~(\ref{Mppdef}) are
replaced by $i\oplus 1$ and $j\oplus 1$ in eq.~(\ref{MQGppdef}) is only
a formal difference. The physical meaning is in fact the same: in both
cases, these denote the particles to the immediate right of particles
$i$ and $j$.

Equation~(\ref{MQGppdef}) is in a form suited to transform the
sum in eq.~(\ref{QAeqAsQG}) in a sum over real flows. The procedure
is identical to that of sect.~\ref{sec:gluflowR} and, as in that
case, stems from the observation that there is a one-to-one 
correspondence between a real flow $\Gamma$, and a pair $(\gamma,j)$,
where $\gamma$ is a Born flow, and $j$ is an integer whose range is
the same as that in eq.~(\ref{MQGppdef}). This follows from the 
existence and uniqueness of the solution of the equation:
\beqn
&&I_+(j)\,\gamma=\Gamma\;\;\;\;\Longleftrightarrow\;\;\;\;
\label{gvsG}
\\*&&\phantom{aaaaaa}
\ldots
(\ldots \gamma(j),\,n+1,\,\gamma(j\oplus 1),\ldots)
\ldots
\nonumber
\\*&&\phantom{aaa}\;=
\ldots
(\ldots \Gamma(j),\,\Gamma(j\oplus 1),\,\Gamma(j\oplus 1\oplus 1),\ldots)
\ldots\,,
\label{gjvsG}
\eeqn
as was the case for its analogue, eq.~(\ref{svsS}) or eq.~(\ref{sjvsS}).
The use of the $\oplus$ operator in eq.~(\ref{gjvsG}) in place of the
ordinary $+$ that appears in eq.~(\ref{sjvsS}) does not underscore any
difference between the two cases, but simply allows one to treat with the
same notation the cases in which $j$ is a gluon or is a quark (note
that the range spanned by $j$ in eq.~(\ref{MQGppdef}) implies that $j$
is never an antiquark). The existence and uniqueness of the solution of 
eq.~(\ref{gjvsG}) can be easily understood e.g.~from the relationships that 
connect real- and Born-level flows according to
eqs.~(\ref{removenpo})--(\ref{tvsTc}).
Furthermore, one has to take into account the fact that for the two
flows $\gamma_a=(\sigma_a,\mu_a,t_a)$ and $\gamma_b=(\sigma_b,\mu_b,t_b)$ 
to be different, it is sufficient that only one of the conditions
$\sigma_a\ne\sigma_b$; $\mu_a\ne\mu_b$; $t_a\ne t_b$ be fulfilled.
By solving eq.~(\ref{gjvsG}) one obtains:
\beqn
\gamma&=&\Gamma_{\cancel{n+1}}\,,
\\
j&=&\iGamma(n+1)\ominus 1\,,
\label{jsolQG}
\eeqn
which also imply that:
\beqn
\gamma(j)&=&\Gamma(\iGamma(n+1)\ominus 1)\,,
\\
\gamma(j\oplus 1)&=&\Gamma(\iGamma(n+1)\oplus 1)\,.
\eeqn
In full analogy with eq.~(\ref{sseqSS}), the arguments above 
are such that:
\beq
\sum_{\gamma\in\flowBgg}
\sum_{p=1}^q
\sum_{j=t_{p-1}+1\ominus 1}^{t_p}
f(\gamma,j)
\;\equiv\;
\sum_{\Gamma\in\flowR}
f\left(\Gamma_{\cancel{n+1}},\iGamma(n+1)\ominus 1\right)\,.
\label{ggeqGG}
\eeq
As a consistency check of eq.~(\ref{ggeqGG}), one can again use
a counting argument as done in the case of gluon amplitudes.
Denoting by ${\cal N}(n,q)$ the number of partitions defined
in eqs.~(\ref{part1}) and~(\ref{part2}), one can prove that:
\beqn
{\cal N}(n,q)&=&\left(
\begin{array}{c}
n+q-1\\
q-1
\end{array}
\right)\,.
\label{Npart}
\eeqn
The number of terms that appear in the sums on the l.h.s.~of
eq.~(\ref{ggeqGG}) is
\beq
q!\,n!\,{\cal N}(n,q)\,(n+q)\,,
\label{numpo}
\eeq
since there are $(n+q)$ different ways of inserting a gluon
in a Born flow ($n+q$ is in fact the number of terms of the sums
over $p$ and $j$ in eq.~(\ref{ggeqGG})). On the r.h.s.~of
eq.~(\ref{ggeqGG}) the number of terms is instead:
\beq
q!\,(n+1)!\,{\cal N}(n+1,q)\,.
\label{numpt}
\eeq
Thanks to eq.~(\ref{Npart}), one sees that the numbers
in eqs.~(\ref{numpo}) and~(\ref{numpt}) coincide.

By using the results above, their analogues for the L-flows, 
and eq.~(\ref{MQGppdef}), one finally obtains the soft limit 
of the matrix element squared at fixed real flows:
\beqn
\ampsqQGRS&=&\sum_{\Gamma,\Gammap\in\flowR}\ampsqQGRS(\Gammap,\Gamma)
\label{QAeqAsQGFR}
\\
\ampsqQGRS(\Gammap,\Gamma)&=&-\gs^2
\Big\{
\eik{\Gammap(\iGammap(n+1)\ominus 1)}{\Gamma(\iGamma(n+1)\ominus 1)}
\label{MQGssMpp2}
\\*&&\phantom{-\gs^2\!\!\!}
-\eik{\Gammap(\iGammap(n+1)\ominus 1)}{\Gamma(\iGamma(n+1)\oplus 1)}
\nonumber\\*&&\phantom{-\gs^2\!\!\!}
-\eik{\Gammap(\iGammap(n+1)\oplus 1)}{\Gamma(\iGamma(n+1)\ominus 1)}
\nonumber\\*&&\phantom{-\gs^2\!\!\!}
+\eik{\Gammap(\iGammap(n+1)\oplus 1)}{\Gamma(\iGamma(n+1)\oplus 1)}\Big\}
\ampsqRQGBgg(\Gammap,\Gamma),
\nonumber
\\
\ampsqRQGBgg(\Gammap,\Gamma)&=&
\brat{\ampRQGR(\Gammap)}
\ket{\ampRQGR(\Gamma)}\,.
\label{MQGppdef2}
\eeqn
The form of eq.~(\ref{MQGssMpp2})
is fully analogous to that of eq.~(\ref{MssMpp2}).
In terms of scalar quantities:
\beq
\ampsqRQGBgg(\Gammap,\Gamma)=
\ampCSQGBgg(\Gammap_{\cancel{n+1}})^\star\,
C(\Gammap,\Gamma)\,
\ampCSQGBgg(\Gamma_{\cancel{n+1}})\,.
\label{MQGppdef2sc}
\eeq

The derivation of the soft limit of the matrix elements given above
shows the complete similarity between the cases of gluon and 
quark-gluon amplitudes. The collinear limits relevant
to the branchings $g\to gg$, $q\to qg$, and $\qb\to\qb g$
are also fully analogous, as one can easily understand by comparing 
eqs.~(\ref{Mcolldef})--(\ref{Mtcolldefz}) with
eqs.~(\ref{MQGcolldef})--(\ref{MtQGcolldefz}).
In other words, in the case of quark-gluon amplitudes one can
follow the procedure that in sect.~\ref{sec:gluflowR} has led me
from eq.~(\ref{QAeqAc}) to eqs.~(\ref{McollFRsum})--(\ref{wMcollFR}).
I refrain from doing that explicitly, and limit myself to
present the final result:
\beqn
\ampsqQGRC&=&\sum_{\Gamma,\Gammap\in\flowR}
\ampsqQGRC(\Gammap,\Gamma)\,,
\label{MQGcollFRsum}
\\
\ampsqQGRC(\Gammap,\Gamma)&=&\frac{\gs^2}{k_s\mydot\knpo}
\,\delta(\Gammap,\Gamma)\Bigg\{
\hat{P}_{\ident_s\ident_s}(z)\ampsqRQGBgg(\Gammap,\Gamma)
\label{MQGcollFR}
\\*&&\!\!\!\!\!\!\!\!\!\!\!\!
+\half\hat{Q}_{\ident_s\ident_s^\star}(z)\,
\Bigg(\frac{\langle k_s k_{n+1}\rangle}{[k_s k_{n+1}]}
\wampsqRmpQGBgg(\Gammap,\Gamma)
+\frac{[k_s k_{n+1}]}{\langle k_s k_{n+1}\rangle}
\wampsqRpmQGBgg(\Gammap,\Gamma)
\Bigg)\Bigg\},
\nonumber
\eeqn
where
\beq
\wampsqRllQGBgg(\Gammap,\Gamma)=
\brat{{\cal A}_{\rm\sss \RED\lambda}^{(2q;n)}(\Gammap)}
\ket{{\cal A}_{\rm\sss \RED\blambda}^{(2q;n)}(\Gamma)},
\eeq
and
\beq
\delta(\Gammap,\Gamma)=\sum_{\alpha=-1,1}\sum_{\beta=-1,1} \alpha\beta\,
\delta\Big(\iGammap(s),\iGammap(n+1)\oplus\alpha\Big)\,
\delta\Big(\iGamma(s),\iGamma(n+1)\oplus\beta\Big).
\label{deltaQGdef}
\eeq
In terms of scalar quantities:
\beq
\wampsqRllQGBgg(\Gammap,\Gamma)=
\ampCSQGBgg_\lambda(\Gammap_{\cancel{n+1}})^\star\,
C(\Gammap,\Gamma)\,
\ampCSQGBgg_{\blambda}(\Gamma_{\cancel{n+1}})\,.
\label{wMQGppdef2sc}
\eeq
The very close correspondence between eqs.~(\ref{MQGssMpp2}) 
and~(\ref{MQGcollFR}), and their counterparts in the case of
gluon amplitudes, implies that the consistency (in the sense of 
eq.~(\ref{MSCdef})) between the soft and collinear limits of
quark-gluon matrix elements can be proved in exactly the same way 
as explained at the end of sect.~\ref{sec:gluflowR}.

One comment is in order as far as the form of eq.~(\ref{deltaQGdef})
is concerned. In the case of a $g\to gg$ branching, i.e.~when
$s=n$, the situation is strictly identical to that explicitly
presented in eqs.~(\ref{M0nn})--(\ref{c2}). However, when one
considers a $q\to qg$ branching, i.e. when $s=-q$, the analogue
of eq.~(\ref{M0nn}) contains only one term, because of
eq.~(\ref{QonAq}). Therefore, the relevant real flows, 
which correspond to those in eqs.~(\ref{c1}) and~(\ref{c2}), are:
\beq
I_+(\igammap(-q))\gammap\,,\;\;\;\;\;\;
I_+(\igamma(-q))\gamma
\;\;\;\longrightarrow\;\;\;
(-q;n+1,\ldots)\,.
\label{c1q}
\eeq
These correspond to the contribution obtained when $\alpha=-1$
and $\beta=-1$ in eq.~(\ref{deltaQGdef}). However, it is easy to
see that if either $\alpha=1$ or $\beta=1$ in eq.~(\ref{deltaQGdef}),
then $\delta(\Gammap,\Gamma)=0$ if $s=-q$. In fact, if
$\delta(\Gammap,\Gamma)=1$ these cases would correspond
to real flows where the $(n+1)^{th}$ gluon occupied the position
at the immediate left of quark $-q$, which is impossible.
Similar arguments apply to the $\qb\to\qb g$ branching, 
i.e. when $s=-2q$. The bottom line is that eq.~(\ref{deltaQGdef})
can indeed be used to describe all branchings considered here,
as claimed before.

\subsubsection{Fixed Born flows\label{sec:QGflowB}}
The results relevant to fixing the Born flows can be 
straightforwardly obtained from sect.~\ref{sec:QGflowR}.
Once again, the analogy with the case of gluon amplitudes
is very close.

The soft limit of the matrix element has already been presented
in eqs.~(\ref{QAeqAsQG})--(\ref{MQGkldef2}). The real-emission-level
quantity whose soft limit is $\ampsqQGRS(\gammap,\gamma)$ will be
a linear combination of the fixed-real-flow matrix elements
$\ampsqQGR(\Gammap,\Gamma)$ defined in eq.~(\ref{MQGRSS}):
\beq
\ampsqQGR(\gammap,\gamma)=
\sum_{\Gamma\in\xi(\gamma)}\sum_{\Gammap\in\xi(\gammap)}
\ampsqQGR(\Gammap,\Gamma)\,,
\label{MQGnpoBorn}
\eeq
for suitable sets of real flows $\xi(\gamma)$ and $\xi(\gammap)$, 
to be defined later.
In terms of scalar quantities, the colour-linked Born's of
eq.~(\ref{MQGkldef2}) read:
\beqn
\ampsqQGBgg_{kl}(\gammap,\gamma)&=&-2\,\ampCSQGBgg(\gammap)^\star\Big[
C\left(I_+(\igammap(k))\,\gammap,\,I_+(\igamma(l))\,\gamma\right)
\left(1-\delta_{\qb\ident_k}\right)
\left(1-\delta_{\qb\ident_l}\right)
\nonumber\\*&&\phantom{-2\,\ampCSQGBgg(\gammap)}\!\!
-C\left(I_+(\igammap(k))\,\gammap,\,I_-(\igamma(l))\,\gamma\right)
\left(1-\delta_{\qb\ident_k}\right)
\left(1-\delta_{q\ident_l}\right)
\nonumber\\*&&\phantom{-2\,\ampCSQGBgg(\gammap)}\!\!
-C\left(I_-(\igammap(k))\,\gammap,\,I_+(\igamma(l))\,\gamma\right)
\left(1-\delta_{q\ident_k}\right)
\left(1-\delta_{\qb\ident_l}\right)
\nonumber\\*&&\phantom{-2\,\ampCSQGBgg(\gammap)}\!\!
+C\left(I_-(\igammap(k))\,\gammap,\,I_-(\igamma(l))\,\gamma\right)
\left(1-\delta_{q\ident_k}\right)
\left(1-\delta_{q\ident_l}\right)
\Big]
\nonumber\\*&&\!\!\times
\ampCSQGBgg(\gamma)\,.
\label{MQGkldef2sc}
\eeqn
The $(1-\delta)$ terms that multiply the colour-flow matrix elements
in the r.h.s.~of eq.~(\ref{MQGkldef2sc}) exclude the contributions
of the $I_-$ and $I_+$ operators in the case when the corresponding
particle is a quark or an antiquark respectively, as dictated
by eqs.~(\ref{QonAq}) and~(\ref{QonAa}).

Equation~(\ref{MQGnpoBorn}) is the analogue of eq.~(\ref{MnpoBorn}), and 
the set $\xi(\gamma)$ is the analogue of its counterpart $\zeta(\sigma)$
in the case of gluon amplitudes, defined in eq.~(\ref{zetaset}).
Since $\zeta(\sigma)$ is the set of all real flows
that can be obtained by acting on a given Born flow $\sigma$
with the operators $I_\pm$ relevant to the computation of the
colour-linked Born's, one can obtain $\xi(\gamma)$ precisely
in the same way. Hence, from eqs.~(\ref{MQGkldef2})--(\ref{QonAa}),
one gets:
\beqn
\xi(\gamma)&=&\bigcup_{l=1}^n\,\Big\{I_+(\igamma(l))\,\gamma\,,\;
I_-(\igamma(l))\,\gamma\Big\}
\bigcup_{l=-q}^{-1}\,\Big\{I_+(\igamma(l))\,\gamma\Big\}
\bigcup_{l=-2q}^{-q-1}\,\Big\{I_-(\igamma(l))\,\gamma\Big\}
\nonumber\\*&=&
\bigcup_{p=1}^q\bigcup_{j=t_{p-1}+1\ominus 1}^{t_p}
\Big\{I_+(j)\,\gamma\Big\},
\label{xiset}
\eeqn
where the last form follows from eq.~(\ref{ImeqIpQG}), but could
equally well be deduced directly from eqs~(\ref{QonAg})--(\ref{MQGppdef}).
The derivation presented in sect.~\ref{sec:QGflowR} implies that:
\beqn
&&\xi(\gamma_1)\,\bigcap\,\xi(\gamma_2)=\emptyset
\;\;\;\;\;\;\;\;
{\rm if}\;\;\;\;\gamma_1\ne\gamma_2\,,
\label{xiI}
\\
&&\bigcup_{\gamma\in\flowBgg}\xi(\gamma)=\flowR\,,
\label{xiU}
\eeqn
as in eqs.~(\ref{zetaI}) and~(\ref{zetaU}). Hence:
\beqn
\ampsqQGR=\sum_{\gamma,\gammap\in\flowBgg}\ampsqQGR(\gammap,\gamma)\,,
\label{MQGnposum}
\eeqn
which confirms that indeed the quantity defined eq.~(\ref{MQGnpoBorn})
has the property that one expects.

In the collinear limit, the matrix elements at fixed Born flows
can be obtained directly from eqs.~(\ref{MQGcolldef})--(\ref{MtQGcolldefz}),
by using the identity of eq.~(\ref{Casimir2}):
\beqn
\ampsqQGRC(\gammap,\gamma)&=&\frac{\gs^2}{k_s\mydot\knpo}\Bigg\{
P_{\ident_s\ident_s}(z)\ampsqQGBgg(\gammap,\gamma)
\label{MQGcollFB2}
\\&&
+\half Q_{\ident_s\ident_s^\star}(z)\,\Bigg(
\frac{\langle k_s k_{n+1}\rangle}{[k_s k_{n+1}]}
\wampsqQGBgg_{-+}(\gammap,\gamma)+
\frac{[k_s k_{n+1}]}{\langle k_s k_{n+1}\rangle}
\wampsqQGBgg_{+-}(\gammap,\gamma)
\Bigg)\Bigg\}\,,
\nonumber
\eeqn
with the reduced matrix elements that have the usual definitions:
\beqn
\ampsqQGBgg(\gammap,\gamma)&=&\brat{\ampQGBgg(\gammap)}
\ket{\ampQGBgg(\gamma)}
\nonumber\\*&=&
\ampCSQGBgg(\gammap)^\star \,C(\gammap,\gamma)\,\ampCSQGBgg(\gamma)\,,
\label{MQGcolldefzsc}
\\
\wampsqQGBgg_{\lambda\blambda}(\gammap,\gamma)&=&
\brat{\ampQGBgg_{\lambda}(\gammap)}
\ket{\ampQGBgg_{\blambda}(\gamma)}
\nonumber\\*&=&
\ampCSQGBgg_{\lambda}(\gammap)^\star \,C(\gammap,\gamma)\,
\ampCSQGBgg_{\blambda}(\gamma)\,.
\label{MtQGcolldefzsc}
\eeqn
The very same arguments as in the case of gluon amplitudes can 
now be applied. Namely, one can construct the sets
\beq
\xi_{\rm\sss C}(\gamma)=\Big\{I_+(\igamma(n))\,\gamma\,,\;
I_-(\igamma(n))\,\gamma\Big\}\,,
\;\;\;\;
\Big\{I_+(\igamma(-q))\Big\}\,,
\;\;\;\;
\Big\{I_-(\igamma(-2q))\Big\}\,,
\label{xisetC}
\eeq
relevant to the $g\to gg$, $q\to qg$, and $\qb\to\qb g$ branchings
respectively. However, one finds again that
\beq
\xi_{\rm\sss C}(\gamma)\,\subseteq\,\xi(\gamma)\,,
\eeq
for the three cases, and that all flows belonging to the sets
\beq
\xi(\gamma)\,\setminus\,\xi_{\rm\sss C}(\gamma)
\eeq
do not induce collinear singularities. Therefore, the matrix elements
defined in eq.~(\ref{MQGnpoBorn}) have the collinear limits given
by eq.~(\ref{MQGcollFB2}), and can thus be used with the latter
and with those of eq.~(\ref{MQGsoftdef2}) for the subtractions
of eq.~(\ref{subt}).

\subsubsection{The $g\to q\qb$ branching at fixed flows\label{sec:flowqgg}}
As was the case for the other collinear branchings discussed so far,
the starting point is the expression of the colour-summed collinear
limits (here, eqs.~(\ref{MQGcollqqdef})--(\ref{tamps0qqdef})), where one
writes the scattering amplitudes using their representations in terms
of Born colour flows. Therefore:
\beqn
&&\ampsqQGRC=\sum_{\gamma,\gammap\in\flowBqq}\ampsqQGRC(\gammap,\gamma)\,,
\label{Mcollqqsumf}
\\
&&\ampsqQGRC(\gammap,\gamma)=\frac{\gs^2}{\kmq\mydot\kmtq}\Bigg\{
\hat{P}_{qg}(z)\zampsqQGBqqqq(\gammap,\gamma)
\label{MQGcollqqflow}
\\*&&\phantom{aaaa}
\!+\half\hat{Q}_{qg^\star}(z)\,
\Bigg(\frac{\langle\kmq\kmtq\rangle}{[\kmq\kmtq]}
\wzampsqQGBqqqqmp(\gammap,\gamma)
+\frac{[\kmq\kmtq]}{\langle\kmq\kmtq\rangle}
\wzampsqQGBqqqqpm(\gammap,\gamma)
\Bigg)\Bigg\},
\nonumber
\eeqn
where
\beqn
\zampsqQGBqqqq(\gammap,\gamma)&=&
\bra{\ampQGBqq(\gammap)}\,\sum_{bc} G_{bc}^\star G_{bc}\;
\ket{\ampQGBqq(\gamma)}\,,
\label{amps0qqflow}
\\
\wzampsqQGBqqqqll(\gammap,\gamma)&=&
\bra{\ampQGBqq_\lambda(\gammap)}\,\sum_{bc} G_{bc}^\star G_{bc}\;
\ket{\ampQGBqq_{\blambda}(\gamma)}\,.
\label{tamps0qqflow}
\eeqn
Hence (using the same arguments as in eq.~(\ref{doublesum})), the 
relevant quantities to compute are the colour vectors:
\beq
G_{bc}\;\ket{\ampQGBqq(\gamma)}
\label{GonA}
\eeq
(or their bra counterparts),
which are the analogues of those in eqs.~(\ref{QonAg})--(\ref{QonAa})
or eq.~(\ref{QonA}). By using the definition of the operator $G_{bc}$
given in eq.~(\ref{Gdef}), one sees that the colour part of eq.~(\ref{GonA})
involves the computation of:
\beqn
&&\sum_{\anpt}\lambda^{\anpt}_{bc}
\Big(\lambda^{a_{\sigma(t_{r-1}+1)}}\ldots
\lambda^{a_{\sigma(\isigma(n+2)-1)}}
\lambda^{\anpt}
\lambda^{a_{\sigma(\isigma(n+2)+1)}}\ldots
\lambda^{a_{\sigma(t_r)}}\Big)_{\amor a_{\mu(-r-q)}}=
\nonumber\\*&&\phantom{aaaaaaa}\half
\Big(\lambda^{a_{\sigma(t_{r-1}+1)}}\ldots
\lambda^{a_{\sigma(\isigma(n+2)-1)}}\Big)_{\amor c}
\Big(\lambda^{a_{\sigma(\isigma(n+2)+1)}}\ldots
\lambda^{a_{\sigma(t_r)}}\Big)_{b a_{\mu(-r-q)}}
\nonumber\\*&&\phantom{aaaa}-\frac{1}{2N}\,\delta_{bc}\,
\Big(\lambda^{a_{\sigma(t_{r-1}+1)}}\ldots
\lambda^{a_{\sigma(\isigma(n+2)-1)}}
\lambda^{a_{\sigma(\isigma(n+2)+1)}}\ldots
\lambda^{a_{\sigma(t_r)}}\Big)_{\amor a_{\mu(-r-q)}}
\,,
\nonumber\\*
\label{Gacts}
\eeqn
where, as the notation suggests, $r$ is the colour line to which
gluon $n+2$ belongs, and the r.h.s.~of eq.~(\ref{Gacts}) has been 
computed using eq.~(\ref{sumijkl}). Therefore, for any Born flow
\beq
\gamma\in\flowBqq\,,\;\;\;\;\;\;\;\;
\gamma=\bigcup_{p=1}^{q-1} \gamma_p\,,
\eeq
eq.~(\ref{Gacts}) suggests to define two operators as follows:
\beq
n+2\in\gamma_r\,,\phantom{aaaaa}
\\
J\gamma=\left(\mathop{\bigcup_{p=1}}_{p\ne r}^{q-1} \gamma_p\right)\;
\bigcup\;\left(J\gamma_r\right)\,,
\phantom{aaaaa}
K\gamma=\left(\mathop{\bigcup_{p=1}}_{p\ne r}^{q-1} \gamma_p\right)\;
\bigcup\;\left(K\gamma_r\right)\,,
\eeq
where by construction $J\gamma, K\gamma\in\flowR$, and
\beqn
J\gamma_r&=&
\Big(\Mr\,;\sigma(t_{r-1}+1),\ldots\sigma(\isigma(n+2)-1);-2q\Big)\;\bigcup
\nonumber\\*&&
\Big(\Mq\,;\sigma(\isigma(n+2)+1),\ldots\sigma(t_r);\mu(-r-q)\Big)\,,
\label{Jrdef}
\\
K\gamma_r&=&\Big(\Mr\,;\sigma(t_{r-1}+1),\ldots\sigma(\isigma(n+2)-1),
\nonumber\\*&&\phantom{\Big(\Mr\,;}
\sigma(\isigma(n+2)+1),\ldots\sigma(t_r);\mu(-r-q)\Big)\;\bigcup\;
\Big(\Mq\,;-2q\Big)\,.
\label{Krdef}
\eeqn
In other words, the operator $J$ splits in two the colour line to which gluon 
$n+2$ belongs. The gluons to the left of gluon $n+2$ now belong to a colour
line in which the colour of the quark of the original line forms an antenna
with the anticolour of antiquark $-2q$ (that emerges from the 
$g\to q\qb$ branching), whereas the gluons to the right of gluon $n+2$
now belong to a colour line in which the colour of quark $-q$ (that emerges 
from the $g\to q\qb$ branching) forms an antenna with the anticolour of the
antiquark of the original line. On the other hand, operator $K$ simply 
removes gluon $n+2$ from the original colour line, and creates the colour
line that connects $-q$ with $-2q$; the latter line does not contain
any gluons. With the definitions above, one arrives at:
\beq
\sum_{bc}G_{bc}\;\ket{\ampQGBqq(\gamma)}=
\half\ket{\ampRQGRqq(J\gamma,\gamma)}-
\half\ket{\ampRQGRqq(K\gamma,\gamma)}\,,
\label{Gonvect}
\eeq
where the definition of eq.~(\ref{ampRQGggdef}) has been extended:
\beqn
&&\ket{\ampRQGRqq(\Gamma,\gamma)}=
\sum_{\setaQGR}
\Lambda\left(\seta,\Gamma\right)
\ampCSQGBqq\left(\gamma\right)
\ket{\amtq,\ldots \anpo}\,.\phantom{aaa}
\label{ampRQGqqdef}
\\*
&&{\rm for~any}\phantom{aaa}\Gamma\in\flowR\,,\;\;\;\;\;\;\;\;
\gamma\in\flowBqq\,.
\eeqn
Note that the factor $1/N$ that appears in the second term on the
r.h.s.~of eq.~(\ref{Gacts}) is indeed contained in the second term 
on the r.h.s.~of eq.~(\ref{Gonvect}) owing to the definition of $K$,
eq.~(\ref{Krdef}). In fact, the latter equation shows that $K\gamma$
features a colour line $(-q;-2q)$ which is also a flavour line,
and therefore:
\beq
\rho(K\gamma)=\rho(\gamma)+1\,.
\eeq
At this points, one plugs eq.~(\ref{Gonvect}) into eqs.~(\ref{amps0qqflow})
and~(\ref{tamps0qqflow}). Then, the sums over Born flows in 
eq.~(\ref{Mcollqqsumf}) can be turned into sums over real flows using
the same technique as was employed in eq.~(\ref{siSident}). To be
definite, I consider the case of R-flows, that of L-flows being
identical. One has
\beqn
&&\sum_{\gamma\in\flowBqq}\Big(f(J\gamma,\gamma)+g(K\gamma,\gamma)\Big)=
\nonumber\\*&&\phantom{aaaa}
\sum_{\gamma\in\flowBqq}\sum_{\Gamma\in\flowR}
\Big(\delta(\Gamma,J\gamma)f(J\gamma,\gamma)+
\delta(\Gamma,K\gamma)g(K\gamma,\gamma)\Big)\,,
\eeqn
where I have used the identities:
\beqn
1&=&\sum_{\Gamma\in\flowR} \!\!\delta(\Gamma,J\gamma)\,,
\label{dGJg}
\\
1&=&\sum_{\Gamma\in\flowR} \!\!\delta(\Gamma,K\gamma)\,,
\label{dGKg}
\eeqn
which hold because, for a given $\gamma$, the flows
$J\gamma$ and $K\gamma$ exist and are unique. The $\delta$ functions
can then be used to get rid of the sum over $\gamma$, which requires
fixing $\Gamma$, and solving:
\beqn
\Gamma&=&J\gamma\,,
\label{GvsJg}
\\
\Gamma&=&K\gamma\,,
\label{GvsKg}
\eeqn
for $\gamma$ (these equations are the analogues of eq.~(\ref{svsS})
or eq.~(\ref{gvsG})).
I start by considering eq.~(\ref{GvsJg}). As one can deduce from 
eq.~(\ref{Jrdef}), eq.~(\ref{GvsJg}) has a solution if and only if
$\Gamma$ belongs to a subset $\flowRJ$ of the set of real flows,
defined as follows:
\beq
\Gamma\equiv(\sigma,\mu,t)\in\flowRJ\subseteq\flowR
\;\;\;\;\Longleftrightarrow\;\;\;\;
\mu(-2q)\ne -2q\,.
\label{flowJdef}
\eeq
From eq.~(\ref{Jrdef}), it is also obvious that when such a solution
exists, it is also unique, and I shall denote it by $\gamma=\iJ\Gamma$.
Therefore:
\beq
\sum_{\gamma\in\flowBqq}\!\!\!f(J\gamma,\gamma)\equiv
\sum_{\gamma\in\flowBqq}\sum_{\Gamma\in\flowR}
\delta(\Gamma,J\gamma)f(J\gamma,\gamma)=
\sum_{\Gamma\in\flowRJ}\!\!\!f(\Gamma,\iJ\Gamma)\,.
\label{sumGgJ}
\eeq
Let me now turn to eq.~(\ref{GvsKg}) which, as can be seen from 
eq.~(\ref{Krdef}), has a solution if and only if $\Gamma$
belongs to a subset $\flowRK$ of the set of real flows,
defined as follows:
\beq
\Gamma\equiv(\sigma,\mu,t)\in\flowRK\subseteq\flowR
\;\;\;\;\Longleftrightarrow\;\;\;\;
\mu(-2q)=-2q~~{\rm and}~~t_{q-1}=t_q\equiv n+1\,.
\label{flowKdef}
\eeq
At variance with the case of the operator $J$, however, the solution
of eq.~(\ref{GvsKg}) is not unique. In fact, the operator $K$ just removes
gluon $n+2$ from the first $q-1$ colour lines, replacing it with the
colour line $(-q;-2q)$. Hence, the Born flows in which all quarks and
gluons have the same relative positions, up to that of gluon $n+2$,
will result in the same real flow $K\gamma$. It is easy to convince
oneself that this happens exactly $(n+q)$ times. I shall therefore 
define the {\em multi-valued} inverse of the operator $K$, which will give
the solutions to eq.~(\ref{GvsKg}), as follows:
\beq
\iKi\Gamma\in\flowBqq\;\;\;\;\;\;{\rm for~any}\;\;\;\;\;\;
1\le i\le n+q\,,\;\;\;\;\;\;
\Gamma\in\flowRK\,.
\eeq
The explicit form of $\iKi\Gamma$ can be worked out from eq.~(\ref{Krdef});
it corresponds to inserting gluon $n+2$ in all possible ways in $\Gamma$,
except in the colour line $(-q,-2q)$, which is removed\footnote{This implies
that $\iK$ could be written in terms of the operators $I_+$ defined for
gluon $n+2$, if need be.}. Putting all this together, one obtains:
\beq
\sum_{\gamma\in\flowBqq}\!\!\!g(K\gamma,\gamma)\equiv
\sum_{\gamma\in\flowBqq}\sum_{\Gamma\in\flowR}
\delta(\Gamma,K\gamma)g(K\gamma,\gamma)=
\sum_{\Gamma\in\flowRK}\sum_{i=1}^{n+q} g(\Gamma,\iKi\Gamma)\,.
\label{sumGgK}
\eeq
Counting arguments can again be given as a consistency check of the
results obtained here. The number of elements in the sets of flows
relevant to the derivation above are:
\beqn
\#\left(\flowBqq\right)&=&(q-1)!\,(n+2)!\,{\cal N}(n+2,q-1)\,,
\\
\#\left(\flowRJ\right)&=&(q-1)\,(q-1)!\,(n+1)!\,{\cal N}(n+1,q)\,,
\\
\#\left(\flowRK\right)&=&(q-1)!\,(n+1)!\,{\cal N}(n+1,q-1)\,,
\eeqn
with ${\cal N}$ given in eq.~(\ref{Npart}). By direct computation,
one shows that
\beqn
\#\left(\flowBqq\right)&=&\#\left(\flowRJ\right)\,,
\\
\#\left(\flowBqq\right)&=&\#\left(\flowRK\right)\;(n+q)\,.
\eeqn
Hence, the same number of terms appear on the two sides of
eqs.~(\ref{sumGgJ}) and~(\ref{sumGgK}). 
Equations~(\ref{Mcollqqsumf})--(\ref{tamps0qqflow}) 
can now be expressed in terms of real flows:
\beqn
&&\ampsqQGRC=\sum_{\Gamma,\Gammap\in\flowRJ\bigcup\flowRK}
\ampsqQGRC(\Gammap,\Gamma)\,,
\label{McollqqsumRf}
\\
&&\ampsqQGRC(\Gammap,\Gamma)=\frac{\gs^2}{4\kmq\mydot\kmtq}\Bigg\{
\hat{P}_{qg}(z)\ampsqQGBqq_{\rm\sss \RED,AB}(\Gammap,\Gamma)
\label{MQGcollqqfinal}
\\*&&\phantom{aaaa}
\!+\half\hat{Q}_{qg^\star}(z)\,
\Bigg(\frac{\langle\kmq\kmtq\rangle}{[\kmq\kmtq]}
\wampsqQGBqq_{\rm\sss \RED,AB,-+}(\Gammap,\Gamma)
+\frac{[\kmq\kmtq]}{\langle\kmq\kmtq\rangle}
\wampsqQGBqq_{\rm\sss \RED,AB,+-}(\Gammap,\Gamma)
\Bigg)\Bigg\},
\nonumber
\eeqn
where, in eq.~(\ref{MQGcollqqfinal}):
\beq
A,\,B\;\in\;\{J,\,K\}\,,\;\;\;\;\;\;\;\;
\Gammap\in\flowRA\,,\;\;\;\;\;\;\;\;
\Gamma\in\flowRB\,.
\eeq
The reduced matrix elements squared are defined as follows:
\beqn
\ampsqQGBqq_{\rm\sss \RED,JJ}(\Gammap,\Gamma)&=&
\brat{\ampRQGRqq(\Gammap,\iJ\Gammap)}
\ket{\ampRQGRqq(\Gamma,\iJ\Gamma)}\,,
\label{MredJJ}
\\
\ampsqQGBqq_{\rm\sss \RED,JK}(\Gammap,\Gamma)&=&-\sum_{j=1}^{n+q}
\brat{\ampRQGRqq(\Gammap,\iJ\Gammap)}
\ket{\ampRQGRqq(\Gamma,\iKj\Gamma)}\,,
\label{MredJK}
\\
\ampsqQGBqq_{\rm\sss \RED,KJ}(\Gammap,\Gamma)&=&-\sum_{i=1}^{n+q}
\brat{\ampRQGRqq(\Gammap,\iKi\Gammap)}
\ket{\ampRQGRqq(\Gamma,\iJ\Gamma)}\,,
\label{MredKJ}
\\
\ampsqQGBqq_{\rm\sss \RED,KK}(\Gammap,\Gamma)&=&
\sum_{i=1}^{n+q}\sum_{j=1}^{n+q}
\brat{\ampRQGRqq(\Gammap,\iKi\Gammap)}
\ket{\ampRQGRqq(\Gamma,\iKj\Gamma)}\,,\phantom{aaaaa}
\label{MredKK}
\eeqn
and similarly for the azimuthal terms
$\wampsqQGBqq_{\rm\sss \RED,AB,\lambda\blambda}$, 
which as usual only require the use of amplitudes at given polarizations.
The matrix elements of eqs.~(\ref{MredJJ})--(\ref{MredKK}) all feature 
the same colour structure, as is implied by eq.~(\ref{ampRQGqqdef}). 
This becomes evident if one expresses them 
in terms of scalar quantities:
\beqn
\ampsqQGBqq_{\rm\sss \RED,JJ}(\Gammap,\Gamma)&=&
\ampCSQGBqq(\iJ\Gammap)^\star \,C(\Gammap,\Gamma)\,\ampCSQGBqq(\iJ\Gamma)\,,
\label{MredJJsc}
\\
\ampsqQGBqq_{\rm\sss \RED,JK}(\Gammap,\Gamma)&=&
-\ampCSQGBqq(\iJ\Gammap)^\star \,C(\Gammap,\Gamma)
\left(\sum_{j=1}^{n+q}\ampCSQGBqq(\iKj\Gamma)\right),
\nonumber\\*
\label{MredJKsc}
\\
\ampsqQGBqq_{\rm\sss \RED,KJ}(\Gammap,\Gamma)&=&
-\left(\sum_{i=1}^{n+q}\ampCSQGBqq(\iKi\Gammap)^\star\right)
C(\Gammap,\Gamma)\,\ampCSQGBqq(\iJ\Gamma)\,,
\nonumber\\*
\label{MredKJsc}
\\
\ampsqQGBqq_{\rm\sss \RED,KK}(\Gammap,\Gamma)&=&
\left(\sum_{i=1}^{n+q}\ampCSQGBqq(\iKi\Gammap)^\star\right)
C(\Gammap,\Gamma)
\nonumber\\*&\times&
\left(\sum_{j=1}^{n+q}\ampCSQGBqq(\iKj\Gamma)\right).
\label{MredKKsc}
\eeqn
As is implied by eq.~(\ref{McollqqsumRf}), a closed flow
$(\Gammap,\Gamma)$ will not induce a singularity if either its 
L-flow or its R-flow (or both) does not belong to the set 
$\flowRJ\bigcup\flowRK$.

Collinear singularities due to $g\to q\qb$ branchings can obviously
occur when the underlying Born amplitude is a pure-gluon one, which
corresponds to $q=1$. The notation used so far is not suited
to describe such a case, the Born amplitudes having been written as 
quark-gluon ones. It is however not difficult to extend what was done
before to $q=1$, since one can formally understand
\beq
\ket{\ampQGBqq(\gamma)}\;\stackrel{q=1}{\longrightarrow}\;
\ket{\amp^{(0;n+2)}(\gamma)}\equiv\ket{\amp^{(n+2)}(\sigma)}\,,
\eeq
with $\sigma$ being the only non-trivial part of the flow $\gamma$ in
this case, namely a permutation of gluon labels (one must also understand
that the permutations are restricted to be non-cyclic only, when $q=1$ is
considered). Equation~(\ref{GonA}) is then still relevant, but its
colour part is not given by eq.~(\ref{Gacts}), but rather by:
\beqn
&&\sum_{\anpt}\lambda^{\anpt}_{bc}
{\rm Tr}\Big(\lambda^{a_{\sigma(1)}}\ldots
\lambda^{a_{\sigma(\isigma(n+2)-1)}}
\lambda^{\anpt}
\lambda^{a_{\sigma(\isigma(n+2)+1)}}\ldots
\lambda^{a_{\sigma(n+2)}}\Big)=
\nonumber\\*&&\phantom{aaaaaaa}\half
\Big(\lambda^{a_{\sigma(\isigma(n+2)+1)}}\ldots
\lambda^{a_{\sigma(n+2)}}
\lambda^{a_{\sigma(1)}}\ldots
\lambda^{a_{\sigma(\isigma(n+2)-1)}}\Big)_{bc}
\nonumber\\*&&\phantom{aaaa}-\frac{1}{2N}\,\delta_{bc}\,
{\rm Tr}\Big(\lambda^{a_{\sigma(1)}}\ldots
\lambda^{a_{\sigma(\isigma(n+2)-1)}}
\lambda^{a_{\sigma(\isigma(n+2)+1)}}\ldots
\lambda^{a_{\sigma(n+2)}}\Big)\,
\,.\phantom{aaaa}
\label{Gactsng}
\eeqn
The second term on the r.h.s.~of eq.~(\ref{Gactsng}) does not belong to 
the set of real flows; hence,
it must give a contribution to the final result equal to zero.
That this is indeed the case can be proved by direct computation.
One starts by extending the definitions of the operators $J$ and $K$
to the case $q=1$:
\beqn
J\gamma&=&
\Big(\Mq\,;\sigma(\isigma(n+2)+1),\ldots\sigma(n+2),
\nonumber\\*&&\phantom{\Big(\Mr\,;}
\sigma(1),\ldots\sigma(\isigma(n+2)-1);-2q\Big)\,,
\label{Jrdefqz}
\\
K\gamma&=&\Big(\sigma(1),\ldots\sigma(\isigma(n+2)-1),
\nonumber\\*&&\phantom{\Big(\Mr\,;}
\sigma(\isigma(n+2)+1),\ldots\sigma(n+2)\Big)\,
\bigcup\,\Big(\Mq\,;-2q\Big)\,.
\label{Krdefqz}
\eeqn
With these, one proceeds exactly as done before. The only difference
is that now
\beqn
\flowRJ=\flowR\,.
\eeqn
As far as $\flowRK$ is concerned, one simply defines it as the set
of flows that have the same form as the one in eq.~(\ref{Krdefqz}). By doing
that, ones arrives again at eqs.~(\ref{McollqqsumRf})--(\ref{MredKK}).
However, the following reduced matrix elements are identically equal 
to zero:
\beq
\ampsq^{(0;n+2)}_{\rm\sss \RED,JK}(\Gammap,\Gamma)=
\ampsq^{(0;n+2)}_{\rm\sss \RED,KJ}(\Gammap,\Gamma)=
\ampsq^{(0;n+2)}_{\rm\sss \RED,KK}(\Gammap,\Gamma)=0\,,
\label{JKeqzero}
\eeq
since they contain one or both of the linear combinations
\beq
\sum_{i=1}^{n+1}\amp^{(0;n+2)}(\iKi\Gammap)=
\sum_{j=1}^{n+1}\amp^{(0;n+2)}(\iKj\Gamma)=0\,,
\eeq
which are equal to zero owing to the dual Ward identity~\cite{Mangano:1990by}.
This proves the fact that the second term on the r.h.s.~of eq.~(\ref{Gactsng}) 
indeed does not contribute to the final result. The collinear limit for the
$g\to q\qb$ branching in the case of a pure-gluon Born is therefore still
given by eqs.~(\ref{McollqqsumRf}) and~(\ref{MQGcollqqfinal}), with the
condition that $\flowRK$ be equal to the empty set, and taking
eq.~(\ref{JKeqzero}) into account.

In analogy with what was done previously, I should now discuss the
case of the $g\to q\qb$ branching at fixed Born flows. However, it
does not appear to be possible to formulate it in a gauge-invariant
manner. This is due to the fact that the operator $K$ does not
have a single-valued inverse. If one must keep the Born flows
$\iKi\Gamma$ separate, rather than summing them as done 
in eqs.~(\ref{MredJK})--(\ref{MredKK}),
one must also ``split'' the contributions to the real flow $\Gamma$
into $(n+q)$ components, in order for each of them to have a limit
proportional to $\Lambda(\iKi\Gamma)$. This may be possible by considering
individual Feynman diagrams, at the price of violating gauge invariance.
It does not seem justified to do so since, contrary to the cases of 
$g\to gg$ and $q\to qg$ branchings, the Altarelli-Parisi kernel
associated with the $g\to q\qb$ branching is ${\cal O}(N^0)$,
which implies that for this branching the two formulations
at fixed Born or real flows are strictly equivalent from the
colour point of view. The above discussion does not apply to the case 
$q=1$, i.e.~when one has a pure-gluon Born since, as was shown before, when 
$q=1$ the $K$ operator does not contribute to the result. But it is 
obvious that in such a case fixing Born flows is identical to fixing
real flows.

\section{Summary of subtraction formulae\label{sec:summ}}
Since the detailed derivations carried out in sects.~\ref{sec:gluons}
and~\ref{sec:quarks} can obscure the final results, in this section
I collect the references to the formulae needed for the implementation
of FKS subtraction at fixed colour configurations or flows. For the
colour-summed real-emission matrix element squared $\ampsq$ and
its limit $\ampsqL$ (be it soft, collinear, or soft-collinear):
\beq
\ampsq\stackrel{{\rm L}}{\longrightarrow}\ampsqL\,,\;\;\;\;\;\;\;\;
{\rm L}={\rm SOFT}\,,\;{\rm COLL}\,,\;{\rm SC}\,,
\eeq
I have used the decompositions:
\beq
\ampsq=\sum_g\ampsq(g)\,,\;\;\;\;\;\;\;\;
\ampsqL=\sum_g\ampsqL(g)\,,
\eeq
with
\beq
\ampsq(g)\stackrel{{\rm L}}{\longrightarrow}\ampsqL(g)\;\;\;\;\;\;\;\;
\forall\,g\,.
\eeq
I have considered three cases, which I list here using the notations
relevant to gluon and to quark-gluon amplitudes respectively. As discussed
in the text, the soft-collinear limit $\ampsqSC(g)$ can be trivially obtained 
by computing either the collinear limit of $\ampsqS(g)$, or the soft limit
of $\ampsqC(g)$; hence, it will not be considered in what follows.
\begin{itemize}
%%%%%%%%%%%%%%%%%%
\item
Fixed colour configurations:
\beq
g=\left(a_1,\ldots \anpo\right)\,,\;\;\;\;\;\;\;\;
g=\left(\amtq,\ldots \anpo\right)\,.
\eeq
The matrix elements $\ampsq(g)$ are defined in eq.~(\ref{MnpoCD})
for gluon amplitudes, and in eq.~(\ref{MQGCD}) for quark-gluon amplitudes.\\
{\em Soft limits} $\ampsqS(g)$. For gluon amplitudes: eqs.~(\ref{MsoftCD})
and~(\ref{Mklscalar}). For quark-gluon amplitudes: eqs.~(\ref{MQGsoftCD})
and~(\ref{MQGklscalar}).
\\
{\em Collinear limits} $\ampsqC(g)$. For gluon amplitudes, 
eqs.~(\ref{McollCD}), (\ref{M0scalar}), and~(\ref{wM0scalar}).
For quark-gluon amplitudes, and $g\to gg$, $q\to qg$, or $\qb\to\qb g$
branchings: eqs.~(\ref{MQGcollCD}), (\ref{MQG0scalar}),
and~(\ref{wMQG0scalar}). For quark-gluon amplitudes, and $g\to q\qb$
branchings: eqs.~(\ref{MQGcollqqCD}), (\ref{MQG0qqscalar}),
and~(\ref{wMQG0qqscalar}).
%%%%%%%%%%%%%%%%%%
\item
Fixed real flows:
\beq
g=\left(\Sigmap,\Sigma\right)\,,\;\;\;\;\;\;\;\;
g=\left(\Gammap,\Gamma\right)\,.
\eeq
The matrix elements $\ampsq(g)$ are defined in eq.~(\ref{MnpoSS})
for gluon amplitudes, and in eq.~(\ref{MQGRSS}) for quark-gluon amplitudes.
The definitions of flows in the two cases are given at the beginning
of sect.~\ref{sec:gluons} and of sect.~\ref{sec:quarks} respectively.
\\
{\em Soft limits} $\ampsqS(g)$. For gluon amplitudes: eqs.~(\ref{MssMpp2})
and~(\ref{Mred}). For quark-gluon amplitudes: eqs.~(\ref{MQGssMpp2})
and~(\ref{MQGppdef2sc}).
\\
{\em Collinear limits} $\ampsqC(g)$. For gluon amplitudes, 
eqs.~(\ref{McollFR}), (\ref{Mred}), and~(\ref{wMcollFRsc}).
For quark-gluon amplitudes, and $g\to gg$, $q\to qg$, or $\qb\to\qb g$
branchings: eqs.~(\ref{MQGcollFR}), (\ref{MQGppdef2sc}),
and~(\ref{wMQGppdef2sc}). For quark-gluon amplitudes, and $g\to q\qb$
branchings:\\ eqs.~(\ref{MQGcollqqfinal})
and~(\ref{MredJJsc})--(\ref{MredKKsc}).
%%%%%%%%%%%%%%%%%%
\item
Fixed Born flows:
\beq
g=\left(\sigmap,\sigma\right)\,,\;\;\;\;\;\;\;\;
g=\left(\gammap,\gamma\right)\,.
\eeq
The matrix elements $\ampsq(g)$ relevant to this case are linear
combinations of those defined at fixed real flows. They are given
in eqs.~(\ref{MnpoBorn}) and~(\ref{zetaset}) for gluon amplitudes,
and in eqs.~(\ref{MQGnpoBorn}) and~(\ref{xiset}) for quark-gluon
amplitudes.\\
{\em Soft limits} $\ampsqS(g)$. For gluon amplitudes: eqs.~(\ref{Msoftdef2})
and~(\ref{Mkldef2sc}). For quark-gluon amplitudes: eqs.~(\ref{MQGsoftdef2})
and~(\ref{MQGkldef2sc}).
\\
{\em Collinear limits} $\ampsqC(g)$. For gluon amplitudes, 
eqs.~(\ref{McollFB2}), (\ref{Mmflowsc}), and~(\ref{wMmflowsc}).
For quark-gluon amplitudes, and $g\to gg$, $q\to qg$, or $\qb\to\qb g$
branchings: eqs.~(\ref{MQGcollFB2})--(\ref{MtQGcolldefzsc}).
For quark-gluon amplitudes, and $g\to q\qb$ branchings: 
this case cannot be treated at fixed Born flows.
\end{itemize}

\section{Born-like contributions\label{sec:Borns}}
I have so far discussed the case of the real-emission matrix elements
and of their local subtraction terms. NLO cross sections also receive other
contributions, which have a Born-like kinematics. These contributions
are due to the Born proper, to the one-loop corrections, and to the
analytically-integrated subtraction terms. In this section, I shall
discuss the treatment of the latter when one fixes the colour
configurations or the colour flows.

In FKS one defines two types of Born-like contributions, which arise from 
the analytical integration of either the soft or the collinear counterterms.
The one of soft origin reads as follows, up to overall trivial factors
(see e.g.~ref.~\cite{Frederix:2009yq}):
\beq
\sum_{k,l=-2q}^n\eikint_{kl}\ampsqQGBgg_{kl}\,,
\label{sigmaS}
\eeq
where I have used the notation of the quark-gluon amplitudes case
for generality. Equation~(\ref{sigmaS}) coincides with 
eq.~(\ref{MQGsoftdef}), except for the fact that the eikonal factors 
$[k,l]$ in the latter equation have been replaced in eq.~(\ref{sigmaS})
by the finite parts of their integrals, denoted by $\eikint_{kl}$.
The eikonals do not play any role in the manipulations carried
out in sects.~\ref{sec:gluons} and~\ref{sec:quarks}. Hence,
one can just use the results for the soft limits of the matrix
elements presented before and summarized in sect.~\ref{sec:summ},
and simply replace $[k,l]$ with $\eikint_{kl}$ there.

The Born-like contribution of collinear origin is proportional
to the Born matrix element squared:
\beq
{\cal Q}\,\ampsqQGBgg\,,
\label{QtimesB}
\eeq
where
\beqn
&&{\cal Q}=\sum_{k=-2q}^n\Bigg[\gamma^\prime(\ident_k)
-\log\frac{s\deltaO}{2Q^2}\left(\gamma(\ident_k)
-2C(\ident_k)\log\frac{2E_k}{\xicut\sqrt{s}}\right)
\nonumber \\*&&\phantom{{\cal Q}=\sum_{k,l=-2q}^n}
+2C(\ident_k)\left(\log^2\frac{2E_k}{\sqrt{s}}-\log^2\xicut\right)
-2\gamma(\ident_k)\log\frac{2E_k}{\sqrt{s}}\Bigg]\delta_{0m_k}\,,
\label{Qcolldef}
\eeqn
and
\beqn
\gamma(g)&=&\frac{11}{6}\CA-\frac{2}{3}\TF N_f\,,
\label{gammag}
\\
\gamma(q)&=&\frac{3}{2}\CF\,,
\label{gammaq}
\\
\gamma^\prime(g)&=&\left(\frac{67}{9}-\frac{2\pi^2}{3}\right)\CA
-\frac{23}{9}\TF N_f\,,
\label{gmmprimeglu}
\\
\gamma^\prime(q)&=&\left(\frac{13}{2}-\frac{2\pi^2}{3}\right)\CF\,.
\label{gmmprimeqrk}
\eeqn
The definitions of the various quantities that appear in eq.~(\ref{Qcolldef})
are irrelevant here; the interested reader may find them in 
ref.~\cite{Frederix:2009yq}. What matters is the definition of the
colour factors, given in eqs.~(\ref{gammag})--(\ref{gmmprimeqrk}).
These basically arise from the integration of the Altarelli-Parisi
kernels: $P_{qq}$ for $\gamma(q)$ and $\gamma^\prime(q)$ ($q\to qg$ 
and $\qb\to\qb g$ branchings); $P_{gg}$ {\em plus} $P_{qg}$ for
$\gamma(g)$ and $\gamma^\prime(g)$ ($g\to gg$ and $g\to q\qb$ branchings).
The sum of $P_{gg}$ and $P_{qg}$ is necessary because $\gamma(g)$ 
and $\gamma^\prime(g)$ are inclusive properties of gluons, and therefore
all possible branchings must be taken into account. However, these
can still be told apart in eqs.~(\ref{gammag}) and~(\ref{gmmprimeglu}),
thanks to the different colour factors ($\CA$ versus $\TF$). This
implies that, in eq.~(\ref{Qcolldef}), one can unambiguously associate
the various terms with either a $q\to qg$ 
or a $\qb\to\qb g$ branching ($\CF$ terms),
or a $g\to gg$ branching ($\CA$ terms), or a $g\to q\qb$ branching 
($\TF$ terms). For each of these branching types, one can then repeat
what was done in sects.~\ref{sec:gluons} and~\ref{sec:quarks} in the
case of collinear limits. This implies starting from re-writing 
eq.~(\ref{QtimesB}) as follows:
\beq
{\cal Q}\,\ampsqQGBgg=
\sum_{k=-2q}^{-1}\CF\Big(\ldots\Big)\ampsqQGBgg
+\sum_{k=1}^{n}\CA\Big(\ldots\Big)\ampsqQGBgg
+\sum_{k=1}^{n}\TF\Big(\ldots\Big)\ampsqQGBgg\,,
\label{QBexp}
\eeq
where the terms in brackets can be easily worked out from
eqs.~(\ref{Qcolldef})--(\ref{gmmprimeqrk}). Their forms are irrelevant
here, except for the fact that they do not contain any colour factors.
At this point, one exploits eq.~(\ref{Casimir2}) in the first two
terms of eq.~(\ref{QBexp}), to replace $\ampsqQGBgg$ with
$\zampsqQGBggk$ defined in eq.~(\ref{MQGcolldefz}) (with $s\to k$ there). 
Analogously, one exploits eq.~(\ref{Casimir5}) in the last term of
eq.~(\ref{QBexp}), to replace $\ampsqQGBgg$ with
$\zampsqQGBqqqq$ defined in eq.~(\ref{amps0qqdef}). 
After these replacements, the matrix elements in eq.~(\ref{QBexp}) 
are in the same form as those that appear in the colour-summed
expressions of the collinear limits. Hence, the same manipulations
as in sects.~\ref{sec:gluons} and~\ref{sec:quarks} can be carried
out here; the final results can be directly obtained from the
expressions of the collinear limits at fixed colour configurations
or real flows, by simply replacing the Altarelli-Parisi kernels
and prefactors there with the expressions in round brackets
that appear in eq.~(\ref{QBexp}). Note that this procedure must
not be carried out if one is interested in fixing Born flows:
the original expression, eq.~(\ref{QtimesB}), is already suited
to that, the only change being the formal replacement:
\beq
\ampsqQGBgg\;\longrightarrow\;\ampsqQGBgg(\gammap,\gamma)\,.
\eeq

\section{Discussion\label{sec:disc}}
Processes with large particle multiplicities have a colour
algebra so involved that its direct computation is impossible,
and one must use alternative methods, such as Monte Carlo (MC) ones.
While it is common to sum over colour configurations with MC
techniques, there is no reason of principle that prevents one
from doing an MC sum over colour flows -- such a strategy is indeed
being considered in the new version of MadGraph/MadEvent~\cite{MGv5}.
At the tree level, matrix elements 
at fixed colour configurations (e.g.~eq.~(\ref{amportho}))
have the advantage over those at fixed flows (e.g.~eq.~(\ref{Mmflowsc}))
of being positive definite, while the latter are in general complex
numbers. However, this advantage is lost beyond the leading order in 
perturbation theory, since subtractions such as those of eq.~(\ref{subt})
will always be involved, thus implying results of either sign.
Also, in an actual computation one will consider the sum
$\ampsq(\sigmap,\sigma)+\ampsq(\sigma,\sigmap)$, which is a real
number, since physical observables
must not depend on the distinction between L- and R-flows.
Furthermore, at any order in perturbation theory, including tree
level, matrix elements at fixed colour flows have an immediate
interpretation in terms of large-$N$ expansion, which is rather indirect
at fixed colour configurations. In particular, it is not necessary
to explicitly compute the colour-flow matrix of eq.~(\ref{CFmatdef})
(or eq.~(\ref{CFQGmatdef})) to determine the largest possible power
of $N$ in any of its elements -- it is sufficient to count
the number of colour loops determined by the closed flow $(\sigmap,\sigma)$
(or $(\gammap,\gamma)$), which is basically instantaneous if performed
by a computer. This fact not only paves the way to a systematic
organization of the computation in terms of increasing powers
of $1/N$, but also suggests a way to save computing time, since
terms with large $1/N$ powers may be computed with a relatively
low statistics.

The results presented in this paper allow one to follow either of
the strategies discussed above in the context of the computation
of NLO observables with FKS subtraction. The approach that uses
fixed colour configurations bases its efficiency on the fast
calculation of colour-dressed amplitudes, at both the real-emission
and Born levels. It should be stressed that, for a given real
matrix element at fixed colours, such e.g.~that of eq.~(\ref{MnpoCD})
for gluon amplitudes, one must compute {\em several} Born-level 
colour-dressed amplitudes -- see eq.~(\ref{Mklscalar}) for the 
soft limit, and eqs.~(\ref{M0scalar}) and~(\ref{wM0scalar}) for 
the collinear limit. It is clear that, for the sake of numerical
stability, the sums over $b_k^\prime$ and $b_l$ in eq.~(\ref{Mklscalar}),
and those over $b_n^\prime$ and $b_n$ in eqs.~(\ref{M0scalar}) 
and~(\ref{wM0scalar}), must be performed exactly, and not with 
MC methods. Furthermore, in the case of the soft limits the sums
over $k$ and $l$ that appear in eq.~(\ref{MsoftCD}) must also be
taken into account. The bottom line is that, if one performs these
sums blindly, a non-negligible complexity creeps back into the game.
For a fast computation of the counterterms, then, an algorithm
is essential that pre-determines which colour configurations 
are associated with tree-level amplitudes that are equal to zero.
It is probably also important to compute the colour matrices
${\cal Q}_{\ident_k\ident_l}$ and ${\cal G}$ once and for all,
since many of their elements will be equal to zero, an information
which can be effectively used to improve the efficiency of the algorithm.
These issues are beyond the scope of this paper, and will not be
discussed any further.

Turning to the fixed-flow approach, the general arguments related
to the $1/N$ expansion apply to both the fixed-real and the fixed-Born 
flow cases. I also stress that, in order to derive the formulae for
the fixed-flow subtraction terms, I have started from the known limiting
behaviours of the {\em squared} matrix elements, the same that I have used 
in the case of fixed colour configurations. In other words, I did not
employ the limits of colour-ordered amplitudes. While the latter may have
provided a more straightforward derivation for fixed real flows, the
approach followed here allows the treatment of the three schemes considered
(fixed colour configurations, Born flows, and real flows) in a common
language. The fact that the results I have arrived at for the matrix 
element limits at fixed real flows can also be derived using the limits of 
dual amplitudes as a starting point constitutes a partial cross check of the 
procedure adopted in this paper.

When fixing the Born flows, the building blocks of the
counterterms are the same ones as those used in the automatic
implementation of the FKS subtraction achieved by 
MadFKS~\cite{Frederix:2009yq}. Furthermore, in this scheme
the simultaneous integration of the real-emission and of
the Born-like and Born contributions can be performed in exactly
the same way as in ref.~\cite{Frederix:2009yq}, where it is discussed 
in detail. On the other hand, from the point of view of computational
complexity, for any pair of Born-level dual amplitudes that determine
the counterterms, one needs to evaluate $n^2$ (in the case of gluon
amplitudes -- see eqs.~(\ref{MnpoBorn}) and~(\ref{zetaset})),
or $(n+q)^2$ (in the case of quark-gluon amplitudes --
see eqs.~(\ref{MQGnpoBorn}) and~(\ref{xiset})), real-emission
dual amplitudes. This has an obvious physical interpretation:
the sets of real flows $\zeta(\sigma)$ and $\xi(\gamma)$ effectively
achieve a block decomposition of the colour-flow matrices
$C(\Sigmap,\Sigma)$ and $C(\Gammap,\Gamma)$ respectively. 
The choice of a Born-level closed flow corresponds to choosing
one of these blocks; the relevant colour algebra at the real-emission
level is then that of the block so determined. In other words, when
performing an MC sum over Born flows, each seed is associated with
a block in the matrix of real colour flows; within this block, the
colour algebra is performed exactly, i.e. without using MC methods.

Let me finally discuss the case of fixed real flows. This scheme
has the nice property that the colour-flow matrix can be factored
out of the subtraction procedure. In other words, all of the four
terms in the linear combination of eq.~(\ref{subt}) are proportional
either to $C(\Sigmap,\Sigma)$ (for gluon amplitudes) or to 
$C(\Gammap,\Gamma)$ (for quark-gluon amplitudes). This implies
that colour never enters into the definition of subtraction terms
if not in a trivial way, which does not happen (owing to the definition
of colour-linked Born's) in the other formulations of the FKS subtraction 
given here, let alone in the colour-summed one. 
A pleasant implication of this fact is the
disappearance of the colour-linked Born's from the soft subtraction
terms (see eqs.~(\ref{MssMpp2}) and~(\ref{MQGssMpp2})). This, which
is nothing but the colourless nature of the subtraction procedure, together 
with the possibility of defining the soft and collinear kinematics so as the
underlying Born kinematics coincide, implies the following equation (which 
I write for quark-gluon amplitudes, for greater generality, and without 
considering the case of the $g\to q\qb$ branching, 
which I shall discuss later):
\beqn
&&\Sfunij\,\ampCSQGR(\Gammap)^\star\,\ampCSQGR(\Gamma)
\label{merged}
\;\;\longrightarrow
\\*&&\phantom{aaaaaaa+}
{\cal K}_{ij}(\Gammap,\Gamma)\,
\ampCSQGBgg(\Gammap_{\cancel{i}})^\star\,
\ampCSQGBgg(\Gamma_{\cancel{i}})
\nonumber\\*&&\phantom{aaaaaa}
+{\cal K}_{ij-+}^{(\varphi)}(\Gammap,\Gamma)\,
\ampCSQGBgg_-(\Gammap_{\cancel{i}})^\star\,
\ampCSQGBgg_+(\Gamma_{\cancel{i}})
\nonumber\\*&&\phantom{aaaaaa}
+{\cal K}_{ij+-}^{(\varphi)}(\Gammap,\Gamma)\,
\ampCSQGBgg_+(\Gammap_{\cancel{i}})^\star\,
\ampCSQGBgg_-(\Gamma_{\cancel{i}})\,,
\nonumber
\eeqn
where the arrow denotes the soft, collinear, or soft-collinear limit.
The kernel ${\cal K}_{ij}$ is thus able to describe the singular behaviour
of the interference between dual amplitudes in these three limits.
The only contribution to the divergences that cannot be included
in ${\cal K}_{ij}$ is the azimuthal-dependent part of the collinear
limit, for which the kernels ${\cal K}_{ij\lambda\blambda}^{(\varphi)}$
need be introduced, owing to the different structure of the factorized
Born-level dual amplitudes (the helicity of the branching parton is
not summed over in the last two terms on the r.h.s.~of 
eq.~(\ref{merged}))\footnote{Alternatively, one can define in a 
straightforward manner a
unique kernel, as a tensor in the space of the helicities of the
branching parton. This is what is done e.g.~in dipole subtraction.}.
As is known, the azimuthal terms vanish upon integration, but must
nevertheless be included in the subtraction, in order to have a local
cancellation of divergences in all phase-space points, and not only
at the integrated level. Needless to say, in eq.~(\ref{merged}) and in
what follows all terms of collinear origin are present only if parton
$j$ is not a massive quark. The explicit form of the kernels that appear
in eq.~(\ref{merged}) can be worked out from eqs.~(\ref{MQGssMpp2})
and~(\ref{MQGcollFR}). One obtains:
\beqn
{\cal K}_{ij}(\Gammap,\Gamma)&=&
-\gs^2\left(\sum_{\alpha=-1,1}\sum_{\beta=-1,1} \alpha\beta\,
\eik{\Gammap(\iGammap(i)\oplus\alpha)}{\Gamma(\iGamma(i)\oplus\beta)}\right)
\Sfunij\,\Bigg|_{\rm\sss SOFT}
\nonumber\\*&+&
\frac{\gs^2}{k_i\mydot k_j}
\,\delta(\Gammap,\Gamma)
\hat{P}_{\ident_j\ident_j}^{(+)}(z)\,\Bigg|_{\rm\sss COLL}\,,
\label{Kijdef}
\\
{\cal K}_{ij-+}^{(\varphi)}(\Gammap,\Gamma)&=&
\frac{\gs^2}{2k_i\mydot k_j}
\,\delta(\Gammap,\Gamma)
\hat{Q}_{\ident_j\ident_j^\star}(z)\,
\frac{\langle k_i k_j\rangle}{[k_i k_j]}\,\Bigg|_{\rm\sss COLL}\,,
\label{Kijmpdef}
\\
{\cal K}_{ij+-}^{(\varphi)}(\Gammap,\Gamma)&=&
\frac{\gs^2}{2k_i\mydot k_j}
\,\delta(\Gammap,\Gamma)
\hat{Q}_{\ident_j\ident_j^\star}(z)\,
\frac{[k_i k_j]}{\langle k_i k_j\rangle}\,\Bigg|_{\rm\sss COLL}\,.
\label{Kijpmdef}
\eeqn
Several comments are in order here. First of all, being clear that
the results obtained in sect.~\ref{sec:quarks} are valid for any
partons, I have relabeled $n+1$ and $s$ as $i$ and $j$ respectively,
consistently with the role of $\Sfunij$ in eq.~(\ref{merged}); it
is however understood that parton $i$ is a gluon. The ${\rm SOFT}$
and ${\rm COLL}$ tags in eqs.~(\ref{Kijdef})--(\ref{Kijpmdef})
imply that the four momenta used to compute the corresponding quantities
are those relevant to the soft and collinear configurations respectively.
In the antenna or dipole formulations, the analogues of these 
configurations are explicitly defined by means of invariants.
In FKS, one does not need an explicit definition; any maps induced
by the real-emission phase-space parametrization will do, provided
that the latter satisfies the condition that the underlying Born kinematics
obtained by taking the soft and collinear limits coincide\footnote{As
was mentioned at the end of sect.~\ref{sec:glusum}, this condition is
necessary for eq.~(\ref{merged}) to hold. It is however not mandatory
in general for FKS subtraction, but has been found to be convenient from
the numerical point of view.}. 
The exactly collinear configurations used in the computation
of the second term on the r.h.s.~of eq.~(\ref{Kijdef}),
and in eqs.~(\ref{Kijmpdef}) and~(\ref{Kijpmdef}), is the reason
why no $\Sfun$ function appear there -- in such a limit, 
$\Sfunij=1$~\cite{Frixione:1995ms}.
The $(+)$ index attached to the Altarelli-Parisi kernel in the second
term on the r.h.s.~of eq.~(\ref{Kijdef}) implies that such a kernel
must be understood as being subtracted by means of
a plus prescription. It is easy to convince
oneself (see e.g.~eq.~(\ref{MSCdef})) that this is sufficient to
take into account the soft-collinear limit (last term in 
eq.~(\ref{subt})). Although in principle the same should be done
for the azimuthal kernels of eqs.~(\ref{Kijmpdef}) and~(\ref{Kijpmdef}),
in practice this is not necessary, owing to the fact that 
$\hat{Q}_{ab^\star}(z)$ vanish in the soft limit. I finally
point out that eqs.~(\ref{merged})--(\ref{Kijpmdef}) are valid also
in the case of $g\to q\qb$ branching, with only changes in notation.
However, because of the vanishing of the soft limit in this case, 
the final result would be identical to that already presented in
eqs.~(\ref{MQGcollqqfinal}) and~(\ref{MredJJsc})--(\ref{MredKKsc}).

I have already shown in sect.~\ref{sec:Borns} that Born-like 
contributions to the NLO cross section can also be cast in the
form of terms with fixed real flows. As far as the Born proper
is concerned, in order to integrate it simultaneously with the
other terms one can use the same trick as that used to transform
eq.~(\ref{QtimesB}) into eq.~(\ref{QBexp}) -- this is possible
since the Born can be split into several components associated
with $\Sfunij$ functions, as explained in ref.~\cite{Frederix:2009yq},
thus effectively defining Born contributions at fixed real flows.
Another possibility is that of exploiting the block structure
of the real colour-flow matrix discussed above: for a given
real flow $\Gamma$, there exists a Born flow $\gamma$ such
that $\Gamma\in\xi(\gamma)$. In this way, for any random choice
of $\Gamma$ in the MC sum over real flows, one also performs
an MC sum over Born flows (which is unbiased, at least as long as
$\Gamma$ is chosen flat).

I have presented the results in the form of unphysical $0\to m$ 
processes for ease of notation. It is clear that they have the
same structure, colour-wise, when two of the final-state particles are
crossed into the initial state. It should be kept in mind that for
physical $2\to m-2$ processes there is contribution to the NLO
cross section which is essentially a Born-like one, except for
the fact that it features an additional integration variable. 
This contribution, called ``degenerate $(n+1)$-body'' in
ref.~\cite{Frederix:2009yq}, has the same
form as the collinear limits discussed at length in this paper,
and can therefore be manipulated in exactly the same way.

\vskip 0.2truecm
\noindent
{\em Antenna and dipole subtractions.} It seems appropriate to conclude
this discussion by stressing the connections between FKS, antenna,
and dipole subtractions, which are best uncovered when a fixed-real-flow
scheme is adopted in the former. Firstly, given the fact that
antenna subtraction is formulated in terms of dual amplitudes
or of quantities closely related to those, 
it is natural that FKS at fixed real flows use the same matrix 
elements limits as in antenna (see e.g.~eq.~(\ref{MQGssMpp2})
and its counterpart obtained in ref.~\cite{Abelof:2011jv}, in the
context of a process-specific computation). As was mentioned before,
this is reassuring, since the factorization formulae one starts
from do not coincide in the two approaches. From the kinematical
point of view, however, FKS and antenna are still fairly different.
In particular, in FKS one of the two radiators that define an antenna
is not needed (the other radiator may be identified with the FKS sister),
and this is reflected in the different choices made for the phase-space
parametrizations (which in FKS do not depend on colour connections).

Secondly, the methods used here to manipulate the colour-linked Born's
(which for example lead one from eqs.~(\ref{QAeqAsQG})--(\ref{MQGkldef2})
to eqs.~(\ref{QAeqAsQGFR})--(\ref{MQGppdef2})) can also be used to
carry out the same operations on dipoles. More specifically, the 
structure of the colour kernel of a dipole ${\cal D}_{ij,k}$ is 
(in the notation of this paper) $\sum_b Q^b(ij)Q^b(k)$, up to an
overall Casimir. The most involved case is when either $i$ or $j$ (or both)
is a gluon -- be it $i$ just to fix the notation. Then, one can identify $ij$ 
and $k$ of the colour kernel above with $l$ and $k$ in eq.~(\ref{MQGsoftdef2}),
$i$ with the FKS parton, and proceed as done in this paper\footnote{When
$j$ is a gluon, ${\cal D}_{ij,k}$ also subtracts the soft singularities
associated with $j$, whereas in FKS the soft singularities of the FKS 
sister are damped. In order to simplify the discussion, I assume here
that such a damping is performed also in dipoles, thus a-symmetrizing
the roles of $i$ and $j$. Many other options are obviously possible.}. 
In this way, a linear combination of dipoles will emerge, analogue to 
that featuring the eikonal factors in eq.~(\ref{Kijdef}):
\beq
\sum_{\alpha=-1,1}\sum_{\beta=-1,1} \alpha\beta\,{\cal D}_{ij,k}
\left(j\to\Gammap(\iGammap(i)\oplus\alpha),
k\to\Gamma(\iGamma(i)\oplus\beta)\right)\,.
\label{dipoles}
\eeq
As was the case for the colour-linked Born's, all the dipoles in
eq.~(\ref{dipoles}) will factorize the same colour-flow matrix element
$C(\Gammap,\Gamma)$. However, different dipoles will factorize dual 
amplitudes (which sandwich helicity-dependent kernels) 
computed with different kinematics, and therefore the analogue 
of eq.~(\ref{merged}) cannot be written in this case. This problem can
be avoided by using variants of the dipole formalism that curb the 
proliferation of reduced kinematics
(see e.g.~refs~\cite{Nagy:2007ty,Chung:2010fx}).

The similarities among the various subtraction formalisms need not
be surprising; as was shown in ref.~\cite{Frixione:2004is}, all
of them must have the same underlying structure, and
differences arise when choices are made for the projections that
map resolved kinematic configurations onto unresolved ones, and
for the definitions of subtraction terms away from the zero-measure
soft and collinear regions. Colours somehow blur the picture,
which becomes clear again if one works in schemes such as the
fixed-real-flow one (which may not be unique in this respect).

\section{Conclusions\label{sec:concl}}
The results of this paper will allow the implementation
of the FKS formalism in a colour-friendly way, since the elementary
ingredients of the subtraction procedure can be defined at fixed
colour configurations or colour flows. This is a necessary condition
in order to be able to perform the colour algebra with Monte Carlo
methods, and thus to tackle the computation of large-multiplicity
processes. The formulation of the subtraction at fixed colour flows
can be used to organize the calculations as a systematic expansion
in $1/N$, since all results given here are exact to all orders
in $N$. When the flows are fixed at the real-emission level,
the colour and Lorentz structures completely decouple, and this
allows one to define colourless kernels that can simultaneously
describe the soft, collinear, or soft-collinear behaviour of
the matrix element squared. In this scheme, one sees more clearly
the connections between the FKS subtraction method and the antenna
and dipoles ones, which I have briefly discussed.

\section*{Acknowledgments}
I am indebted to Fabio Maltoni for the countless discussions
we have had on this matter (and on many others). I thank Michelangelo
Mangano for patiently listening to, and answering, my many questions.
Correspondence with Thomas Gehrmann and Rikkert Frederix is also 
gratefully acknowledged. This manuscript has benefited from comments
by Fabio, Michelangelo, and Rikkert.

\appendix
\section{Conventions for colour matrices\label{sec:conv}}
Contrary to the usual conventions, in this paper I normalize the
Gell-Mann matrices in the same way as the SU$(N)$ generators. Hence:
\beqn
[\lambda^a,\lambda^b]&=&i\sum_{c=1}^{N^2-1}f^{abc}\lambda^c\,,
\\
{\rm Tr}\left(\lambda^a\lambda^b\right)&=&\TF\delta^{ab}\,,
\\
\sum_{a=1}^{N^2-1}\left(\lambda^a\lambda^a\right)_{ij}&=&\CF\delta_{ij}\,,
\\
\sum_{a,b=1}^{N^2-1}f^{abc}f^{abd}&=&\CA\delta^{cd}\,,
\label{ffsum}
\\
\sum_{a=1}^{N^2-1}
\lambda^a_{ij}\lambda^a_{kl}&=&\half\left(\delta_{il}\delta_{jk}-
\frac{1}{N}\delta_{ij}\delta_{kl}\right),
\label{sumijkl}
\eeqn
with the usual colour factors
\beqn
\TF&=&\half\,,
\\
\CF&=&\frac{N^2-1}{2N}\,,
\\
\CA&=&N\,.
\eeqn

\section{Colour operators, flows, and colour conservation\label{sec:appa}}
As shown in sects.~\ref{sec:glusum} and~\ref{sec:QGsum} for gluon 
and quark-gluon amplitudes respectively, the consistency between the 
soft and collinear limits of the matrix elements follows from the 
colour-conservation identities of eqs.~(\ref{colcons}) and~(\ref{colconsQG}).
These identities are in turn a consequence of expressing the scattering 
amplitudes in terms of dual amplitudes, as done in eqs.~(\ref{mgCDamp}) 
and~(\ref{mgCDampQG}), as I shall show in this appendix.

An important by-product of this proof is the fact that the
colour operators $Q^b$, while defined in a natural way in the
spaces of colour configurations as in eqs.~(\ref{Qdef})--(\ref{Qreprglu}) 
and~(\ref{Qmatdef2})--(\ref{Qaqdef}), can also be easily interpreted 
in terms of flows. I shall therefore introduce operators acting on
flows that have an explicit correspondence with $Q^b$, and
which will be instrumental in deriving a formulation of the
FKS subtraction alternative to that at fixed colour configurations.

\subsection{Gluon amplitudes\label{sec:appglu}}
Since eq.~(\ref{colcons}) is an identity in the colour space
of $n+1$ gluons, it can be rewritten as follows
\beq
X\equiv\sum_{k=1}^n\, \bra{a_1,\ldots \anpo}Q^b(k)\ket{\ampn}=0
\;\;\;\;\;\;\;\;
\forall\;\setanpo\,.
\label{colcons2}
\eeq
Note that eq.~(\ref{colcons2}) is equivalent to saying that colour 
conservation applies to $(n+1)$-gluon amplitudes at fixed colour 
configurations. Using eq.~(\ref{ampflow}), one can rewrite
\beqn
X&=&\sum_{\sigma\in P_n^\prime} X_\sigma\,,
\\
X_\sigma&=&\sum_{k=1}^n\, \bra{a_1,\ldots \anpo}Q^b(k)\ket{\ampn(\sigma)}\,.
\label{Xsigdef}
\eeqn
Therefore, if one can prove that $X_\sigma=0$ for an arbitrary $\sigma$,
eq.~(\ref{colcons2}) will follow (the converse is obviously not true,
and thus $X_\sigma=0$ is a stronger conditions than $X=0$).
By using eqs.~(\ref{Aflowdef}), (\ref{Qdef}), and~(\ref{Qreprglu}), 
one obtains
\beq
X_\sigma=Y_\sigma\,\ampCSn\left(\sigma\right)\,,
\eeq
where
\beqn
Y_\sigma&=&\sum_{k=1}^n\sum_{b_k} \left(T^{\anpo}\right)_{a_kb_k}
{\rm Tr}\Big(\lambda^{a_{\sigma(1)}}\ldots
\lambda^{a_{\sigma(\isigma(k)-1)}}\lambda^{b_k}
\lambda^{a_{\sigma(\isigma(k)+1)}}
\ldots\lambda^{a_{\sigma(n)}}\Big)
\\
&=&\sum_{k=1}^n
{\rm Tr}\Big(\lambda^{a_{\sigma(1)}}
\ldots\lambda^{a_{\sigma(\isigma(k)-1)}}
\Big[\lambda^{a_k},\lambda^{\anpo}\Big]\lambda^{a_{\sigma(\isigma(k)+1)}}
\ldots\lambda^{a_{\sigma(n)}}\Big)\,.
\label{Y2}
\eeqn
Hence, in order to prove that $X_\sigma=0$, one must prove that $Y_\sigma=0$.
I start from observing that eq.~(\ref{Y2}) is a linear combination of
traces obtained by inserting $\lambda^{\anpo}$ into the traces that appear
in $\ket{\ampn(\sigma)}$. Such insertions can be conveniently 
represented in terms of flows. This can be done by defining
the following operators:
\beqn
I_+(i)\sigma&\equiv&I_+(i)(\sigma(1),\ldots\sigma(n))=
(\sigma(1),\ldots\sigma(i),n+1,\ldots\sigma(n))\,,
\label{Ipdef}
\\
I_-(i)\sigma&\equiv&I_-(i)(\sigma(1),\ldots\sigma(n))=
(\sigma(1),\ldots n+1,\sigma(i),\ldots\sigma(n))\,.
\label{Imdef}
\eeqn
In other words, $I_+(i)$ ($I_-(i)$) inserts the number $n+1$ into
the list defined by $\sigma$ after (before) the $i^{th}$ member of the list.
One can now rewrite eq.~(\ref{Y2}) as follows, using the shorthand
notation of eq.~(\ref{Lambdashort}):
\beq
Y_\sigma=\sum_{k=1}^n \Big\{
\Lambda\left(I_+(\isigma(k))\sigma\right)-
\Lambda\left(I_-(\isigma(k))\sigma\right)
\Big\}
\label{Y22}
\eeq
A comparison of eq.~(\ref{Y22}) with eq.~(\ref{Xsigdef}) shows the
relationship between the $Q^b$ and the $I_\pm$ operators:
\beq
Q^b\;\;\longleftrightarrow\;\;I_+ - I_-\,.
\label{QvsII}
\eeq
In keeping with the fact that $Q^b$ creates a colour state associated
with gluon $n+1$, the operators $I_\pm$ transform a $n$-gluon flow
into an $(n+1)$-gluon flow. A technical difference between the
operators $Q^b$ and $I_\pm$ is that the argument of the former is
a particle label, whereas the arguments of the latter are the positions
in the list of particles that defines the flow. This is justified by the
fact that when manipulating flows it is more convenient to deal with 
positions rather than particle labels -- singularity configurations 
are due to particles that are adjacent in a flow. It is obvious that
summing over particle labels is equivalent to summing over positions
in a flow. Hence, from eq.~(\ref{Y22}):
\beq
Y_\sigma=\sum_{i=1}^n \Big\{
\Lambda\left(I_+(i)\sigma\right)-
\Lambda\left(I_-(i)\sigma\right)
\Big\}
\label{Y3}
\eeq
Furthermore
\beqn
\sum_{i=1}^n \Lambda\left(I_-(i)\sigma\right)
&=&
\Lambda\left(I_-(1)\sigma\right)+
\sum_{i=2}^n \Lambda\left(I_-(i)\sigma\right)
\nonumber\\
&=&
\Lambda\left(I_+(n)\sigma\right)+
\sum_{i=2}^n \Lambda\left(I_+(i-1)\sigma\right)
\nonumber\\
&=&
\Lambda\left(I_+(n)\sigma\right)+
\sum_{i=1}^{n-1} \Lambda\left(I_+(i)\sigma\right)
\nonumber\\
&=&\sum_{i=1}^n \Lambda\left(I_+(i)\sigma\right)\,,
\label{Y4}
\eeqn
where the various manipulations follow from the invariance of the
trace under cyclic permutations, and from the property
\beq
I_-(i)\sigma=I_+(i-1)\sigma\,,\;\;\;\;\;\;\;\;
2\le i\le n\,,
\label{ImeqIp}
\eeq
which trivially follows from the definitions of $I_\pm(i)$. By replacing
eq.~(\ref{Y4}) into eq.~(\ref{Y3}) one proves that $Y_\sigma=0$.
As discussed at the beginning of this section, this not only 
proves eq.~(\ref{colcons}), but also that 
\beq
\sum_{k=1}^n Q^b(k)\ket{\ampn(\sigma)}=0\,.
\label{colcons3}
\eeq
In other words, colour conservation also holds at fixed Born flows;
this is far from surprising from the physics viewpoint.

\subsection{Quark-gluon amplitudes\label{sec:appQG}}
In the case of quark-gluon amplitudes, the proof of eq.~(\ref{colconsQG}) 
proceeds through proving the analogue of eq.~(\ref{colcons2}), i.e.:
\beq
X\equiv\sum_{k=-2q}^n\,\bra{\amtq,\ldots \anpo}Q^b(k)\ket{\ampQGBgg}=0
\;\;\;\;\;\;\;\;
\forall\;\setaQGR\,.
\label{colcons2QG}
\eeq
Following what is done in the case of gluon amplitudes, one writes
\beqn
X&=&\sum_{\gamma\in\flowBgg} X_\gamma\,,
\\
X_\gamma&=&\sum_{k=-2q}^n\,\bra{\amtq,\ldots \anpo}\,Q^b(k)\,
\ket{\ampQGBgg(\gamma)}\,.
\eeqn
As was done before, I shall show that the condition
$X_\gamma=0$ holds, thereby proving eq.~(\ref{colcons2QG})
as well. Using eq.~(\ref{QGAflowdef}), 
one gets
\beq
X_\gamma=Y_\gamma\,\ampCSQGBgg(\gamma)\,,
\eeq
where now
\beq
Y_\gamma=
\sum_{\setcQGBgg}\!\!\Lambda\left(\setc,\gamma\right)
\sum_{k=-2q}^n\,\bra{\amtq,\ldots \anpo}\,Q^b(k)\,
\ket{c_{-2q},\ldots c_n}\,.
\label{Ygdef}
\eeq
Owing to the factorized form of the colour structure, 
eq.~(\ref{Lambdagq}), one can associate
a colour structure with each colour line:
\beqn
\Lambda\left(\seta,\gamma\right)&=&N^{-\rho(\gamma)}\,\prod_{p=1}^q
\Lambda\left(\seta,\gamma_p\right)\,,
\label{Cstructall}
\\
\Lambda\left(\seta,\gamma_p\right)&=&
\Big(\lambda^{a_{\sigma(t_{p-1}+1)}}\ldots
\lambda^{a_{\sigma(t_p)}}\Big)_{a_{-p}a_{\mu(-p-q)}}\,.
\label{Cstructline}
\eeqn
Using the definition of the operator $Q^b(k)$, eq.~(\ref{Ygdef}) 
then becomes:
\beqn
Y_\gamma&=&N^{-\rho(\gamma)}\,
\sum_{p=1}^q \left(\mathop{\prod_{r=1}^q}_{r\ne p}
\Lambda\left(\seta,\gamma_r\right)\right)\,Y_{\gamma,p}\,,
\label{Ydecomp}
\\
Y_{\gamma,p}&=&\sum_{k\in\gamma_p}\sum_{c_k}
\left(Q^{\anpo}(k)\right)_{a_kc_k}
\Lambda\left(\seta_{i\ne k},c_k,\gamma_p\right)\,.
\label{Ypdef}
\eeqn
Equation~(\ref{Ypdef}) states formally that the action of the 
operators $Q^b$ onto the scattering amplitude can be conveniently
rewritten colour line per colour line.
It also implies that in order to prove $X_\gamma=0$ 
a sufficient condition is that of proving $Y_{\gamma,p}=0$ for all $p$,
which is what I set out to do now. First, by using eqs.~(\ref{Qreprglu}),
(\ref{Qqdef}) and~(\ref{Qaqdef}), one obtains:
\beqn
&&\sum_{c_k}\left(Q^{\anpo}(k)\right)_{a_kc_k}
\Lambda\left(\seta_{i\ne k},c_k,\gamma_p\right)=
\nonumber\\*&&\phantom{aaaaa}
\Big(\lambda^{a_{\sigma(t_{p-1}+1)}}
\ldots\lambda^{a_{\sigma(\isigma(k)-1)}}
\Big[\lambda^{a_k},\lambda^{\anpo}\Big]\lambda^{a_{\sigma(\isigma(k)+1)}}
\ldots\lambda^{a_{\sigma(t_p)}}\Big)_{a_{-p}a_{\mu(-p-q)}}\,,
\label{QCglu}
\\
&&\sum_{c_k}\left(Q^{\anpo}(k)\right)_{a_kc_k}
\Lambda\left(\seta_{i\ne k},c_k,\gamma_p\right)=
\Big(\lambda^{\anpo}\lambda^{a_{\sigma(t_{p-1}+1)}}
\ldots\lambda^{a_{\sigma(t_p)}}\Big)_{a_{-p}a_{\mu(-p-q)}}\,,
\label{QCqrk}
\\
&&\sum_{c_k}\left(Q^{\anpo}(k)\right)_{a_kc_k}
\Lambda\left(\seta_{i\ne k},c_k,\gamma_p\right)=
-\Big(\lambda^{a_{\sigma(t_{p-1}+1)}}\ldots\lambda^{a_{\sigma(t_p)}}
\lambda^{\anpo}\Big)_{a_{-p}a_{\mu(-p-q)}}\,,\phantom{aaaaa}
\label{QCaqrk}
\eeqn
for the cases when $k$ is a gluon, a quark, or an antiquark respectively.
Note that in the latter two cases the requirement that $k\in\gamma_p$
implies $k=-p$ and $k=\mu(-p-q)$ respectively. Precisely as in the case 
of gluon amplitudes, eqs.~(\ref{QCglu})--(\ref{QCaqrk}) suggest the
use of operators acting on flows. For consistency with what was done
before, the arguments of these operators will have to be the positions
of the particles that appear in flows. My conventions for such positions
are the following. The colour line of eq.~(\ref{qgflowp})
\beq
\gamma_p=\Big(\Mp\,;\sigma(t_{p-1}+1),\ldots\sigma(t_p);\mu(-p-q)\Big)
\eeq
corresponds to positions:
\beq
(\Mp\,;t_{p-1}+1,\ldots t_p;-p-q)\,.
\label{plist}
\eeq
This implies that $-p$ is adjacent to $t_{p-1}+1$, and that $-p-q$
is adjacent to $t_p$. It is therefore convenient to define the
operations $\oplus~1$ and $\ominus~1$, that will serve to move across 
the list in eq.~(\ref{plist}):
\beqn
i\oplus 1&=&\left\{
\begin{array}{ll}
t_{p-1}+1 &\phantom{aaaaaa}i=-p\,,\\
i+1  &\phantom{aaaaaa}t_{p-1}+1\le i<t_p\,,\\
-p-q &\phantom{aaaaaa}i=t_p\,,
\end{array}
\right.
\label{opdef}
\\
i\ominus 1&=&\left\{
\begin{array}{ll}
-p   &\phantom{aaaaaaaaa}i=t_{p-1}+1\,,\\
i-1  &\phantom{aaaaaaaaa}t_{p-1}+1<i\le t_p\,,\\
t_p  &\phantom{aaaaaaaaa}i=-p-q\,,
\end{array}
\right.
\label{omdef}
\\
i\oplus 1\ominus 1&=&i\ominus 1\oplus 1=i\,.
\\
i\oplus -1&=&i\ominus 1\,.
\eeqn
The definitions in eqs.~(\ref{opdef}) and~(\ref{omdef}) imply that
one can move continuously across a given colour line, but cannot
pass continuously from line $\gamma_p$ to line $\gamma_{p\pm 1}$.
This is consistent with the physical interpretation of colour lines,
which from the colour viewpoint are disconnected from each other,
and with the fact that when studying singularities at fixed flows
the emphasis is on adjacent particles. 
Equations~(\ref{plist})--(\ref{omdef}) are meant to hold for all 
colour lines $\gamma_p$ belonging to a given flow $\gamma$.
Note that in the case of eq.~(\ref{emptyline}), i.e. when there
are no gluons on colour line $\gamma_p$, $-p$ and $-p-q$ are
contiguous, and therefore
\beqn
\left.
\begin{array}{rl}
(-p)\oplus 1&=-p-q\\
(-p-q)\ominus 1&=-p
\end{array}
\phantom{aaaa}
\right\}&&
{\rm if~}t_{p-1}=t_p\,.
\eeqn
Finally, I shall denote by $\igamma(k)$ the position of particle
$k$ in flow $\gamma$ (exactly as $\isigma(k)$ denotes the position
of gluon $k$ in flow $\sigma$). Thus, eqs.~(\ref{QCglu})--(\ref{QCaqrk}) 
can be rewritten as follows:
\beqn
&&\sum_{c_k}\left(Q^{\anpo}(k)\right)_{a_kc_k}
\Lambda\left(\seta_{i\ne k},c_k,\gamma_p\right)=
\nonumber\\*&&\phantom{aaaaaaaaaa}
\Lambda\left(\seta,I_+(\igamma(k))\gamma_p\right)-
\Lambda\left(\seta,I_-(\igamma(k))\gamma_p\right)\,,
\phantom{aaaaa}
\label{QCglut}
\\
&&\sum_{c_k}\left(Q^{\anpo}(k)\right)_{a_kc_k}
\Lambda\left(\seta_{i\ne k},c_k,\gamma_p\right)=
\Lambda\left(\seta,I_+(\igamma(k))\gamma_p\right)\,,
\label{QCqrkt}
\\
&&\sum_{c_k}\left(Q^{\anpo}(k)\right)_{a_kc_k}
\Lambda\left(\seta_{i\ne k},c_k,\gamma_p\right)=
-\Lambda\left(\seta,I_-(\igamma(k))\gamma_p\right)\,.
\label{QCaqrkt}
\eeqn
As in the case of gluon amplitudes, the result of $I_+(i)$ ($I_+(i)$) 
acting on the list associated with the flow is that of inserting the 
number $n+1$  after (before) the $i^{th}$ member of the list.
Given the factorized form of the flow, eq.~(\ref{qgflow}), the 
operators $I_\pm(i)$ can be equivalently understood as acting on the colour 
flow $\gamma$, or on the colour line $\gamma_p$ which includes the $i^{th}$
position. The conventions adopted before imply that:
\beqn
I_+(\igamma(k))&=&I_+(-p)\;\;\;\;\;\;\;\;\;\phantom{-q}\;
k=-p\,,
\\
I_-(\igamma(k))&=&I_-(-p-q)\;\;\;\;\;\;\;\;
k=\mu(-p-q)\,.
\eeqn
Putting all this together, one obtains:
\beq
Y_{\gamma,p}=\sum_{i=t_{p-1}+1\ominus 1}^{t_p}
\Lambda\left(\seta,I_+(i)\gamma_p\right)\;
-\sum_{i=t_{p-1}+1}^{t_p\oplus 1}
\Lambda\left(\seta,I_-(i)\gamma_p\right)\,.
\label{Ypres}
\eeq
By construction, the analogue of eq.~(\ref{ImeqIp}) holds:
\beq
I_-(i)\gamma_p=I_+(i\ominus 1)\gamma_p\,,\;\;\;\;\;\;\;\;
t_{p-1}+1\le i\le t_p\oplus 1\,.
\label{ImeqIpQG}
\eeq
Hence
\beqn
\sum_{i=t_{p-1}+1}^{t_p\oplus 1}
\Lambda\left(\seta,I_-(i)\gamma_p\right)
&=&
\sum_{i=t_{p-1}+1}^{t_p\oplus 1}
\Lambda\left(\seta,I_+(i\ominus 1)\gamma_p\right)
\\*
&=&
\sum_{i=t_{p-1}+1\ominus 1}^{t_p}
\Lambda\left(\seta,I_+(i)\gamma_p\right)\,,
\label{mcont}
\eeqn
where in the last equation I relabeled the sum variable $i\to i\ominus 1$.
By plugging eq.~(\ref{mcont}) into eq.~(\ref{Ypres}) one gets finally
gets $Y_{\gamma,p}=0$. It is easy to convince oneself that the derivation
above also applies to the case of a colour line with no gluons attached,
since in such a case:
\beq
I_+(-p)\gamma_p=I_-(-p-q)\gamma_p\;\;\;\;\;\;\;\;
{\rm if~}t_{p-1}=t_p\,.
\eeq
To take this fact into account, one may perform the following
formal replacement:
\beq
\sum_{i=t_{p-1}+1\ominus 1}^{t_p}f(i)\;\;\;
\longrightarrow\;\;\;
\left(1-\delta_{t_{p-1}t_p}\right)
\sum_{i=t_{p-1}+1\ominus 1}^{t_p}f(i)
\;+\delta_{t_{p-1}t_p}\delta_{i(-p)}f(i)\,.
\label{zerocell}
\eeq
In practice, the replacement in eq.~(\ref{zerocell}) will always be 
understood.
Given that the arguments above apply to any $p$, one has indeed
proved eq.~(\ref{colconsQG}), and also that
\beq
\sum_{k=-2q}^n Q^b(k)\ket{\ampn(\gamma)}=0\,,
\label{colcons3QG}
\eeq
analogously to what happens in the case of gluon amplitudes.

\section{A simple example: $0\to u\ub d\db g$\label{sec:ex}}
\begin{figure}[htb!]
  \begin{center}
        \epsfig{file=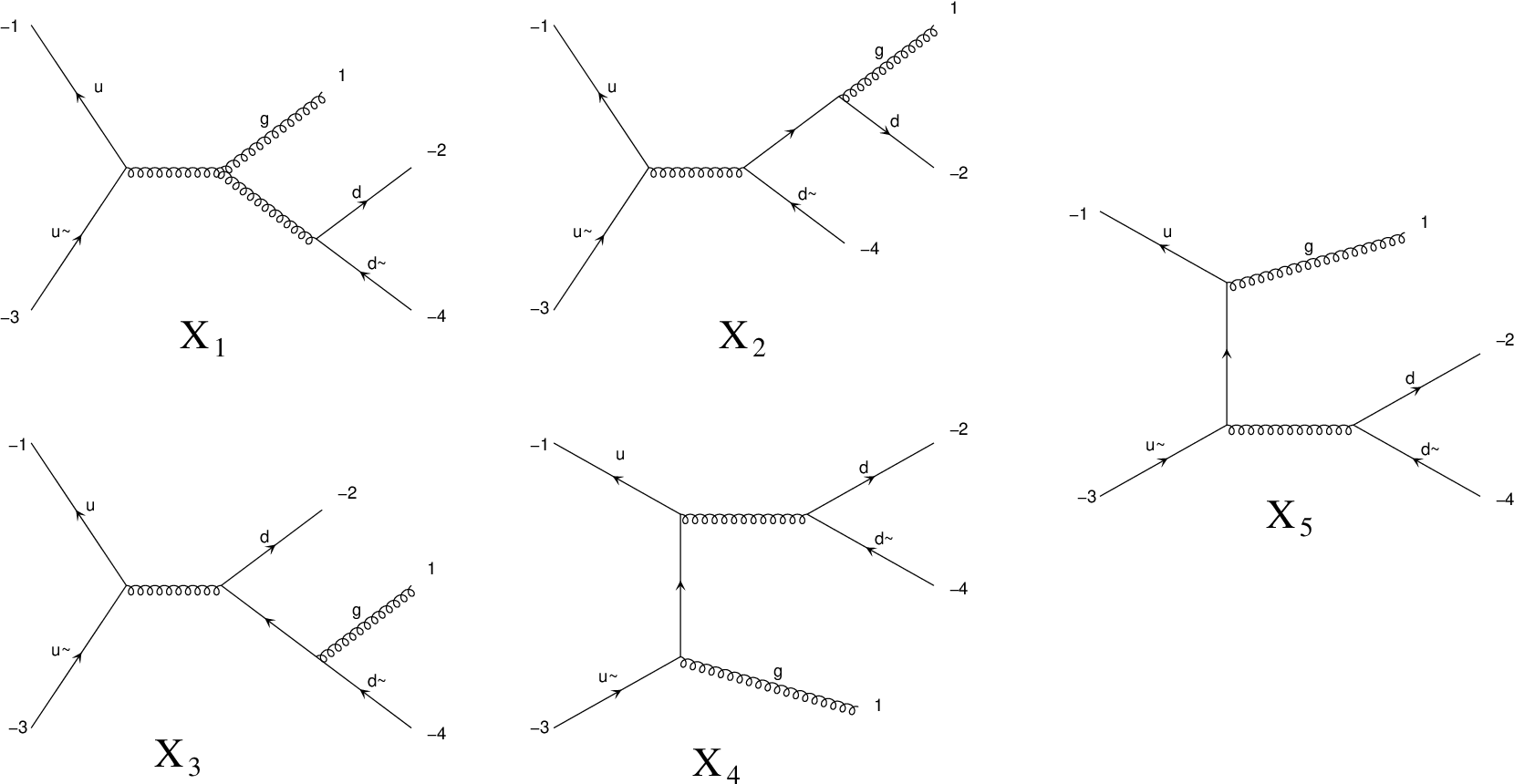, width=0.9\textwidth}
  \end{center}
  \vspace{-20pt}
  \caption{Diagrams that contribute to $0\to u\ub d\db g$.}
      \label{fig:diagX}
\end{figure}

In this appendix I shall illustrate the practical application of
some of the techniques presented in this paper, by considering
the real-emission process
\beq
0\to u(-1)\ub(-3) d(-2)\db(-4) g(1)\,,
\label{aprocR}
\eeq
where the parton labelling follows the conventions introduced
in sect.~\ref{sec:quarks}. The diagrams that contribute to the
process of eq.~(\ref{aprocR}) are depicted in fig.~\ref{fig:diagX}.
A straightforward calculation leads one to writing the amplitude,
whose general form is given in eq.~(\ref{mgCDampQG}), as follows:
\beq
\ampX=\sum_{i=1}^4 \Lambda(\Gamma_i)\ampXCS(\Gamma_i)\,,
\eeq
where the colour flows are
\beqn
&&\Gamma_1=(-1;-4)(-2;1;-3)\,,\phantom{aaaaaa}
\Gamma_2=(-1;1;-4)(-2;-3)\,,
\label{GF12}
\\
&&\Gamma_3=(-1;-3)(-2;1;-4)\,,\phantom{aaaaaa}
\Gamma_4=(-1;1;-3)(-2;-4)\,,
\label{GF34}
\eeqn
which, according to eq.~(\ref{Lambdagq}), correspond to:
\beqn
&&\Lambda(\Gamma_1)=\delta_{a_{-1}a_{-4}}
\lambda_{a_{-2}a_{-3}}^{a_1}\,,\phantom{aaaaaa\frac{1}{\NC}}
\Lambda(\Gamma_2)=\delta_{a_{-2}a_{-3}}
\lambda_{a_{-1}a_{-4}}^{a_1}\,,
\\
&&\Lambda(\Gamma_3)=\frac{1}{\NC}\delta_{a_{-1}a_{-3}}
\lambda_{a_{-2}a_{-4}}^{a_1}\,,\phantom{aaaaaa}
\Lambda(\Gamma_4)=\frac{1}{\NC}\delta_{a_{-2}a_{-4}}
\lambda_{a_{-1}a_{-3}}^{a_1}\,,
\eeqn
and the dual amplitudes are defined in terms of Feynman diagrams
(stripped of colour factors):
\beqn
&&2\ampXCS(\Gamma_1)=-iX_1+X_2+X_4\,,\phantom{aaaaaa}
2\ampXCS(\Gamma_2)=iX_1+X_3+X_5\,,
\label{XCS12}
\\
&&2\ampXCS(\Gamma_3)=-X_2-X_3\,,\phantom{aaaaaa-iX_1}
2\ampXCS(\Gamma_4)=-X_4-X_5\,.
\label{XCS34}
\eeqn
By direct computation, the colour-flow matrix of eq.~(\ref{CFQGmatdef})
\beq
C^{(X)}_{ij}\equiv C(\Gamma_i,\Gamma_j)
\eeq
turns out to read:
\beqn
&&C^{(X)}_{11}=C^{(X)}_{22}=LC\,,
\label{CXLC}
\\
&&C^{(X)}_{12}=C^{(X)}_{21}=C^{(X)}_{34}=C^{(X)}_{43}=0\,,
\label{CXzero}
\\
&&C^{(X)}_{ij}=SC\,,\phantom{aaaaaaaa~aaaaaaaaaaaa}
{\rm all~other~}i,j\,,
\label{CXSC}
\eeqn
with
\beq
LC=\left(\NC^3-\NC\right)/2\,,\phantom{aaaaaaaa}
SC=\NC/2-1/(2\NC)\,.
\label{LCdef}
\eeq
%%%%%%%%%%%%%%%%%%%%%%%%%%%%%%%%%%%%%%%%%%%%%%%%%%%%%%%%%%%%%%%%%%%%%%%%%
%%%%%%%%%%%%%%%%%%%%%%%%%%%%%%%%%%%%%%%%%%%%%%%%%%%%%%%%%%%%%%%%%%%%%%%%%
Therefore, only the closed flows $(\Gamma_1,\Gamma_1)$ and
$(\Gamma_2,\Gamma_2)$ give leading-colour contributions to the cross
section, the other closed flows being subleading in colour, and
associated with the same power of $\NC$. In turn, this suggests
to define:
\beq
p(\Gammap,\Gamma)=\#~{\rm colour~loops}-\rho(\Gammap)-\rho(\Gamma)\,,
\label{pdef}
\eeq
where the number of colour loops is that identified by the closed
flow $(\Gammap,\Gamma)$, and the function $\rho$ has been defined
in eq.~(\ref{rhodef}). The larger $p(\Gammap,\Gamma)$, the more
leading in colour the contribution of the closed flow $(\Gammap,\Gamma)$ 
to the cross section. In the example I am considering in this
appendix, one has $p=3$ for the closed flows in eq.~(\ref{CXLC}),
and $p=1$ for those in eq.~(\ref{CXSC}). I stress that the computation
of $p$ can be carried out without calculating traces of Gell-Mann
matrices, since both the number of colour loops and the value of $\rho$
can be obtained directly from the colour flows as given in
eqs.~(\ref{qgflow}) and~(\ref{qgflowp}).

%%%%%%%%%%%%%%%%%%%%%%%%%%%%%%%%%%%%%%%%%%%%%%%%%%%%%%%%%%%%%%%%%%%%%%%%%
\begin{figure}[htb!]
  \begin{center}
  \begin{minipage}{0.4\textwidth}
        \epsfig{file=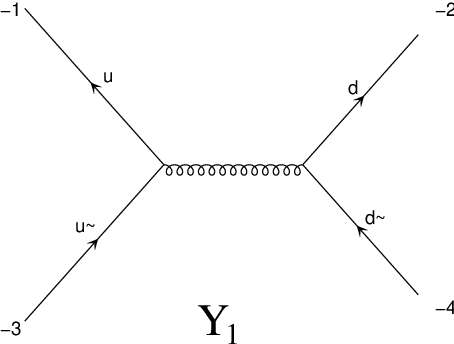, width=0.6\textwidth}
  \end{minipage}
  \begin{minipage}{0.59\textwidth}
        \epsfig{file=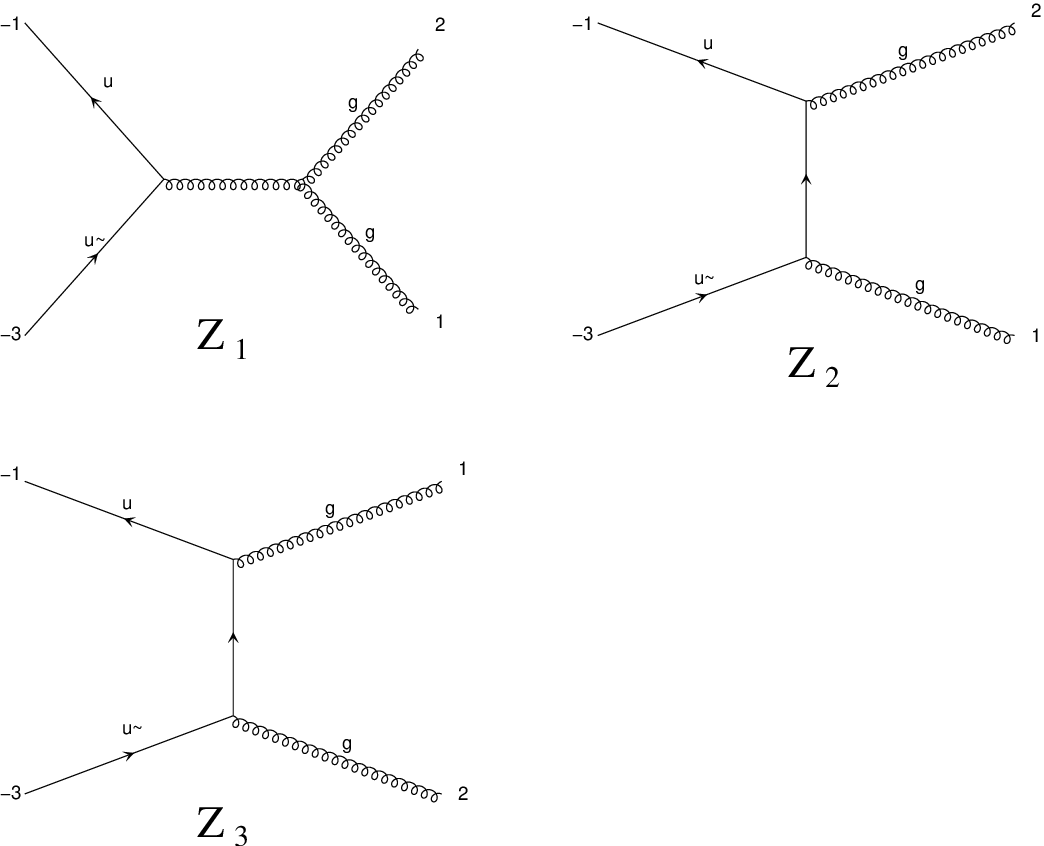, width=\textwidth}
  \end{minipage}
  \end{center}
  \caption{Diagrams that contribute to $0\to u\ub d\db$ and $0\to u\ub gg$.}
      \label{fig:diagYZ}
\end{figure}
%%%%%%%%%%%%%%%%%%%%%%%%%%%%%%%%%%%%%%%%%%%%%%%%%%%%%%%%%%%%%%%%%%%%%%%%%
I now turn to discussing the calculation of the soft and of the 
$g\!\parallel\! d$, $d\!\parallel\!\db$ collinear limits of the process of 
eq.~(\ref{aprocR}),
which are representative of all possible singular configurations. Since I
shall consider analytical results here, the treatment at fixed colour
configurations is not particularly illuminating, and I shall rather
concentrate on that at fixed (Born or real) colour flows.
The following underlying Born processes will be relevant for the
study of the limits:
\beqn
&&0\to u(-1)\ub(-3) d(-2)\db(-4)\,,
\label{aprocB1}
\\
&&0\to u(-1)\ub(-3) g(1) g(2)\,,
\label{aprocB2}
\eeqn
whose contributing Feynman diagrams are depicted in fig.~\ref{fig:diagYZ}.
The amplitude that corresponds to eq.~(\ref{aprocB1}) is:
\beq
\ampY=\sum_{i=1}^2 \Lambda(\gammaY_i)\ampYCS(\gammaY_i)\,,
\eeq
with:
\beqn
&&\gammaY_1=(-1;-4)(-2;-3)\,,\phantom{aaaaaa}
\gammaY_2=(-1;-3)(-2;-4)
\label{gBF12}
\\
&&\Lambda(\gammaY_1)=\delta_{a_{-1}a_{-4}}
\delta_{a_{-2}a_{-3}}\,,\phantom{aaaaaa}
\Lambda(\gammaY_2)=\frac{1}{\NC}\delta_{a_{-1}a_{-3}}
\delta_{a_{-2}a_{-4}}\,,
\\
&&2\ampYCS(\gammaY_1)=Y_1\,,\phantom{aaaaaaaaaaaaaa}\,
2\ampYCS(\gammaY_2)=-Y_1\,.
\eeqn
Hence, the colour-flow matrix:
\beq
C^{(Y)}_{ij}\equiv C(\gammaY_i,\gammaY_j)
\eeq
reads:
\beqn
&&C^{(Y)}_{11}=\NC^2\,,
\label{CY11}
\\
&&C^{(Y)}_{12}=C^{(Y)}_{22}=1\,,
\label{CY12}
\eeqn
consistently with the fact that $p(\gammaY_1,\gammaY_1)=2$,
and $p(\gammaY_1,\gammaY_2)=p(\gammaY_2,\gammaY_2)=0$.

In the case of the process of eq.~(\ref{aprocB2}) the amplitude reads:
\beq
\ampZ=\sum_{i=1}^2 \Lambda(\gammaZ_i)\ampZCS(\gammaZ_i)\,,
\eeq
with:
\beqn
&&\gammaZ_1=(-1;1,2;-3)\,,\phantom{aaaaaaaaaa}
\gammaZ_2=(-1;2,1;-3)\,,
\\
&&\Lambda(\gammaZ_1)=
\left(\lambda^{a_1}\lambda^{a_2}\right)_{a_{-1}a_{-3}}\,,\phantom{aaaaaa}
\Lambda(\gammaZ_2)=
\left(\lambda^{a_2}\lambda^{a_1}\right)_{a_{-1}a_{-3}}\,,
\\
&&\ampZCS(\gammaZ_1)=Z_3-iZ_1\,,\phantom{aaaaaaaaaa}\;
\ampZCS(\gammaZ_2)=Z_2+iZ_1\,.
\label{ZCS12}
\eeqn
Hence, the colour-flow matrix:
\beq
C^{(Z)}_{ij}\equiv C(\gammaZ_i,\gammaZ_j)
\eeq
reads:
\beqn
&&C^{(Z)}_{11}=C^{(Z)}_{22}=\NC^3/4-\NC/2+1/(4\NC)\,,
\\
&&C^{(Z)}_{12}=-\NC/4+1/(4\NC)\,,
\eeqn
consistently with the fact that 
$p(\gammaZ_1,\gammaZ_1)=p(\gammaZ_2,\gammaZ_2)=3$
and $p(\gammaZ_1,\gammaZ_2)=1$.

~\\
\noindent
{\em Soft limit, fixed real flows.} This is given in eq.~(\ref{MQGssMpp2}).
I list below the result for each closed flow, omitting the factors
$-\gs^2 C^{(X)}_{ij}$. One obtains\footnote{Because of eq.~(\ref{Mflowcplx}),
I shall consider here explicitly only closed flows $(\Gamma_i,\Gamma_j)$
with $i\le j$.}:
\beqn
&&(\Gamma_1,\Gamma_1)\;\;\;\longrightarrow\;\;\;
-2\,\eik{-2}{-3}\;\ampYCS(\gammaY_1)^\star\ampYCS(\gammaY_1)\,,
\label{SfFR11}
\\
&&(\Gamma_2,\Gamma_2)\;\;\;\longrightarrow\;\;\;
-2\,\eik{-1}{-4}\;\ampYCS(\gammaY_1)^\star\ampYCS(\gammaY_1)\,,
\label{SfFR22}
\\
&&(\Gamma_1,\Gamma_3)\;\;\;\longrightarrow\;\;\;
\left(-\eik{-2}{-4}-\eik{-2}{-3}+\eik{-3}{-4}\right)
\ampYCS(\gammaY_1)^\star\ampYCS(\gammaY_2)\,,
\label{SfFR13}
\\
&&(\Gamma_1,\Gamma_4)\;\;\;\longrightarrow\;\;\;
\left(\eik{-1}{-2}-\eik{-2}{-3}-\eik{-1}{-3}\right)
\ampYCS(\gammaY_1)^\star\ampYCS(\gammaY_2)\,,
\label{SfFR14}
\\
&&(\Gamma_2,\Gamma_3)\;\;\;\longrightarrow\;\;\;
\left(\eik{-1}{-2}-\eik{-1}{-4}-\eik{-2}{-4}\right)
\ampYCS(\gammaY_1)^\star\ampYCS(\gammaY_2)\,,
\label{SfFR23}
\\
&&(\Gamma_2,\Gamma_4)\;\;\;\longrightarrow\;\;\;
\left(-\eik{-1}{-3}-\eik{-1}{-4}+\eik{-3}{-4}\right)
\ampYCS(\gammaY_1)^\star\ampYCS(\gammaY_2)\,,
\label{SfFR24}
\\
&&(\Gamma_3,\Gamma_3)\;\;\;\longrightarrow\;\;\;
-2\,\eik{-2}{-4}\;\ampYCS(\gammaY_2)^\star\ampYCS(\gammaY_2)\,,
\label{SfFR33}
\\
&&(\Gamma_4,\Gamma_4)\;\;\;\longrightarrow\;\;\;
-2\,\eik{-1}{-3}\;\ampYCS(\gammaY_2)^\star\ampYCS(\gammaY_2)\,,
\label{SfFR44}
\eeqn
where the eikonal factor $[k,l]$ has been introduced in eq.~(\ref{eikdef}).
According to eqs.~(\ref{CXLC})--(\ref{LCdef}), at the leading colour
only eq.~(\ref{SfFR11}) and~(\ref{SfFR22}) contribute to the soft limit,
the other closed flows being colour-suppressed (or identically equal
to zero). 

~\\
\noindent
{\em Collinear limit $g\!\parallel\! d$, fixed real flows.} This is given
in eq.~(\ref{MQGcollFR}). In that equation, the kernels are flow-independent;
hence, the only flow-specific information is contained in 
$\delta(\Gammap,\Gamma)$, and in the reduced matrix elements.
Using the definition of the former, given in eq.~(\ref{deltaQGdef}),
it is easy to obtain the following list of closed flows which give
a non-zero contribution in the collinear limit:
\beqn
&&(\Gamma_1,\Gamma_1)\;\;\;\longrightarrow\;\;\;
\ampYCS(\gammaY_1)^\star\ampYCS(\gammaY_1)\,,
\label{ClFR11}
\\
&&(\Gamma_1,\Gamma_3)\;\;\;\longrightarrow\;\;\;
\ampYCS(\gammaY_1)^\star\ampYCS(\gammaY_2)\,,
\label{ClFR13}
\\
&&(\Gamma_3,\Gamma_3)\;\;\;\longrightarrow\;\;\;
\ampYCS(\gammaY_2)^\star\ampYCS(\gammaY_2)\,,
\label{ClFR33}
\eeqn
where I have also indicated the reduced amplitudes that enter
the limit of each closed-flow amplitude squared. Equation~(\ref{ClFR11})
is the only leading-colour contribution. As was already discussed at
length in the main text, eqs.~(\ref{SfFR11})--(\ref{SfFR44}) are
consistent with eqs.~(\ref{ClFR11})--(\ref{ClFR33}) in the sense 
of eq.~(\ref{MSCdef}). Here, the reader can check with an explicit
example what has been claimed after eq.~(\ref{eiksum}) (which applies
identically to the case of quark-gluon amplitudes). Namely, that 
despite the fact that individual eikonal factors may diverge in
a given collinear limit, the corresponding interferences of amplitudes 
may still be non-divergent in that limit. This is the case here
for eqs.~(\ref{SfFR14}) and~(\ref{SfFR23}), which do not have
a $g\!\parallel\! d$ collinear divergence owing to the cancellations
that occur among pairs of eikonals which contain the momentum
of $d$, and differ by the sign in front of them.

~\\
\noindent
{\em Collinear limit $d\!\parallel\!\db$, fixed real flows.} This is given
in eq.~(\ref{MQGcollqqfinal}). Also in this case, the kernels are 
flow-independent; the only flow-specific information one needs to
derive is which closed flows give a non-zero contribution, and the
reduced matrix elements they are associated with. In order to do
this, one first determines to which class of flows the real-emission
ones belong. Using the definitions given in eqs.~(\ref{flowJdef})
and~(\ref{flowKdef}), one obtains:
\beqn
&&\Gamma_1\,,\Gamma_2\;\in\;{\cal F}^{(J)}\,,
\\
&&\Gamma_4\;\in\;{\cal F}^{(K)}\,,
\\
&&\Gamma_3\;\notin\;{\cal F}^{(J)}\cup{\cal F}^{(K)}\,.
\eeqn
Using the definitions of the $J$ and $K$ operators,
eqs.~(\ref{Jrdef}) and~(\ref{Krdef}), it is easy to see that:
\beqn
&&J\gammaZ_1=\Gamma_2\,,\phantom{aaaaaaaa}\,
J\gammaZ_2=\Gamma_1\,,
\label{JonZ}
\\
&&K\gammaZ_1=\Gamma_4\,,\phantom{aaaaaaaa}
K\gammaZ_2=\Gamma_4\,.
\label{KonZ}
\eeqn
Hence, from eqs.~(\ref{MredJJsc})--(\ref{MredKKsc}):
\beqn
&&(\Gamma_1,\Gamma_1)\;\;\;\longrightarrow\;\;\;
\ampZCS(\gammaZ_2)^\star\ampZCS(\gammaZ_2)\,,
\label{CqqFR11}
\\
&&(\Gamma_2,\Gamma_2)\;\;\;\longrightarrow\;\;\;
\ampZCS(\gammaZ_1)^\star\ampZCS(\gammaZ_1)\,,
\label{CqqFR22}
\\
&&(\Gamma_1,\Gamma_4)\;\;\;\longrightarrow\;\;\;
-\ampZCS(\gammaZ_2)^\star
\left(\ampZCS(\gammaZ_1)+\ampZCS(\gammaZ_2)\right)\,,
\label{CqqFR14}
\\
&&(\Gamma_2,\Gamma_4)\;\;\;\longrightarrow\;\;\;
-\ampZCS(\gammaZ_1)^\star
\left(\ampZCS(\gammaZ_1)+\ampZCS(\gammaZ_2)\right)\,,
\label{CqqFR24}
\\
&&(\Gamma_4,\Gamma_4)\;\;\;\longrightarrow\;\;\;
\left(\ampZCS(\gammaZ_1)+\ampZCS(\gammaZ_2)\right)^\star
\left(\ampZCS(\gammaZ_1)+\ampZCS(\gammaZ_2)\right)\,,
\label{CqqFR44}
\eeqn
where only eqs.~(\ref{CqqFR11}) and~(\ref{CqqFR22}) contribute
at the leading colour.

~\\
\noindent
{\em Soft limit, fixed Born flows.} One starts by constructing the
linear combinations of matrix elements with well-defined limits,
which is equivalent to determining the sets defined in eq.~(\ref{xiset}).
Using eq.~(\ref{GF12}), ~(\ref{GF34}), and~(\ref{gBF12}), one obtains:
\beqn
\xi\!\left(\gammaY_1\right)&=&\left\{\Gamma_1,\Gamma_2\right\},
\\
\xi\!\left(\gammaY_2\right)&=&\left\{\Gamma_3,\Gamma_4\right\}.
\eeqn
Therefore, the relevant matrix elements will be (see eq.~(\ref{MQGnpoBorn})):
\beqn
\ampsq\left(\gammaY_1,\gammaY_1\right)&=&
\ampsq(\Gamma_1,\Gamma_1)+\ampsq(\Gamma_2,\Gamma_2)\,,
\label{m11}
\\
\ampsq\left(\gammaY_1,\gammaY_2\right)&=&
\ampsq(\Gamma_1,\Gamma_3)+\ampsq(\Gamma_1,\Gamma_4)+
\ampsq(\Gamma_2,\Gamma_3)+\ampsq(\Gamma_2,\Gamma_4)\,,
\label{m12}
\\
\ampsq\left(\gammaY_2,\gammaY_2\right)&=&
\ampsq(\Gamma_3,\Gamma_3)+\ampsq(\Gamma_4,\Gamma_4)\,,
\label{m22}
\eeqn
where I have exploited the fact that $\ampsq(\Gamma_1,\Gamma_2)=0$,
$\ampsq(\Gamma_3,\Gamma_4)=0$ owing to eq.~(\ref{CXzero}).
According to eqs.~(\ref{CY11}) and~(\ref{CY12}), only 
eq.~(\ref{m11}) will give a leading-colour contribution.
The soft limits of these quantities can be computed using
eqs. eq.~(\ref{MQGsoftdef2}) and~(\ref{MQGkldef2sc}).
One obtains:
\beqn
&&\left(\gammaY_1,\gammaY_1\right)\;\;\;\longrightarrow\;\;\;
-LC\,\left(\eik{-1}{-4}+\eik{-2}{-3}\right)\,,
\label{SfFB11}
\\
&&\left(\gammaY_1,\gammaY_2\right)\;\;\;\longrightarrow\;\;\;
-SC\,\Big(\eik{-1}{-2}-\eik{-1}{-3}-\eik{-1}{-4}
\nonumber\\*&&
\phantom{\left(\gammaY_1,\gammaY_2\right)\;\;\;\longrightarrow\;\;\;}
\phantom{-SC}
-\eik{-2}{-3}-\eik{-2}{-4}+\eik{-3}{-4}\Big)\,,
\label{SfFB12}
\\
&&\left(\gammaY_2,\gammaY_2\right)\;\;\;\longrightarrow\;\;\;
-SC\,\left(\eik{-1}{-3}+\eik{-2}{-4}\right)\,,
\label{SfFB22}
\eeqn
where I have omitted overall terms 
\mbox{$\gs^2\,\ampYCS(\gammaY_i)^\star\ampYCS(\gammaY_j)$}, and 
$LC$ and $SC$ are given in eq.~(\ref{LCdef}). It is immediate
to see that the results of eqs.~(\ref{SfFB11})--(\ref{SfFB22})
are consistent with those of eqs.~(\ref{SfFR11})--(\ref{SfFR44}).
It should also be stressed that the fact that a single colour
factor factorizes in eqs.~(\ref{SfFB11})--(\ref{SfFB22}) is
accidental, and due to the relative simplicity of the process
being considered here. In general, at fixed Born flows different
(combinations of) eikonals will be multiplied by different colour
factors; the only way in which the colour fully factorizes in the
infrared limits is by fixing real flows.

~\\
\noindent
{\em Collinear limit $g\!\parallel\! d$, fixed Born flows.} As one
can see from eqs.~(\ref{MQGcollFB2})--(\ref{MtQGcolldefzsc}), this
case is trivial: the same closed flow appears on both sides of
eq.~(\ref{MQGcollFB2}). Note that, while all matrix elements
in eqs.~(\ref{m11})--(\ref{m22}) will be collinear divergent,
this behaviour is driven by the first term on each of the r.h.s.'s
of these equations -- this has been discussed in general in
the main text (see e.g.~eq.~(\ref{xisetC}) for quark-gluon amplitudes).

~\\
\noindent
{\em Collinear limit $d\!\parallel\!\db$, fixed Born flows.} As was
discussed in sect.~\ref{sec:flowqgg}, one cannot treat a $g\to q\qb$ 
branching by fixing the Born flows. The present example allows one
to further this point. The problem arises from the fact that the $K$
operator has a multi-valued inverse, see eq.~(\ref{KonZ}) -- it therefore
becomes ambiguous whether to associate the contribution due to $\Gamma_4$ 
with that of $\gammaZ_1=\iK_1\Gamma_4$ or that of $\gammaZ_2=\iK_2\Gamma_4$.
To proceed, one must look at the diagrammatic level. From eq.~(\ref{XCS34})
and fig.~\ref{fig:diagX}, one sees that diagrams $X_4$ and $X_5$, that
contribute to $\ampXCS(\Gamma_4)$, correspond in the collinear limit 
to diagrams $Z_2$ and $Z_3$ of fig.~\ref{fig:diagYZ}, respectively.
Now, $Z_2$ enters $\ampZCS(\gammaZ_2)$, and $Z_3$ enters
$\ampZCS(\gammaZ_1)$ (see eq.~(\ref{ZCS12})). Therefore, in order
to match the collinear behaviour, the $X_4$ and $X_5$ parts of
$\ampXCS(\Gamma_4)$ should be assigned to $\ampZCS(\gammaZ_2)$
and $\ampZCS(\gammaZ_1)$ respectively, which is not a gauge-invariant
procedure.

\end{document}